\pdfoutput=1

\documentclass[11pt,twoside,a4paper,cmspaper,final,collab]{cms-tdr}
\def\svnVersion{494689P}\def\cmsCernNoTag{CERN-EP-2018-263}\def\cmsCernDate{\today}\def\cmsMessage{Published in the European Physical Journal C as \href{http://dx.doi.org/10.1140/epjc/s10052-019-6909-y}{\doi{10.1140/epjc/s10052-019-6909-y.}}}

\begin{document}\cmsNoteHeader{HIG-17-031}

\hyphenation{had-ron-i-za-tion}
\hyphenation{cal-or-i-me-ter}
\hyphenation{de-vices}
\RCS$HeadURL: svn+ssh://svn.cern.ch/reps/tdr2/papers/HIG-17-031/trunk/HIG-17-031.tex $
\RCS$Id: HIG-17-031.tex 494139 2019-04-28 11:39:13Z alverson $

\newlength\cmsFigWidth
\ifthenelse{\boolean{cms@external}}{\setlength\cmsFigWidth{0.85\columnwidth}}{\setlength\cmsFigWidth{0.4\textwidth}}
\ifthenelse{\boolean{cms@external}}{\providecommand{\cmsLeft}{upper\xspace}}{\providecommand{\cmsLeft}{left\xspace}}
\ifthenelse{\boolean{cms@external}}{\providecommand{\cmsRight}{lower\xspace}}{\providecommand{\cmsRight}{right\xspace}}
\providecommand{\NA}{\ensuremath{\text{---}}\xspace}
\newlength\cmsTabSkip\setlength{\cmsTabSkip}{1ex}

\newcommand{\pT}{\pt}
\newcommand{\mH}{\ensuremath{m_{\PH}}\xspace}
\newcommand{\mll}{\ensuremath{m_{\ell\ell}}\xspace}
\newcommand{\bb}{\ensuremath{\PQb\PQb}\xspace}
\newcommand{\hbb}{\ensuremath{\PH\to\bb}\xspace}
\newcommand{\hinv}{\ensuremath{\PH\to\text{invisible}}\xspace}
\newcommand{\pTH}{\ensuremath{\pT(\PH)}\xspace}
\newcommand{\pTV}{\ensuremath{\pT(\mathrm{V})}\xspace}
\newcommand{\wlnbb}{\ensuremath{\PW(\ell\cPgn)\PH(\bb)}\xspace}
\newcommand{\zllbb}{\ensuremath{\cPZ(\ell\ell)\PH(\bb)}\xspace}
\newcommand{\znnbb}{\ensuremath{\cPZ(\cPgn\cPgn)\PH(\bb)}\xspace}
\newcommand{\tautau}{\ensuremath{\Pgt\Pgt}\xspace}
\newcommand{\mumu}{\ensuremath{\Pgm\Pgm}\xspace}
\newcommand{\htt}{\ensuremath{\PH\to\tautau}\xspace}
\newcommand{\hmm}{\ensuremath{\PH\to\mumu}\xspace}
\newcommand{\taue}{\ensuremath{\Pe}\xspace}
\newcommand{\taum}{\ensuremath{\Pgm}\xspace}
\newcommand{\thth}{\ensuremath{\tauh\tauh}\xspace}
\newcommand{\tetm}{\ensuremath{\taue\taum}\xspace}
\newcommand{\teth}{\ensuremath{\taue\tauh}\xspace}
\newcommand{\tmth}{\ensuremath{\taum\tauh}\xspace}
\newcommand{\gamgam}{\ensuremath{\PGg\PGg}\xspace}
\newcommand{\hgg}{\ensuremath{\PH\to\gamgam}\xspace}
\newcommand{\zz}{\ensuremath{\cPZ\cPZ}\xspace}
\newcommand{\hzz}{\ensuremath{\PH\to\zz}\xspace}
\newcommand{\hzzllll}{\ensuremath{\hzz^{(\ast)}\to 4\ell}\xspace}
\newcommand{\ww}{\ensuremath{\PW\PW}\xspace}
\newcommand{\hww}{\ensuremath{\PH\to\ww}\xspace}
\newcommand{\hwwlnln}{\ensuremath{\hww^{(\ast)}\to\ell\cPgn\ell\cPgn}\xspace}
\newcommand{\emu}{\ensuremath{\Pe\PGm}\xspace}
\newcommand{\mue}{\ensuremath{\PGm\Pe}\xspace}
\newcommand{\mmmm}{\ensuremath{4\Pgm}\xspace}
\newcommand{\eeee}{\ensuremath{4\Pe}\xspace}
\newcommand{\eemm}{\ensuremath{2\Pe2\Pgm}\xspace}
\newcommand{\mmee}{\ensuremath{2\Pgm2\Pe}\xspace}
\newcommand{\tth}{\ensuremath{\cPqt\cPqt\PH}\xspace}
\newcommand{\tH}{\ensuremath{\cPqt\PH}\xspace}
\newcommand{\tHW}{\ensuremath{\cPqt\PH\PW}\xspace}
\newcommand{\tHq}{\ensuremath{\cPqt\PH\Pq}\xspace}
\newcommand{\tthbb}{\ensuremath{\cPqt\cPqt\PH(\bb)}\xspace}
\newcommand{\vhbb}{\ensuremath{\mathrm{V}\PH(\bb)}\xspace}
\newcommand{\hlep}{\ensuremath{\PH\to\text{leptons}}\xspace}
\newcommand{\ggh}{\ensuremath{\Pg\Pg\PH}\xspace}
\newcommand{\vbf}{\ensuremath{\mathrm{VBF}}\xspace}
\newcommand{\vh}{\ensuremath{\mathrm{V}\PH}\xspace}
\newcommand{\wh}{\ensuremath{\PW\PH}\xspace}
\newcommand{\zh}{\ensuremath{\cPZ\PH}\xspace}
\newcommand{\ggZH}{\ensuremath{\Pg\Pg\cPZ\PH}\xspace}
\newcommand{\kV}{\ensuremath{\kappa_{\mathrm{V}}}\xspace}
\newcommand{\BR}{\ensuremath{\mathcal{B}}\xspace}
\newcommand{\SM}{\ensuremath{\mathrm{SM}}\xspace}
\newcommand{\BRbsm}{\ensuremath{\BR_{\mathrm{BSM}}}\xspace}
\newcommand{\BRinv}{\ensuremath{\BR_{\text{inv}}}\xspace}
\newcommand{\BRundet}{\ensuremath{\BR_{\text{undet}}}\xspace}
\newcommand{\hzzD}{\ensuremath{\mathcal{D}^{\text{kin}}_\text{bkg}}\xspace}
\renewcommand{\arraystretch}{1.1}
\newcommand{\QQHLNU}{\ensuremath{\sigma_{\PH+\PW(\ell\nu)}}\xspace}
\newcommand{\VHHQQ}{\ensuremath{\sigma_{\PH+\mathrm{V}(\mathrm{qq})}}\xspace}
\newcommand{\STggH}{\ensuremath{\sigma_{\ggh+\bb\PH}\xspace}}
\newcommand{\STqqH}{\sigma_{\vbf}\xspace}
\newcommand{\STttH}{\sigma_{\tth+\PQt\PH}\xspace}
\newcommand{\QQGGHLL}{\ensuremath{\sigma_{\PH+\PZ(\ell\ell/\nu\nu)}}\xspace}

\providecommand{\cmsTable}[1]{\resizebox{\textwidth}{!}{#1}}
\ifthenelse{\boolean{cms@external}}{\providecommand{\cmsTableX}[2]{#2}}{\providecommand{\cmsTableX}[2]{\resizebox{#1}{!}{#2}}}

\cmsNoteHeader{HIG-17-031}
\title{Combined measurements of Higgs boson couplings in proton-proton collisions at $\sqrt{s}=13\TeV$}

\date{\today}

\abstract{
Combined measurements of the production and decay rates of the Higgs boson, as well as its couplings to vector bosons and fermions, are presented. The analysis uses the LHC proton-proton collision data set recorded with the CMS detector in 2016 at $\sqrt{s}=13\TeV$, corresponding to an integrated luminosity of 35.9$\fbinv$. The combination is based on analyses targeting the five main Higgs boson production mechanisms (gluon fusion, vector boson fusion, and associated production with a $\PW$ or $\cPZ$ boson, or a top quark-antiquark pair) and the following decay modes: $\PH\to\PGg\PGg$, $\cPZ\cPZ$, $\PW\PW$, $\Pgt\Pgt$, $\PQb\PQb$, and $\Pgm\Pgm$. Searches for invisible Higgs boson decays are also considered. The best-fit ratio of the signal yield to the standard model expectation is measured to be $\mu=1.17\pm0.10$, assuming a Higgs boson mass of $125.09\GeV$. Additional results are given for various assumptions on the scaling behavior of the production and decay modes, including generic parametrizations based on ratios of cross sections and branching fractions or couplings. The results are compatible with the standard model predictions in all parametrizations considered. In addition, constraints are placed on various two Higgs doublet models.
}

\hypersetup{
pdfauthor={CMS Collaboration},
pdftitle={Combined measurements of Higgs boson couplings in proton-proton collisions at sqrt s = 13 TeV},
pdfsubject={CMS},
pdfkeywords={CMS, Higgs}}

\maketitle

\section{Introduction}
\label{sec:intro}

Understanding the mechanism behind electroweak symmetry breaking (EWSB) remains one of the main objectives of the physics program
at the CERN LHC. In the standard model (SM) of particle physics~\cite{Glashow1961579,Weinberg19671264,Salam1968367,tHooft:1972fi}, EWSB
is realized through the addition of a complex scalar doublet field. A salient feature of this is the prediction of one physical,
neutral, scalar particle,
the Higgs boson (\PH)~\cite{Englert:1964et,Higgs:1964ia,Higgs:1964pj,Guralnik:1964eu,Higgs:1966ev,Kibble:1967sv}.
The Higgs scalar field can also account for the fermion masses through Yukawa interactions~\cite{Weinberg19671264,Nambu:1961fr}.
The Higgs boson was discovered by the ATLAS and CMS Collaborations~\cite{Aad:2012tfa,Chatrchyan:2012ufa,CMSLong2013}, and is the subject of much study.
The Yukawa coupling strengths are free parameters in the SM and do not explain the observed pattern of fermion masses.
Furthermore, it is not understood why the Higgs boson mass is
near the electroweak scale, since it is not protected in the SM from large quantum
corrections~\cite{PhysRevD.3.1818,PhysRevD.14.1667,WEINBERG1979387,tHooft:1979rat,PhysRevD.20.2619}. This has led to the
development of many beyond the SM (BSM) theories that can alter the properties of the
Higgs boson~\cite{DIMOPOULOS1981150,WITTEN1981513,ARKANIHAMED1998263,PhysRevLett.83.3370,ARKANIHAMED2001232}.
Precision measurements of the properties of the Higgs boson are therefore an important test of the SM.

\begin{figure*}[hbtp]
\centering
\includegraphics[width=0.4\textwidth]{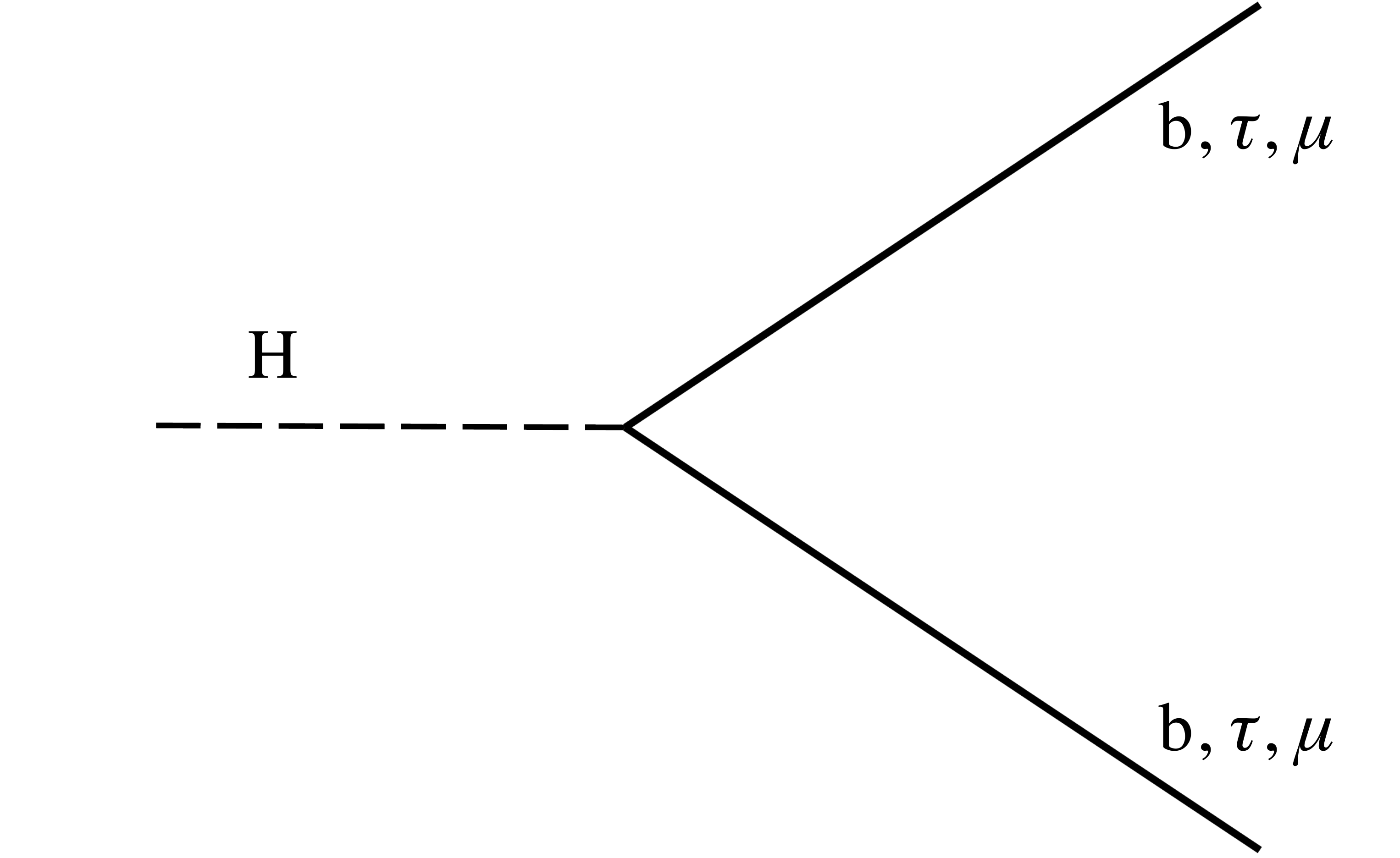}
\includegraphics[width=0.4\textwidth]{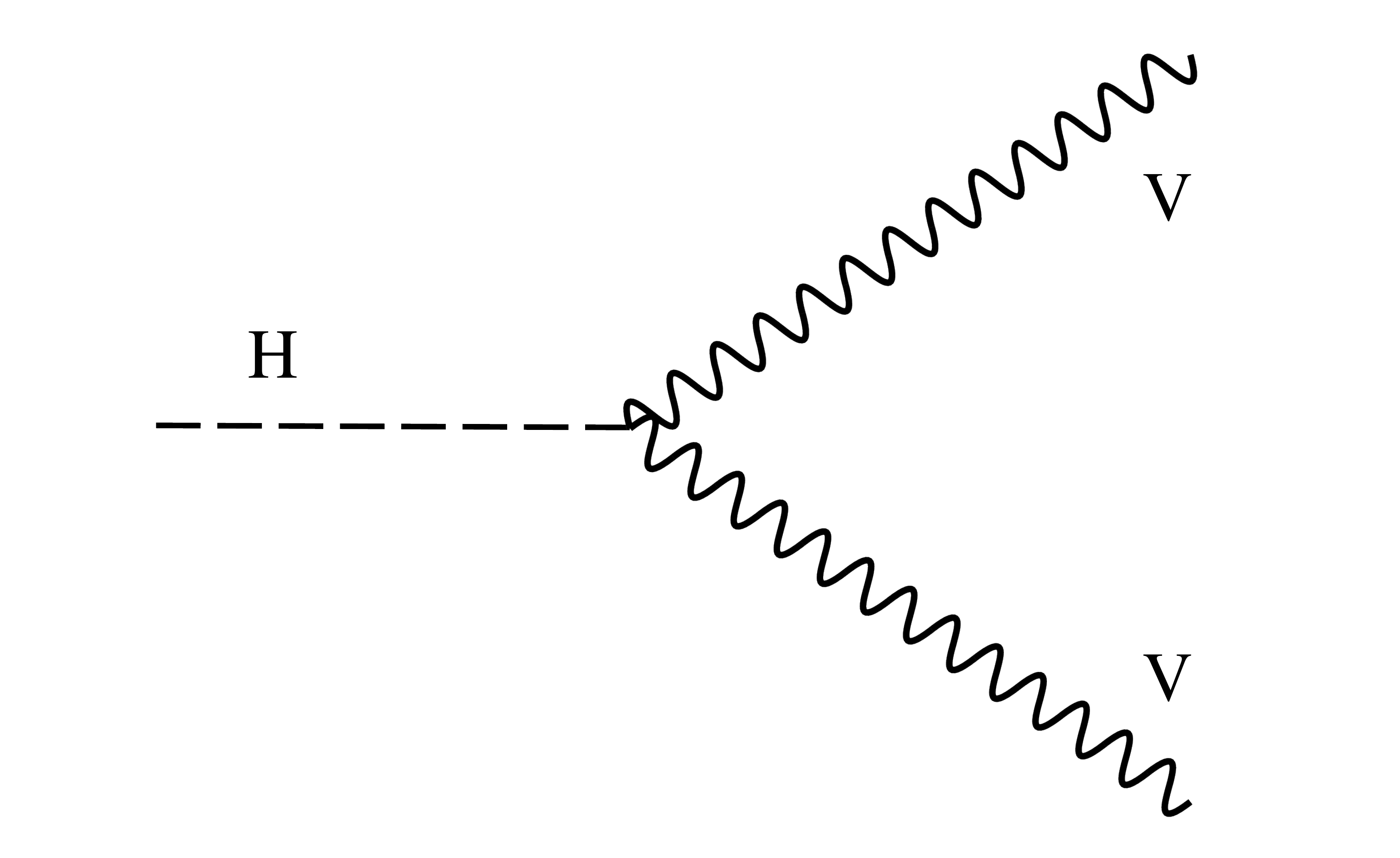}\\
\includegraphics[width=0.4\textwidth]{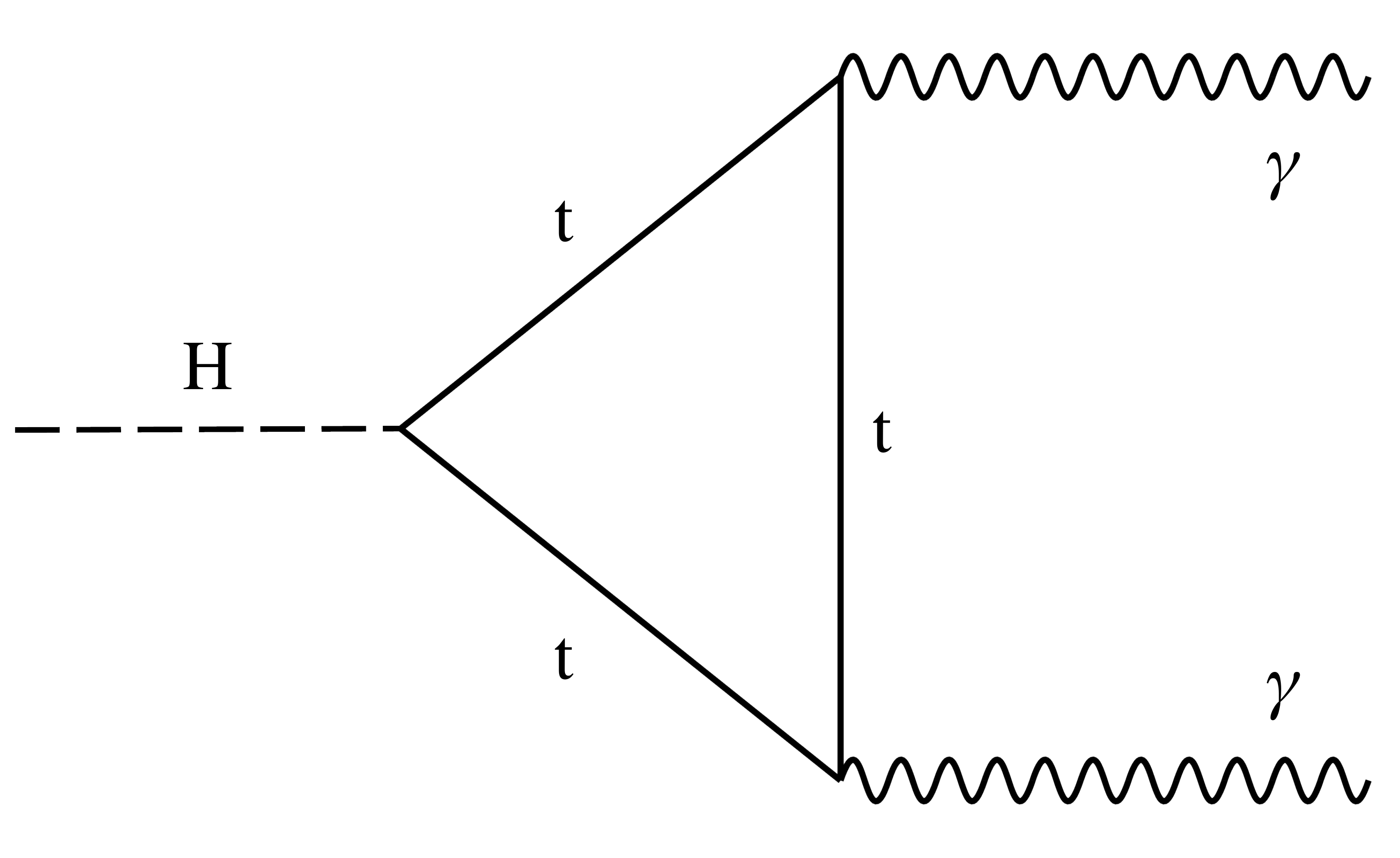}
\includegraphics[width=0.4\textwidth]{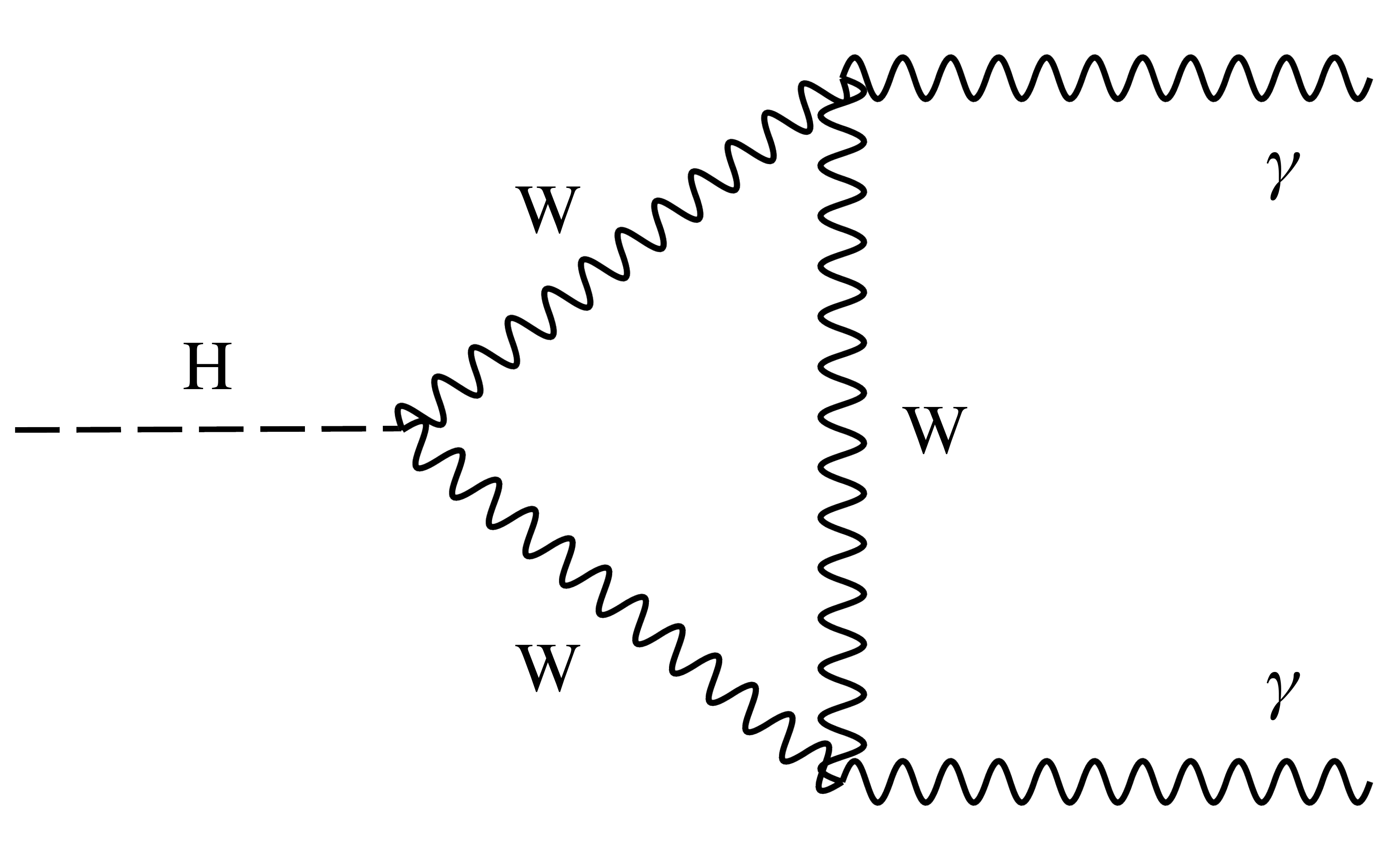}
\caption[]{Examples of leading-order Feynman diagrams for Higgs boson decays in the
\hbb, \htt, and \hmm (upper left);
 \hzz and \hww (upper right);
and \hgg (lower) channels. }
\label{fig:decay}
\end{figure*}

This paper describes combined measurements of the Higgs boson production rates, decay rates, and couplings using
analyses of $\sqrt{s} = 13\TeV$ proton--proton collision data recorded with the CMS detector in 2016. The data set corresponds to an integrated
luminosity of 35.9\fbinv. The following decay channels are included in the combination:
\hgg, \hzz, \hww, \htt, \hbb, and \hmm,
as shown in Fig.~\ref{fig:decay}.
Here and in what follows, we do not distinguish between particles and antiparticles in our notations of production and decay processes.
Searches for invisible decays of the Higgs boson, which are predicted to be considerably enhanced
by several BSM theories~\cite{Belanger:2001am,Giudice:2000av,Dominici:2009pq,Bonilla:2015uwa}, are also considered for selected measurements.
The data samples considered for each decay channel are ensured to have negligible overlap to avoid introducing nontrivial correlations.

\begin{figure*}[hbtp!]
\centering
\includegraphics[width=0.4\textwidth]{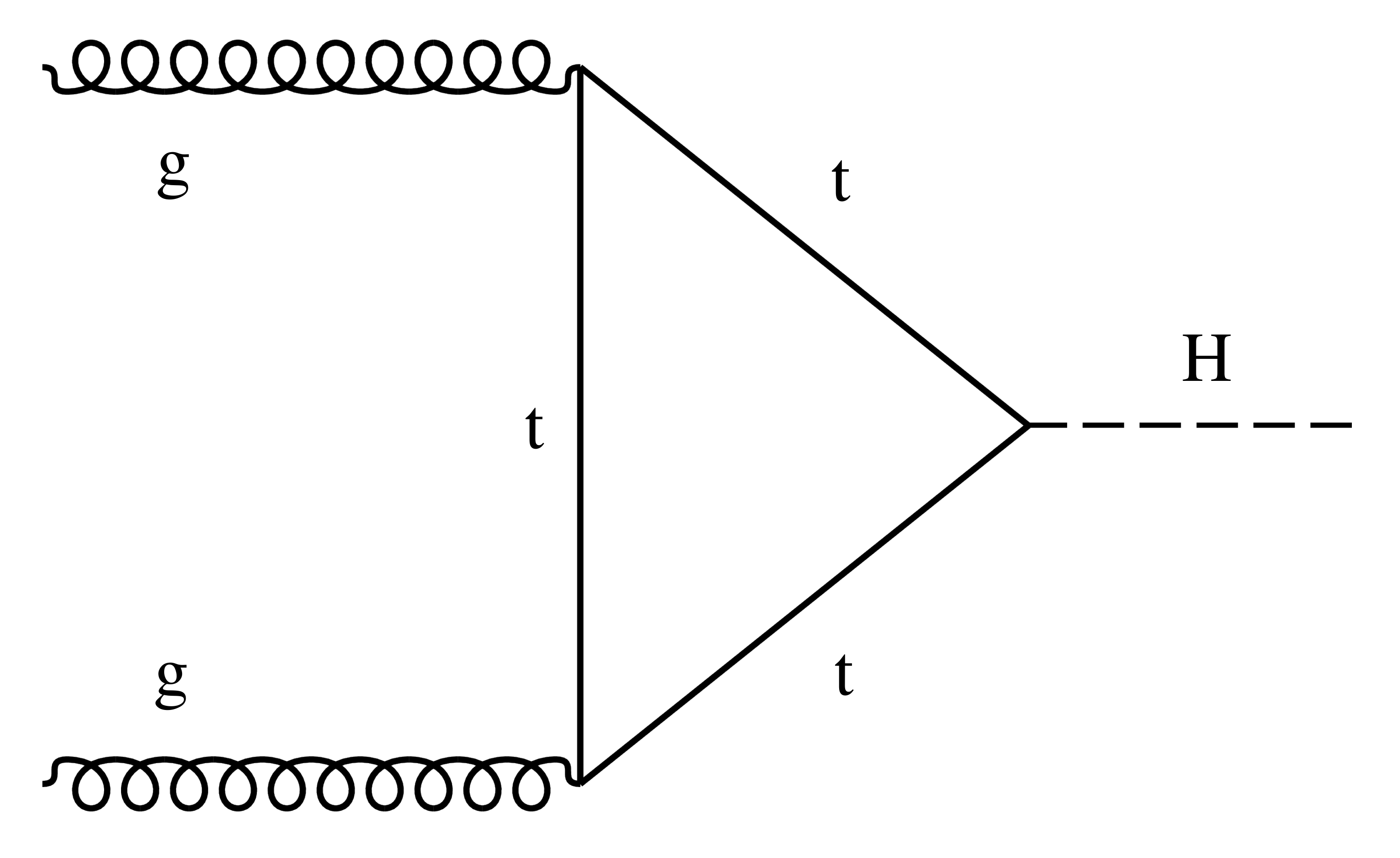}
\includegraphics[width=0.4\textwidth]{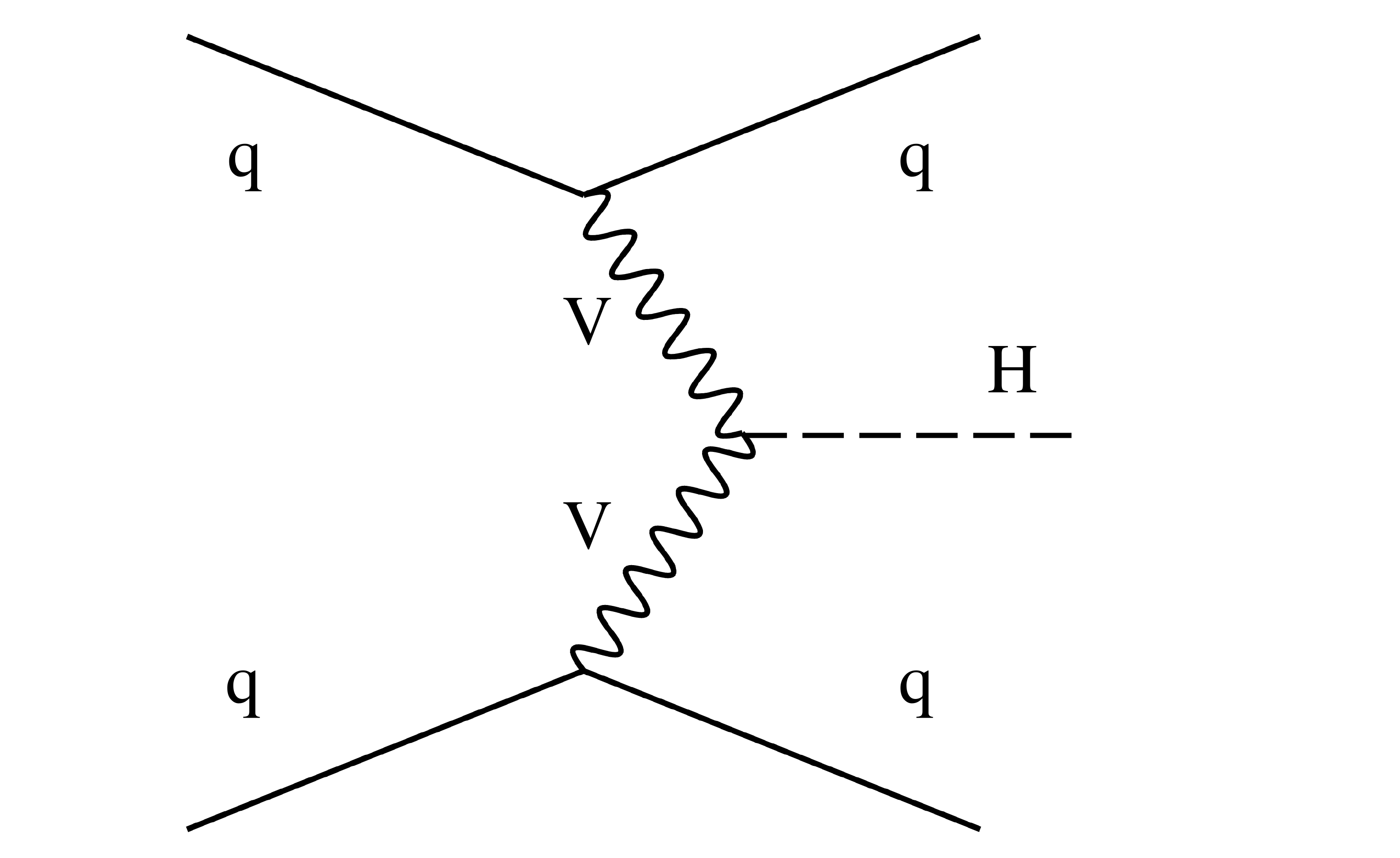}\\
\includegraphics[width=0.4\textwidth]{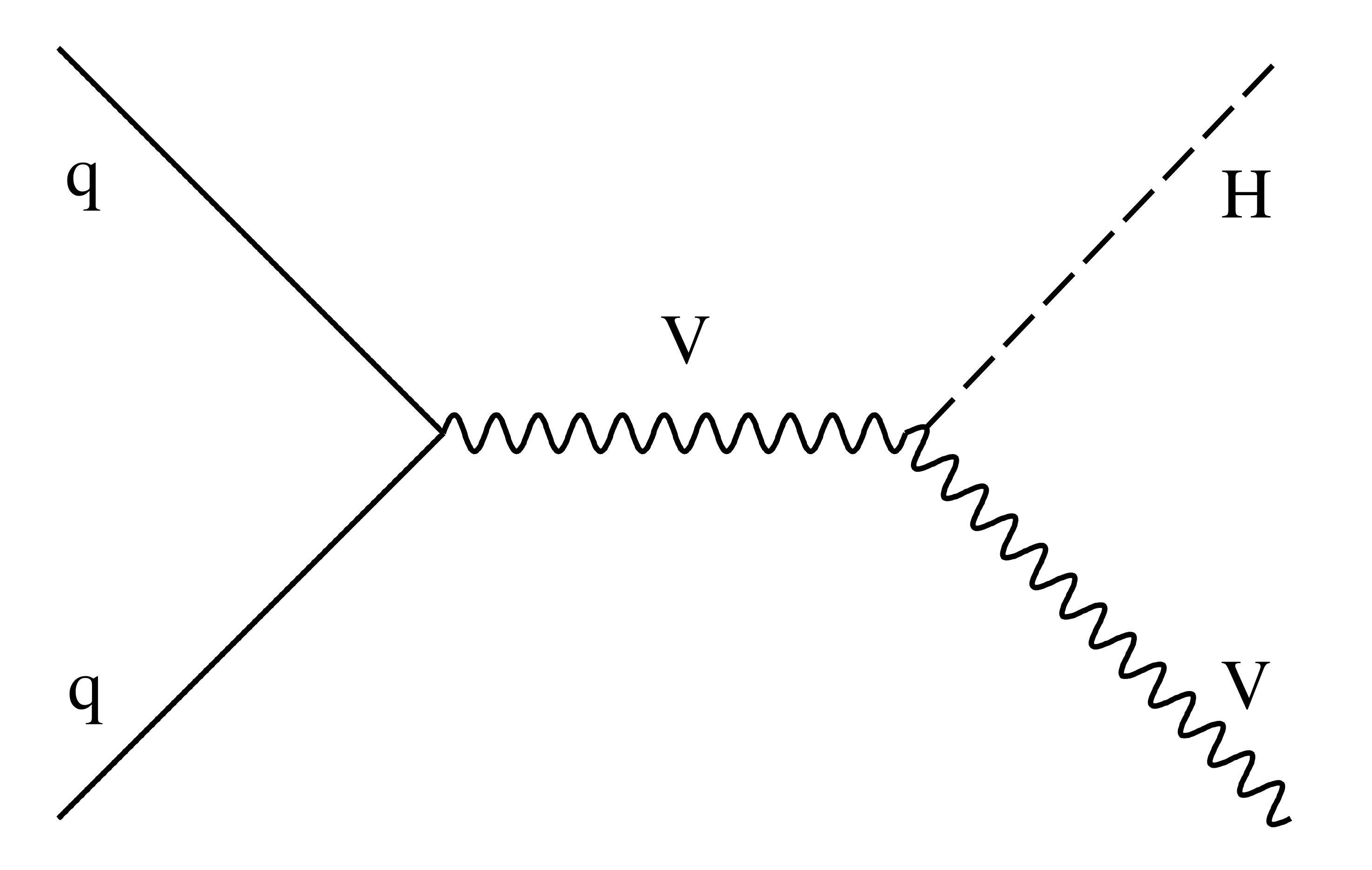}
\includegraphics[width=0.4\textwidth]{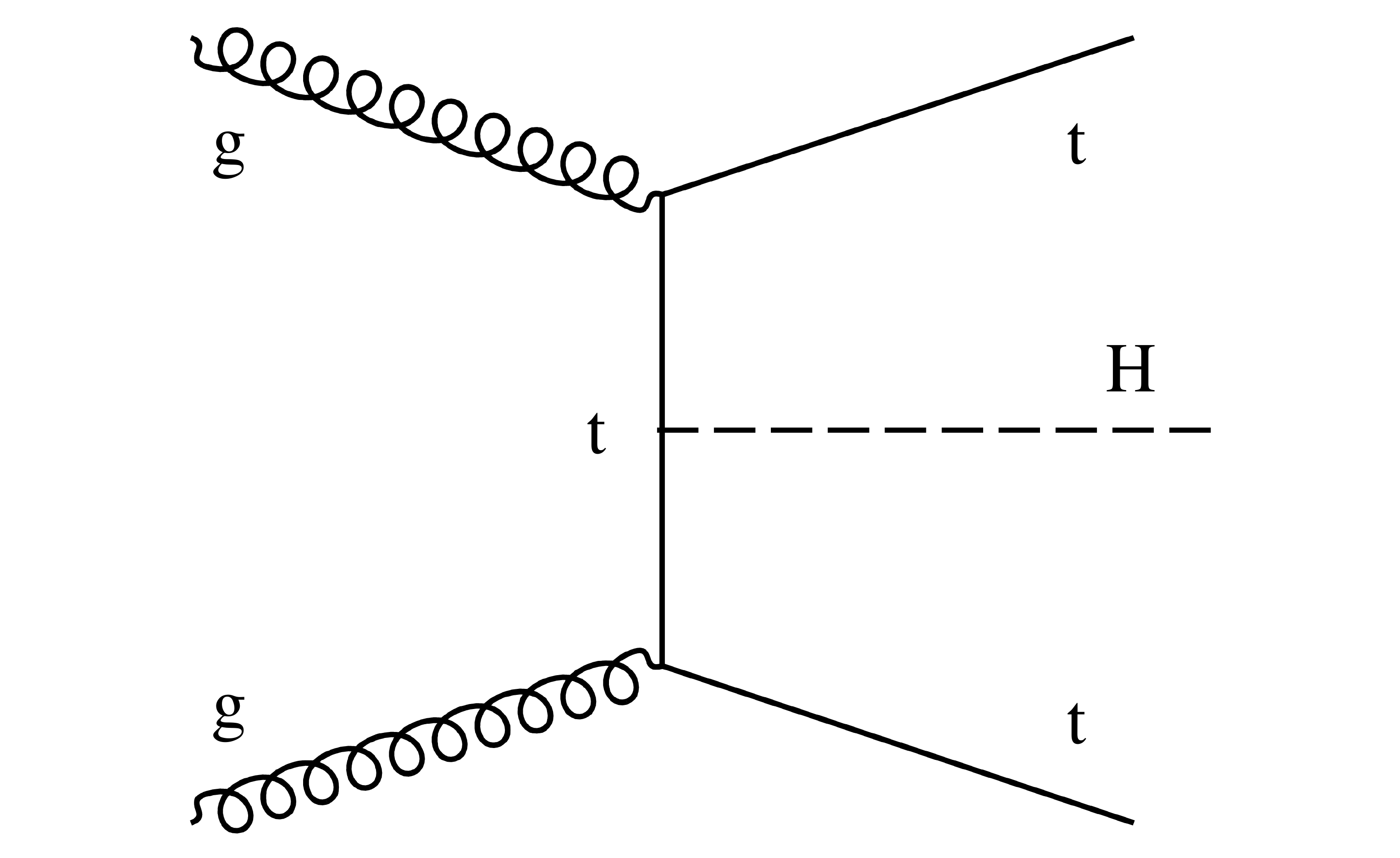}
\caption{Examples of leading-order Feynman diagrams for the \ggh (upper left), \vbf (upper right), \vh (lower left), and \tth (lower right) production modes. }
\label{fig:feynman_diag_main}
\end{figure*}

\begin{figure*}[hbtp]
\centering
\includegraphics[width=0.42\textwidth]{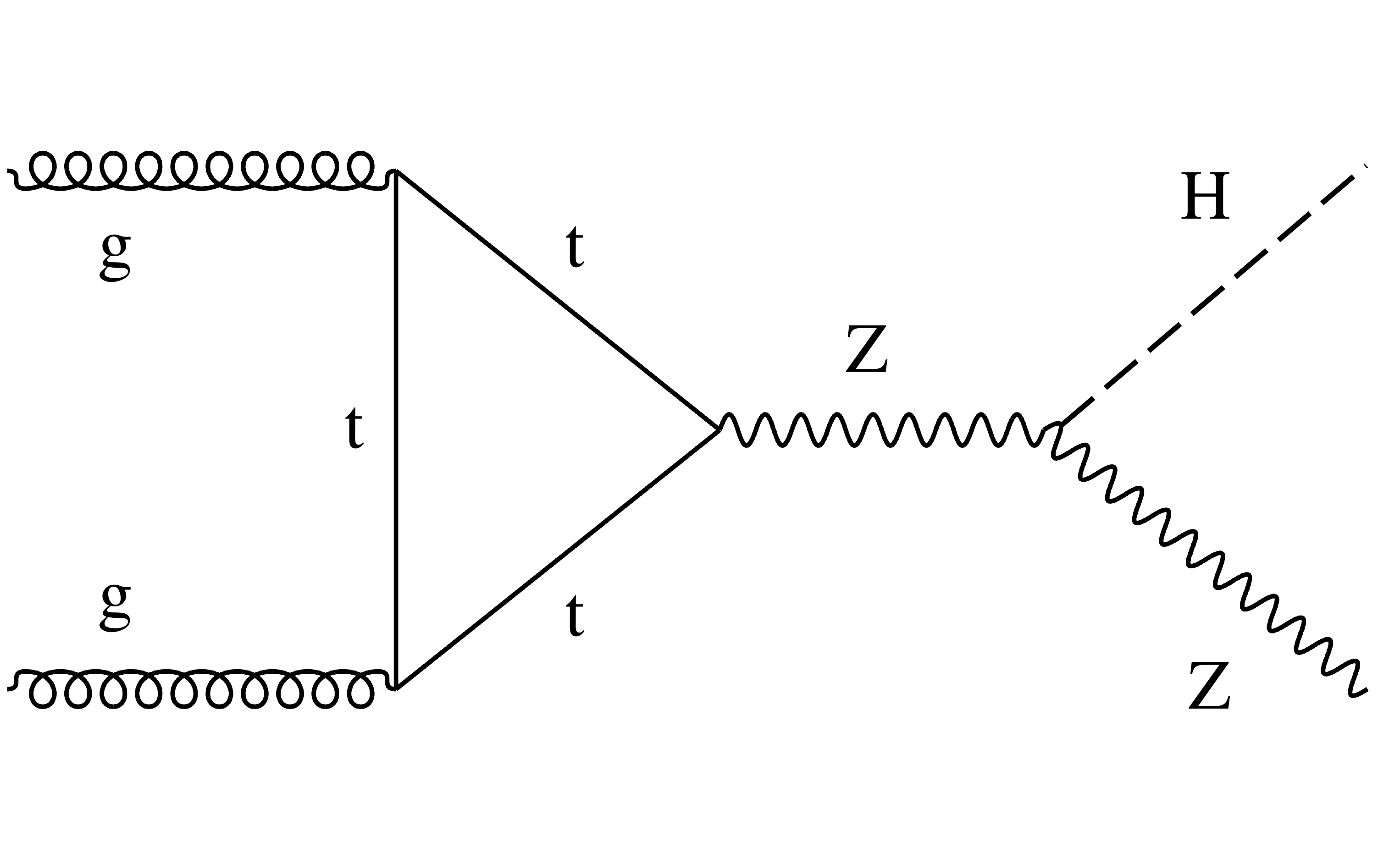}
\includegraphics[width=0.42\textwidth]{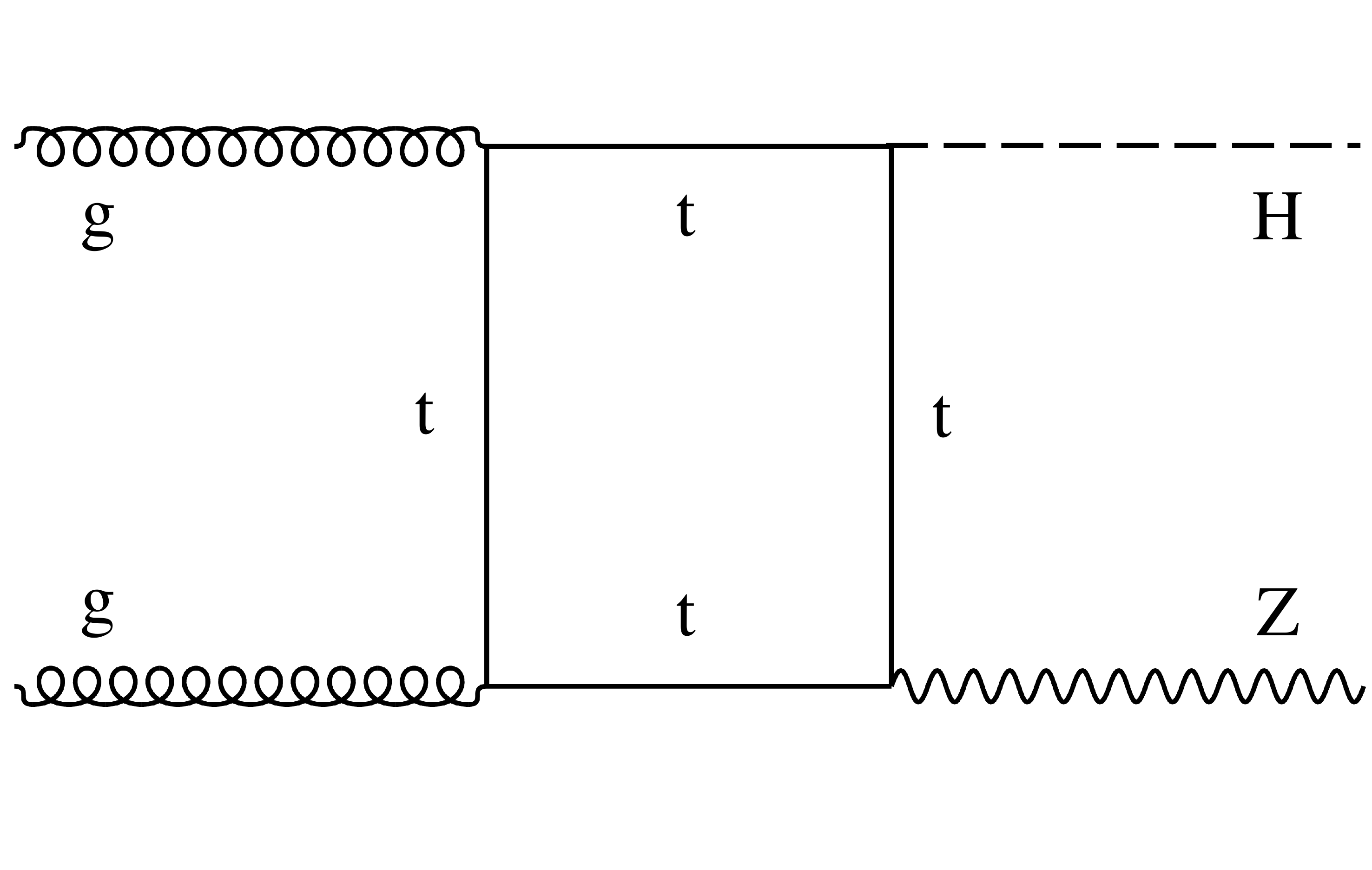}
\caption{Examples of leading-order Feynman diagrams for the $\Pg\Pg\to\zh$ production mode. }
\label{fig:feynam_diag_ZH}
\end{figure*}

\begin{figure*}[hbtp]
\centering
\includegraphics[width=0.4\textwidth]{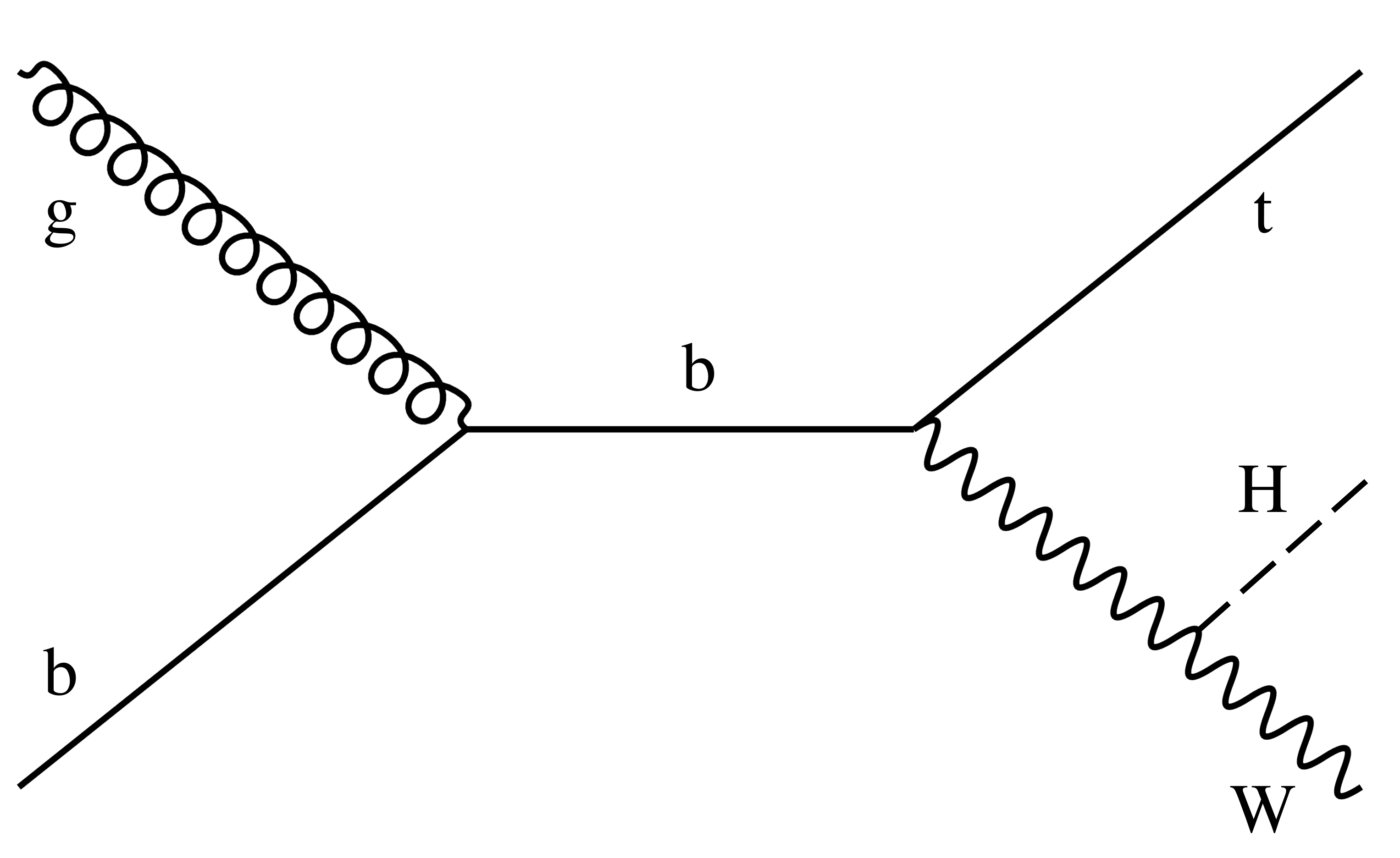}
\includegraphics[width=0.4\textwidth]{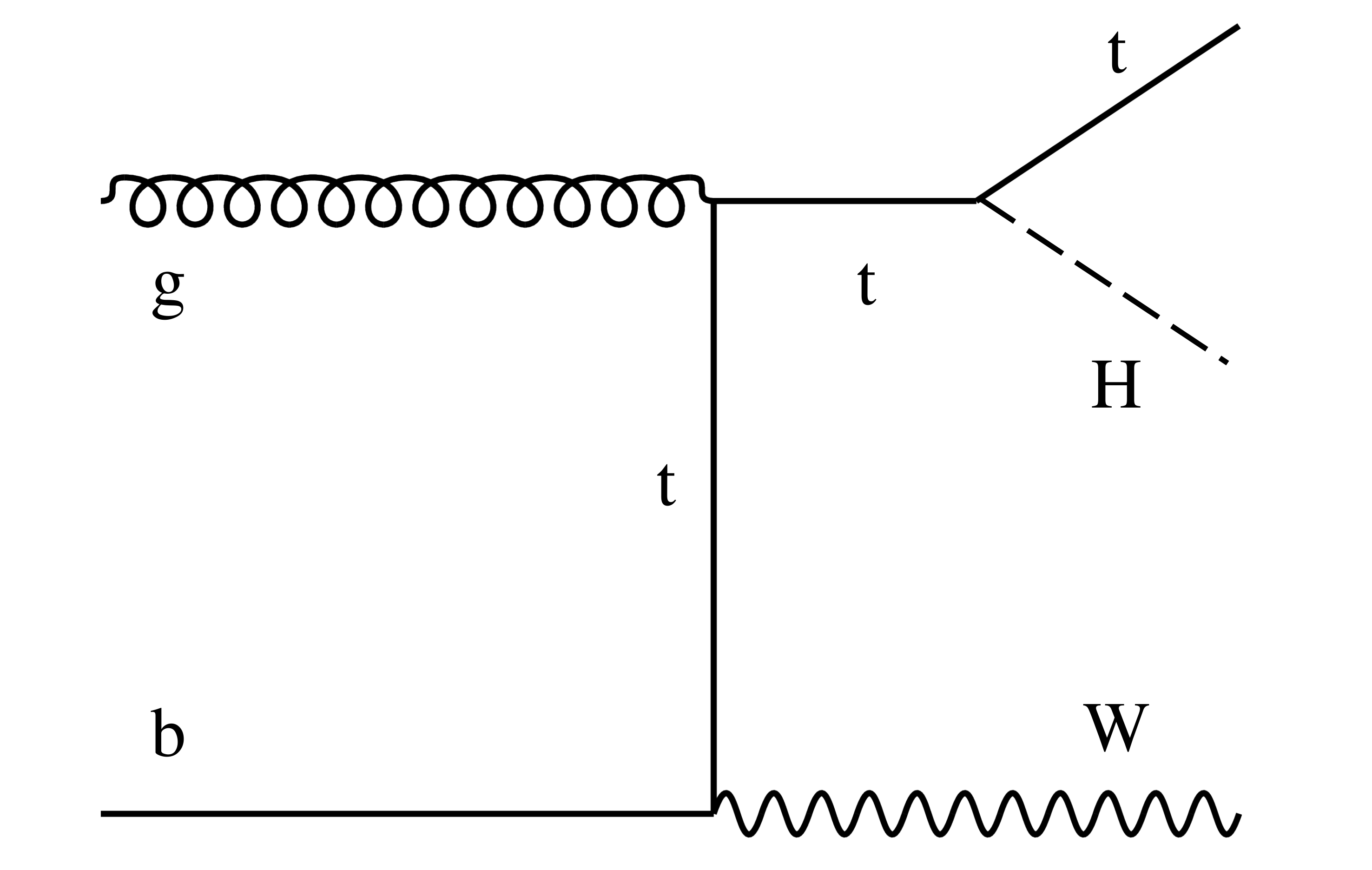}\\
\includegraphics[width=0.4\textwidth]{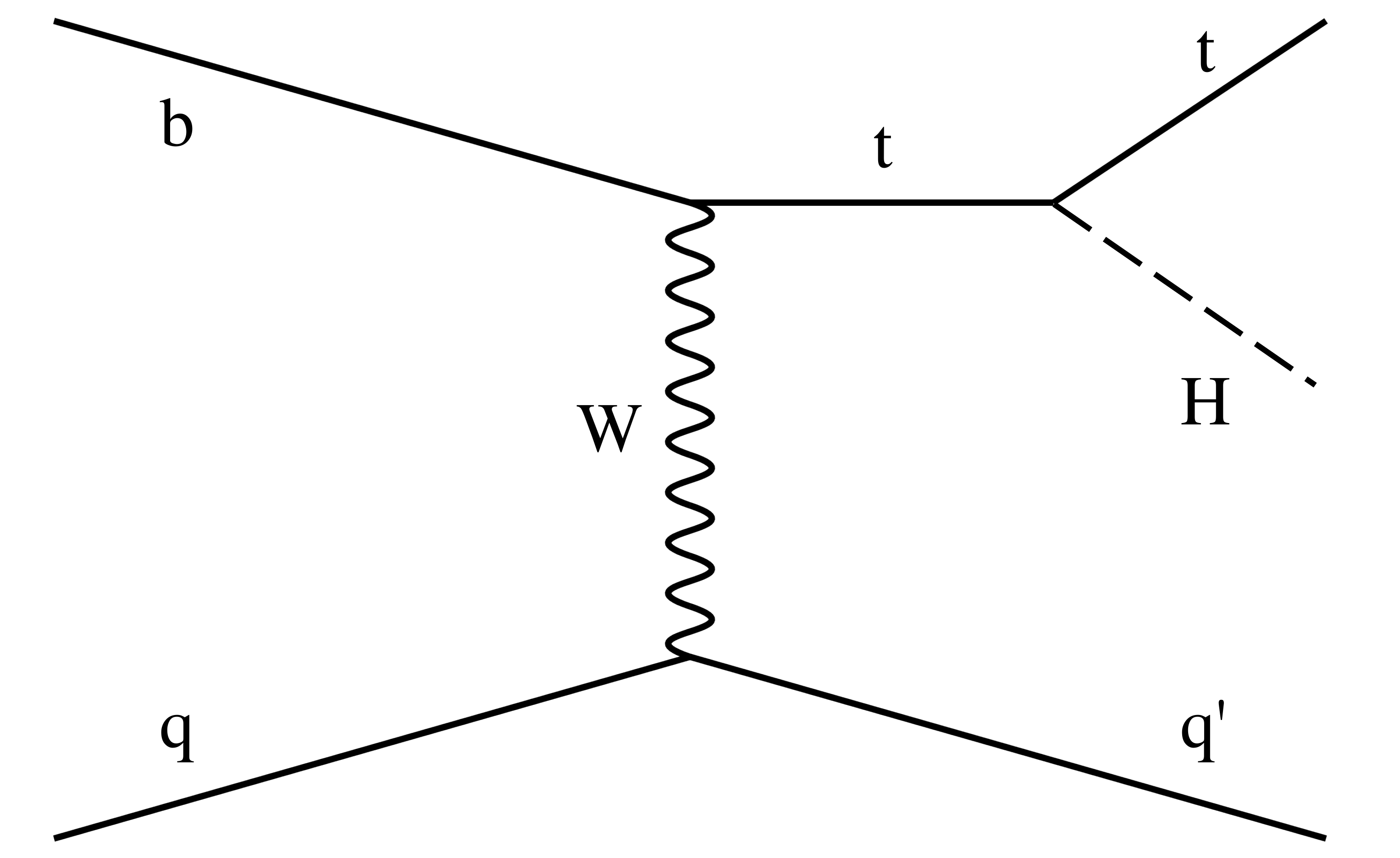}
\caption{Examples of leading-order Feynman diagrams for \tH production via the \tHW (upper left and right) and \tHq (lower) modes. }
\label{fig:feynman_tH}
\end{figure*}

The analyses included in this combination target production via gluon fusion (\ggh), vector boson fusion (\vbf), associated production with a
vector boson (\vh, V$=\PW$ or \cPZ),
and associated production with a pair of top quarks (\tth).
 The prediction for \ggh production has advanced to next-to-next-to-next-to-leading order (N$^{3}$LO) in perturbative quantum chromodynamics (QCD)~\cite{Anastasiou:2015ema,Anastasiou2016} and next-to-leading order (NLO) for electroweak (EW) corrections, reducing its uncertainty from ${}^{+7.6\%}_{-8.1\%}$ (next-to-NLO) to ${}^{+4.6\%}_{-6.7\%}$.
The calculations of the \vbf and \vh cross sections are performed at next-to-NLO QCD and NLO EW accuracy, while the calculation of the \tth cross section is performed at NLO QCD and NLO EW accuracy.
The updated theoretical predictions used for the various production and decay modes in this paper can be found in Refs.~\cite{Anastasiou:2015ema,Anastasiou2016,Ciccolini:2007jr,Ciccolini:2007ec,Bolzoni:2010xr,Bolzoni:2011cu,Brein:2003wg,Ciccolini:2003jy,Beenakker:2001rj,Beenakker:2002nc,Dawson:2002tg,Dawson:2003zu,Yu:2014cka,Frixione:2014qaa,Demartin:2015uha,Frixione:2015zaa,Demartin:2016axk,Denner:2011mq,Djouadi:1997yw,hdecay2,Bredenstein:2006rh,Bredenstein:2006ha,Boselli:2015aha,Actis:2008ts} and are summarized in Ref.~\cite{YR4}.
Examples of leading-order (LO) Feynman diagrams for these production processes can be seen in Figs.~\ref{fig:feynman_diag_main} and~\ref{fig:feynam_diag_ZH}.
In addition to the five main production processes, the contributions due to Higgs boson production in
association with a single top quark (\tH) and either a $\PW$ boson (\tHW) or a quark (\tHq), as shown in Fig.~\ref{fig:feynman_tH}, are included in the analyses that have some sensitivity to them.

For certain measurements in this paper, such as \ggh production and \hgg decay, the interference between the diagrams that contribute to the process is considered.
In addition, the \tH\ cross section is small in the SM, being approximately 14\% of the \tth~cross section,
 due to the destructive interference between the diagrams shown in Fig.~\ref{fig:feynman_tH}, which involve the coupling of the Higgs boson to $\PW$ bosons (\tHW process)
and top quarks (\tHq process). This interference becomes constructive, however, when the relative sign between these couplings is negative, and so the \tH~process is
sensitive to the relative sign of the $\PH\PW\PW$ and \tth couplings.

The ATLAS and CMS Collaborations have published combined measurements of Higgs boson production rates, decay rates, and couplings
with the $\sqrt{s}=7$ and $8\TeV$ LHC Run 1 data~\cite{ATLASRun1,CMSRun1}.
A combination of the Run 1 ATLAS and CMS analyses has also been performed~\cite{ATLASCMSRun1}. All results were found to
be in agreement, within their uncertainties, with the predictions of the SM.
In this paper, due to the larger integrated luminosity and increased signal
cross section at $\sqrt{s} = 13\TeV$, the measured precision for several parameters of
interest has significantly increased with respect to Ref.~\cite{ATLASCMSRun1}.
In particular, the predicted cross sections for the dominant \ggh production mode and the \tth production mode increase
by factors of approximately 2.3 and 3.8, respectively, between
$\sqrt{s} = 8$ and $13\TeV$. In addition, some of the theoretical predictions have improved, as mentioned earlier.

This paper is organized as follows: A brief description of the CMS detector is given in Section~\ref{sec:detector}, Section~\ref{sec:analyses} provides a summary of the various analyses
included in the combination, and Section~\ref{sec:differences} describes the modifications made to these analyses to ensure a common signal and uncertainty
model. Section~\ref{sec:comb_procedure} outlines the statistical procedure used to derive the results, and
Section \ref{sec:systematic_uncertainties} outlines the treatment of the systematic uncertainties. Section~\ref{sec:signalstrength}
reports the results of the signal parametrizations in terms of signal strength modifiers and fiducial cross sections, while Section \ref{sec:couplings} describes
the results obtained from an alternative set of signal parametrizations in terms of Higgs boson couplings. Section~\ref{sec:2hdm} details interpretations in terms of various two Higgs doublet models.
The paper is summarized in Section~\ref{sec:summary}.

\section{The CMS detector}
\label{sec:detector}

The central feature of the CMS apparatus is a superconducting solenoid of 6\unit{m} internal diameter, providing a magnetic field of 3.8\unit{T}. Within the solenoid volume are a silicon pixel and strip tracker, a lead tungstate crystal electromagnetic calorimeter, and a brass and scintillator hadron calorimeter, each composed of a barrel and two endcap sections. Forward calorimeters extend the pseudorapidity coverage provided by the barrel and endcap detectors. Muons are detected in gas-ionization chambers embedded in the steel flux-return yoke outside the solenoid.  A more detailed description of the CMS detector, together with a definition of the coordinate system used and the relevant kinematic variables, can be found in Ref.~\cite{Chatrchyan:2008zzk}.

\section{Analyses included in the combination}
\label{sec:analyses}

In this section, the individual analyses included in the combination are briefly described.
More detailed information on each analysis can be found in the corresponding references.
Many of the analyses split their primary data sample in multiple event categories with specific signatures that enhance the discrimination
power between different Higgs boson production processes. This is achieved through selections that require the presence of
additional leptons or jets, as expected in the decay of a $\PW$ or $\cPZ$ boson in the \wh and \zh modes, or in top quark decays in the \tth mode,
and that exploit the distinctive kinematic properties of the final state objects, such as the presence of two jets with a large separation in pseudorapidity
$\Delta\eta_{\text{jj}}$, and a large invariant mass $m_{\text{jj}}$, in the \vbf topology. In some categories, the kinematic features of an event as a
whole are used to select particular production processes. For example, requiring a large missing transverse momentum \ptmiss, defined as the magnitude of the negative vector sum over
the transverse momenta \pt of all particles reconstructed in an event, targets \zh production in which the $\cPZ$ boson decays to neutrinos.
The event categories within and amongst the individual analyses are constructed to ensure a negligible level of overlap (i.e. the same event entering more than one category).
In many cases, this is accomplished by synchronizing the object (e.g. electron, muon, tau, or jet) identification definitions and imposing strict requirements on the number of
reconstructed objects. In other cases, the orthogonality is ensured by imposing opposing requirements on higher level observables formed using multiple objects.
For rare cases where potential overlap is not explicitly removed, the lists of selected data events were checked and found to contain a negligible number of duplications.
In total, up to 265 event categories are considered, and there are over 5500 nuisance
parameters corresponding to various sources of experimental and theoretical systematic uncertainty.
A summary of the production and decay modes, which are described in more detail in the following sections, is shown in Table~\ref{tab:channels}.

\begin{table*}[tp!]
\centering
\topcaption[]{
Summary of the event categories in the analyses included in this combination.
The first column indicates the decay channel and the second column indicates the production mechanism
targeted by an analysis.
The third column provides the total number of categories per production tag, excluding control regions.
Notes on the expected fractions of different Higgs signal production and decay modes with respect to the total signal yield in the given category are given in the fourth column.
Where the numbers do not sum to 100\%,
the remaining contributions are from other signal production and decay processes.
Finally, where relevant, the fifth column specifies the approximate expected relative mass resolution for the SM Higgs boson.
}
\label{tab:channels}

\cmsTable{

\begin{tabular}{clccc}
\hline

Decay tags  & Production tags  & Number of  & \multirow{2}{*}{~~~~~~Expected signal fractions~~~~~~}  & \multirow{2}{*}{Mass resolution} \\
  &  & categories  &  &  \\ \hline

\multicolumn{1}{l}{$\hgg$,~~~Section~\ref{sec:hgg}}  &   &   &   &  \\ \hline

  & Untagged  & 4  & 74--91\% $\ggh$  &  \\

  & $\vbf$  & 3  & 51--80\% $\vbf$  &  \\

  & $\vh$ hadronic  & 1  & 25\% $\wh$, 15\% $\zh$  &  \\

  & $\wh$ leptonic  & 2  & 64--83\% $\wh$  &  \\

  & $\zh$ leptonic  & 1  & 98\% $\zh$  &  \\

  & $\vh$ $\ptmiss$  & 1  & 59\% $\vh$  &  \\

\multirow{-7}{*}{$\gamgam$}  & $\tth$  & 2  & 80--89\% $\tth$, $\approx$8\% $\tH$  & \multirow{-7}{*}{$\approx$1--2\%} \\ \hline

\multicolumn{5}{l}{$\hzzllll$,~~~Section~\ref{sec:hzz}}  \\ \hline

  & Untagged  & 3  & $\approx$95\% $\ggh$  &  \\

  & $\vbf$ 1, 2-jet  & 6  & $\approx$11--47\% $\vbf$  &  \\

  & $\vh$ hadronic  & 3  & $\approx$13\% $\wh$, $\approx$10\% $\zh$  &  \\

  & $\vh$ leptonic  & 3  & $\approx$46\% $\wh$  &  \\

  & $\vh$ $\ptmiss$  & 3  & $\approx$56\% $\zh$  &  \\

\multirow{-6}{*}{$\mmmm$, $\eemm/\mmee$, $\eeee$}  & $\tth$  & 3  & $\approx$71\% $\tth$  & \multirow{-6}{*}{$\approx$1--2\%} \\ \hline

\multicolumn{5}{l}{$\hwwlnln$,~~~Section~\ref{sec:hww}} \\ \hline

  & $\ggh$ 0, 1, 2-jet  & 17  & $\approx$55--92\% $\ggh$, up to $\approx$15\% $\htt$  &  \\

\multirow{-2}{*}{$\emu/\mue$}  & $\vbf$ 2-jet  & 2  & $\approx$47\% $\vbf$,  up to $\approx$25\% $\htt$  &  \\

$\Pe\Pe$+$\Pgm\Pgm$  & $\ggh$ 0, 1-jet  & 6  & $\approx$84--94\% $\ggh$  &  \\

$\emu$+jj  & $\vh$ 2-jet  & 1  & 22\% $\vh$, 21\% $\htt$  &  \\

$3\ell$  & $\wh$ leptonic  & 2  & $\approx$80\% $\wh$, up to 19\% $\htt$  &  \\

$4\ell$  & $\zh$ leptonic  & 2  & 85--90\% $\zh$, up to 14\% $\htt$  & \multirow{-6}{*}{$\approx$20\%} \\ \hline

\multicolumn{1}{l}{$\htt$,~~~Section~\ref{sec:htt}}  &   &   &   &  \\ \hline

  & 0-jet  & 4  & $\approx$70--98\% $\ggh$, 29\% $\hww$ in $\tetm$  &  \\

  & $\vbf$  & 4  & $\approx$35--60\% $\vbf$, 42\% $\hww$ in $\tetm$   &  \\

\multirow{-3}{*}{$\tetm, \teth, \tmth, \thth$}  & Boosted  & 4  & $\approx$48--83\% $\ggh$, 43\% $\hww$ in $\tetm$   & \multirow{-3}{*}{$\approx$10--20\%} \\ \hline

\multicolumn{5}{l}{$\vh$ production with $\hbb$,~~~Section~\ref{sec:vhbb}} \\ \hline

$\znnbb$  & $\zh$ leptonic  & 1  & $\approx$100\% $\vh$, 85\% $\zh$  & \\

$\wlnbb$  & $\wh$ leptonic  & 2  & $\approx$100\% $\vh$, $\approx$97\% $\wh$  &  \\

  & Low-$\pTV$ $\zh$ leptonic  & 2  & $\approx$100\% $\zh$, of which $\approx$20\% $\ggZH$  &  \\

\multirow{-2}{*}{$\zllbb$}  & High-$\pTV$ $\zh$ leptonic  & 2  & $\approx$100\% $\zh$, of which $\approx$36\% $\ggZH$  & \multirow{-4}{*}{$\approx$10\%} \\ \hline

\multicolumn{5}{l}{Boosted $\PH$ Production with $\hbb$,~~~Section~\ref{sec:boostedhbb}} \\ \hline

$\bb$  & $\pTH$ bins  & 6  & \multicolumn{1}{c}{$\approx$72--79\% $\ggh$}  & \multicolumn{1}{c}{$\approx$10\%} \\ \hline

\multicolumn{5}{l}{$\tth$ production with $\PH\to\mathrm{leptons}$,~~~Section~\ref{sec:tthlep}}  \\ \hline

$2\ell$ss  &  & 10  & $\ww/\tautau\approx 4.5$, $\approx$5\% $\tH$  &  \\

$3\ell$  &  & 4  & $\ww:\tautau:\zz\approx 15:4:1$, $\approx$5\% $\tH$  &  \\

$4\ell$  &  & 1  & $\ww:\tautau:\zz\approx 6:1:1$, $\approx$3\% $\tH$  &  \\

$1\ell$$+2\tauh$  &  & 1  & 96\% $\tth$ with $\htt$, $\approx$6\% $\tH$  &  \\

$2\ell$ss$+1\tauh$  &  & 2  & $\tautau:\ww\approx 5:4$, $\approx$5\% $\tH$  &  \\

$3\ell$$+1\tauh$  & \multirow{-6}{*}{$\tth$}  & 1  & $\tautau:\ww:\zz\approx 11:7:1$, $\approx$3\% $\tH$  & \multirow{-6}{*}{} \\ \hline

\multicolumn{5}{l}{$\tth$ production with $\hbb$,~~~Section~\ref{sec:tthbb}} \\ \hline

  & $\ttbar\to$ jets  & 6  & $\approx$83--97\% $\tth$ with $\hbb$  &  \\

  & $\ttbar\to 1\ell$+jets  & 18  & $\approx$65--95\% $\tth$ with $\hbb$, up to 20\% $\hww$  &  \\

\multirow{-3}{*}{$\bb$}  & $\ttbar\to 2\ell$+jets  & 3  &  $\approx$84--96\% $\tth$ with $\hbb$   & \multirow{-3}{*}{} \\ \hline

\multicolumn{5}{l}{Search for $\hmm$,~~~Section~\ref{sec:hmm}} \\ \hline

$\mumu$  & S/B bins  & 15  & \multicolumn{1}{c}{ 56--96\% $\ggh$, 1--42\% $\vbf$}  & \multicolumn{1}{c}{$\approx$1--2\%} \\ \hline

\multicolumn{5}{l}{Search for invisible $\PH$ decays,~~~Section~\ref{sec:hinv}} \\ \hline

  & $\vbf$  & 1  & 52\% $\vbf$, 48\% $\ggh$  & \\

  & $\ggh$ + $\geq1$ jet  & 1  & 80\% $\ggh$, 9\% $\vbf$  &  \\

  & $\vh$ hadronic  & 1  & 54\% $\vh$, 39\% $\ggh$  &  \\

\multirow{-4}{*}{$\text{Invisible}$}  & $\zh$ leptonic  & 1  & $\approx$100\% $\zh$, of which 21\% $\ggZH$  &  \\ \hline

\hline

\end{tabular}
}\end{table*}

\subsection{\texorpdfstring{\hgg}{Higgs to diphoton}}
\label{sec:hgg}

The \hgg analysis~\cite{Sirunyan:2018ouh} provides good sensitivity to nearly all Higgs boson production processes.
Since the \hgg decay proceeds mainly through \PW- and top-loop processes, interference effects make its branching fraction
sensitive to the relative sign of the fermion and vector boson couplings.
The analysis measures a narrow signal peak in the diphoton invariant mass ($m_{\gamgam}$) spectrum
over a smoothly falling continuum background, originating mainly from prompt,
nonresonant diphoton production, or from events where at least one jet is misidentified as an isolated photon.

Exclusive event categories are defined using dedicated selections based on additional reconstructed
objects to separate the different Higgs boson production mechanisms.
The presence of additional leptons, \ptmiss, or jets is used to classify events
into one of the following categories: \tth leptonic, \tth hadronic, \zh leptonic, \wh leptonic, loose \vh leptonic with
low \ptmiss requirement, \vbf, \vh \ptmiss, and \vh hadronic.
The \vbf category is divided into three subcategories of increasing purity against \ggh production.
Finally, the remaining events are divided into four untagged categories with increasing signal purity. 

In each event class, the background in the signal region (SR) is estimated from a fit
to the observed $m_{\gamgam}$ distribution in data.
The dominant experimental uncertainties in the measurement of the rate of Higgs boson production in the \hgg decay channel are
related to the modeling of the electromagnetic shower shape observables used in the photon identification and the
background shape parametrization.

\subsection{\texorpdfstring{\hzz}{Higgs to ZZ}}
\label{sec:hzz}

Despite the \hzzllll ($\ell=\Pe$ or $\Pgm$) decay having the lowest branching fraction
of the decay channels considered, it also has the lowest background contamination,
resulting in very good sensitivity to production processes with large cross sections, such as \ggh.
It is also the most important decay channel in constraining the $\PH\cPZ\cPZ$ coupling.
The \hzzllll~\cite{HIG16041} analysis measures a narrow four-lepton invariant mass peak
over a small continuum background. The dominant irreducible background in this analysis
is due to nonresonant \zz production with both $\cPZ$ bosons decaying to a pair of charged leptons,
and is estimated from simulation.
The \eeee, \mmmm, and \eemm/\mmee decay channels are treated separately to better model the different mass resolutions and
background rates arising from jets misidentified as leptons.

To separate the different Higgs boson production mechanisms,
the following categories are defined on the basis of the presence of jets, $\PQb$-tagged jets, leptons, \ptmiss,
and various matrix element discriminants that make use of the information about the additional objects: \vbf (1- and 2-jet), \vh hadronic,
\vh leptonic, \tth, \vh \ptmiss, and untagged categories.

In the \hzzllll analysis, the dominant experimental uncertainties are
related to the lepton efficiencies and the determination of the \cPZ+jets background from data.

\subsection{\texorpdfstring{\hww}{Higgs to WW}}
\label{sec:hww}

The \hwwlnln analysis~\cite{HIG-16-042} profits from the fact that the \hww decay mode has one of the largest branching
fractions and has a relatively low-background final state. As a result, this decay channel has very good sensitivity to
most production processes, in particular \ggh and \vbf.
Imposing tight lepton identification criteria and requiring the absence of $\PQb$-tagged jets helps to reduce the misidentified lepton
and top quark backgrounds, respectively.
Several event categories with varying signal-to-background ratios are defined to improve the sensitivity to the signal.
Events are selected that contain two leptons, denoted $2\ell$, which may be of different or same flavor.
The different-flavor $\Pe\Pgm$ decay channel dominates the sensitivity since it has the largest branching fraction and
is the least contaminated by backgrounds.
The same-flavor $\Pe\Pe$ and \mumu final states are also considered, although their sensitivity is
limited by the contamination from Drell--Yan (DY) background events with misreconstructed \ptmiss.
Given the large background contribution from $\PQt\PQt$ production in both the
different-flavor and same-flavor final states, events are further categorized into categories with 0, 1, and 2 associated jets,
with the 0-jet category dominating the overall sensitivity. In addition, events are further categorized on the basis of the
\pt of the subleading lepton, since the background from misidentified leptons is larger in the low-\pt region.
In the different-flavor final state, dedicated 2-jet
categories are included to enhance the sensitivity to VBF and \vh production mechanisms.

The analysis also includes categories that are sensitive to the associated production of the Higgs boson with
a vector boson that decays leptonically. Two $3\ell$ categories that are sensitive to \wh production are defined by
requiring the presence of a total of three leptons (electrons or muons). The two are distinguished by whether or not they contain a pair of
leptons with the same flavor and opposite sign. Events with four charged leptons, in which one pair is consistent with a $\cPZ$ boson decay,
are separated into two categories depending on whether the remaining pair consists of same-flavor leptons or not. These $4\ell$ categories are
sensitive to the \zh production mode.
The signal extraction method depends on the event category.

When measuring the rate of Higgs boson production in the \hww decay channel,
the dominant experimental uncertainties arise from the determination of the top quark pair,
$\PW\PW$ and DY backgrounds from data, and the uncertainties related to the
\pt and $\eta$ dependent lepton reconstruction and identification efficiencies.

\subsection{\texorpdfstring{\htt}{Higgs boson to tautau}}
\label{sec:htt}

The \htt analysis~\cite{HIG16043} benefits from a relatively large branching fraction and a reasonable
mass resolution of $\approx$10--20\%, providing competitive sensitivity to both the \ggh and \vbf production processes.
It also provides the best sensitivity for the direct measurement of a fermionic Higgs boson coupling.
The analysis utilizes the four most sensitive \tautau
final states: \tetm, \teth, \tmth, and \thth, where \tauh denotes a hadronically decaying \Pgt lepton.
In the analysis of each \tautau decay channel, events are divided into three categories
labeled 0-jet, boosted, and VBF.

The VBF category requires the presence of two additional jets with large $m_{\text{jj}}$ and $\Delta\eta_{\text{jj}}$,
designed to increase the purity of \vbf events.
The 0-jet category does not have
much sensitivity to the signal, but is useful
to constrain systematic uncertainties in the background model.
The boosted category
contains all remaining events,
and is binned as a function of \pT of the \tautau system to increase the sensitivity to \ggh production.
There is a nonnegligible contribution from the \hww
process in some categories, and this is treated consistently as an \hww signal in this combined measurement.

The \ptmiss and \tauh energy scale uncertainties are the dominant experimental uncertainties in the measurement of the Higgs boson production rate in
the \htt decay channel, followed by the uncertainties in the determination of the $\cPZ(\tautau)+$jets background from data.

\subsection{\texorpdfstring{\vh production with \hbb}{Associated VH production with Higgs to bb} }
\label{sec:vhbb}

{\tolerance=800 The \hbb decay has the largest expected branching fraction in the SM (58.1\% for $\mH=125.09\GeV$)
and a reasonable mass resolution of 15\%. By requiring \vh production it is possible
to increase the signal purity with respect to the inclusive case for which the background from QCD
multijet production is dominant.
The analysis of the \hbb decay targeting \vh production (\vhbb)~\cite{HIG16044} provides the
best sensitivity to the \wh and \zh processes
as well as to the $\bb\PH$ coupling. Selected events are categorized based on the presence of
two $\PQb$-tagged jets, and two (\zllbb), one (\wlnbb) or
no (\znnbb) electrons or muons in the final state. The \zllbb categories are subdivided into
low-boost ($50 < \pT(\cPZ) < 150\GeV$) and high boost ($\pT(\cPZ) > 150\GeV$) regions.
Events selected in the \znnbb category are further required to have $\ptmiss > 170\GeV$.\par}

The main backgrounds come from $\cPZ$ or $\PW$ boson production in association with light- and heavy-flavor (LF and HF) jets, as well as
from top quark pair and diboson production.
The dominant experimental uncertainties in
in this analysis are related to the determination of these backgrounds, and uncertainties in the $\PQb$ tagging discriminator shapes and efficiencies.

\subsection{\texorpdfstring{Boosted \PH production with \hbb}{Boosted H production with Higgs to bb} }
\label{sec:boostedhbb}

The \hbb decay is also measured in an analysis that targets inclusive production of the Higgs boson~\cite{HIG17010}, exploiting
the higher signal to background ratio at high \pTH (the transverse momentum of the Higgs boson).
The decay products of a high-\pt \hbb system are reconstructed using the anti-\kt algorithm~\cite{Cacciari:2008gp,Cacciari:2011ma} with a distance parameter of 0.8 (AK8 jet),
and the soft-drop algorithm~\cite{Dasgupta:2013ihk,Larkoski:2014wba} is used to reconstruct the jet mass $m_{\rm SD}$, which peaks at the Higgs boson mass for signal events.
Events containing substantial \ptmiss, or identified and isolated electrons, muons or \Pgt  leptons
are vetoed to reduce the background contributions from vector boson production and top quark processes.

The main background component, QCD multijet production, is estimated from
a signal-depleted Control Region (CR).
The selected events are divided according to the jet \pt into six bins of increasing width from 450 GeV to 1\TeV.

The dominant experimental uncertainties in this analysis are the
uncertainties related to the extrapolation of the QCD multijet and top quark pair backgrounds
from the CRs.

\subsection{\texorpdfstring{\tth production}{Higgs boson production in association with top quarks} }

Measurements of the rate of the \tth production process provide a direct test of the Higgs boson's coupling to top quarks.
A recent measurement by CMS combining the $\sqrt{s}=7$, $8$ and $13\TeV$ datasets was able to establish the first $5\sigma$ observation of the \tth production process~\cite{ttHComb}.
Dedicated analyses targeting the \hlep~\cite{Sirunyan:2018shy} and \hbb~\cite{Sirunyan:2018mvw,Sirunyan:2018ygk} decay channels using $\sqrt{s}=13\TeV$ data are described in this section.

\subsubsection{\texorpdfstring{\tth production with \hlep}{Higgs boson production in association with top quarks and H to leptons} }
\label{sec:tthlep}

The analysis of \tth production with \hlep~\cite{Sirunyan:2018shy} is mainly sensitive to the Higgs boson decaying
to \tautau, \ww or \zz with electrons, muons and/or \tauh in the final state.
This analysis provides the best sensitivity to the \tth production process.
The main irreducible backgrounds come from $\PQt\PQt$V and diboson production.
Reducible backgrounds containing misidentified leptons or leptons with misidentified charge are estimated from CRs in data.
Events are categorized according to their lepton content.
The light-lepton ($\Pe$/$\Pgm$) categories are defined as:

\begin{itemize}
\item {2$\ell$ss}: Events with two leptons having the same sign and at least four additional jets. A veto on the presence of hadronic tau decays is applied. Further categories based on lepton charge, flavor and the number of $\PQb$-tagged jets are defined within this class. 
\item {3$\ell$}: Events containing three leptons, with the sum of lepton charges equal to $\pm$1, and at least two additional jets of which one or two are $\PQb$ tagged. 
\item {4$\ell$}: Events with four leptons, with an explicit veto on \hzzllll events as these are selected by the analysis described in Section~\ref{sec:hzz}. 
\end{itemize}

The \tauh categories, which require the presence of hadronically decaying \Pgt leptons, are defined as:

\begin{itemize}
\item {1$\ell$+2$\tauh$}: Events with two oppositely charged $\tauh$ candidates and an additional $\Pe$/$\Pgm$, at least three additional jets, and at least one $\PQb$-tagged jet.
\item {2$\ell$ss+1$\tauh$}: Events containing three leptons, with sum of lepton charges equal to $\pm$1, and at least two additional jets of which one or two are $\PQb$ tagged. These events are further sorted into two subcategories based on whether or not all of the jets expected in the \tth process are reconstructed.
\item {3$\ell$+1$\tauh$}: Events with three light leptons, one \tauh and at least two additional jets and one $\PQb$-tagged jet.
\end{itemize}

In the $\Pe$/$\Pgm$ and \tauh categories, the dominant experimental uncertainties on the measurement of the rate of Higgs boson production in the \tth mode
are related to the lepton reconstruction efficiencies, and the estimation of the reducible background contributions from data.

\subsubsection{\texorpdfstring{\tth production with \hbb}{Higgs boson production in association with top quarks and Higgs to bb} }
\label{sec:tthbb}
There are two analyses that target the associated production of the Higgs boson with a pair of top quarks
in the \hbb decay mode~\cite{Sirunyan:2018mvw,Sirunyan:2018ygk}.
The leptonic analysis requires at least one lepton to be present in the final state, from the $\PQt\PQt$ decay system,
 while the hadronic analysis selects events in the all-hadronic final state.
These analyses provide good sensitivity to the \tth production process and improve the precision in the measurement of
the $\bb\PH$ coupling.

In the leptonic analysis, events are sorted into the $1\ell$
or $2\ell$ classes, depending on the presence of one or two well-identified leptons.
Events are further categorized based on the number of reconstructed jets (Nj) and the
number of jets that are tagged as $\PQb$ jets (Nb) in each event.
The largest background is due to top quark pair production with additional jets that contain heavy flavor hadrons.
In the $1\ell$ class, three categories are used: 4j $\geq$3$\PQb$, 5j $\geq$3$\PQb$, and 6j $\geq$3$\PQb$.
In each category a multi-classification deep neural network (DNN)~\cite{SCHMIDHUBER201585} is used to define six classes on the basis of the most probable
event hypothesis for each event, yielding a total of 18 categories.
In the $2\ell$ class, there are two jet categories: $\geq$4j 3$\PQb$ and $\geq$4j $\geq$4$\PQb$.
The $\geq$4j $\geq$4$\PQb$ category is further divided into two subcategories.

The all-hadronic final-state analysis selects events that contain at least seven jets, at least three of which are tagged as $\PQb$ jets.
These events are divided into seven categories: 7j 3$\PQb$, 7j $\geq$4$\PQb$, 8j 3$\PQb$, 8j $\geq$4$\PQb$, $\geq$9j 3$\PQb$, and $\geq$9j $\geq$4$\PQb$.
Events containing electrons or muons are vetoed to maintain an orthogonal selection to the leptonic final state analysis.
The dominant background is QCD multijet production, with other important backgrounds coming from $\PQt\PQt$+jets processes.

The dominant experimental uncertainties in the measurement of the rate of \tth production with \hbb decay in the leptonic and all-hadronic final states
are due to uncertainties in the determination of the $\PQt\PQt\bb$ backgrounds and $\PQb$ tagging efficiencies. In the all-hadronic final state, the uncertainty in the
determination of the QCD multijet background also has a significant contribution to the overall systematic uncertainty.

\subsection{\texorpdfstring{Search for \hmm}{Search for Higgs to dimuon} }
\label{sec:hmm}

The \hmm search~\cite{HIG-17-019} is the only analysis included here that is sensitive to the coupling of the Higgs boson to second-generation fermions.
The analysis searches for a narrow peak in the dimuon invariant mass ($m_{\mumu}$) spectrum above a large continuum
background from DY production of muon pairs.
Events are categorized using variables that are uncorrelated with $m_{\mumu}$, in order
to avoid introducing an irregular shape in the background spectrum.
Variables that distinguish between the \ggh and \vbf signals, and the DY and $\PQt\PQt$ backgrounds,
are used to define event categories with varying signal-to-background ratios.
The categories are further divided based on the momentum of the muon with the
largest $\abs{\eta}$ in the dimuon pair, to exploit the differences in the $m_{\mumu}$ resolution.
Since there are more variables associated with \vbf production that can be used to separate signal and background,
the events with the highest BDT output value are most compatible with that process.

In each event category, the background is estimated from a fit
to the observed $m_{\mumu}$ distribution.
As in the \hgg analysis, the parameters of the functions used to describe the background contribute to the statistical
uncertainty in the measurements. This is the dominant source of uncertainty in constraining the rate of Higgs boson decay
in the \hmm decay channel. The observed upper limit on the cross section times branching fraction of \hmm obtained in Ref.~\cite{HIG-17-019} is 2.93 times the SM value.

\subsection{\texorpdfstring{Search for $\PH\to$ invisible}{Search for Higgs to invisible} }
\label{sec:hinv}

The direct search for the Higgs boson decaying into particles that cannot be detected provides a constraint on the invisible Higgs boson branching fraction ($\BRinv$), which is predicted to be enhanced in BSM scenarios~\cite{Belanger:2001am,Giudice:2000av,Dominici:2009pq,Bonilla:2015uwa,SHROCK1982250}.
The search is performed using events with large \ptmiss, and containing
additional particles consistent with Higgs boson production via the \vbf~\cite{HIG-17-023},
\zh with $\cPZ\to\ell\ell$~\cite{EXO16052}, \vh with the $\PW$ or $\cPZ$ boson decaying hadronically, or \ggh modes~\cite{EXO16048}.

Events selected in the \vbf category are required to contain two jets, with a large $m_{\text{jj}}$ and a large $\Delta\eta_{\text{jj}}$. The
\vh hadronic and \ggh categories comprise events containing either a high-\pt AK8 jet, consistent with a boosted, hadronically decaying vector boson, or a jet
from initial-state radiation, reconstructed in the fiducial volume of the tracker.
The dominant backgrounds in these categories are due to the $\cPZ(\nu\nu)$+$\,\text{jets}$ and $\PW(\ell\nu)$+$\,\text{jets}$ processes. These are estimated from dedicated
lepton and photon CRs in data.
In all three categories, the dominant uncertainties are related to the extrapolation
of the lepton and photon CRs to determine the $\cPZ(\nu\nu)$+$\,\text{jets}$ and $\PW(\ell\nu)$+$\,\text{jets}$ backgrounds in the SR.

{\tolerance=800 The \zh leptonic category is defined by selecting events that contain a pair of oppositely charged electrons or muons consistent with the
decay of a $\cPZ$ boson.
The dominant backgrounds arise from $\cPZ(\ell\ell)\cPZ(\nu\nu)$ and $\PW(\ell\nu)\cPZ(\nu\nu)$ diboson production and are estimated using a combination of CRs in
data containing additional leptons, and simulated events.
The dominant uncertainty in this category is related to theoretical uncertainties in the
higher-order corrections used in the simulation of these backgrounds.\par}

The observed upper limit on the branching fraction of \hinv assuming SM Higgs production rates is 26\%~\cite{HIG-17-023}.
As described later in Section~\ref{sec:couplings}, the \hinv analyses are included only in models for which a nonzero invisible branching fraction of
the Higgs boson is considered.

\section{Modifications to the input analyses}
\label{sec:differences}

This section describes the changes in each analysis, as implemented for this combination, compared to their respective publications.

\subsection{Gluon fusion modeling}
\label{ssec:pth}

In order to consistently combine the various analyses, it is necessary to use the same theoretical predictions for the signal.
The most significant difference between the input analyses is the modeling of the dominant \ggh production mode
in the \hzz, \htt, \hgg, and \hww decay channels. The published results in these analyses used different generators with next-to-leading order matrix
elements merged with parton showering (NLO+PS). In order to synchronize these analyses and take advantage of the most accurate
simulation of \ggh available, a reweighting is applied. Gluon fusion events are generated using the \POWHEG 2.0~\cite{Bagnaschi:2011tu,Nason:2004rx,powheg,powheg3},
\MGvATNLO version 2.2.2~\cite{Alwall:2014hca,Frederix2012}, and {\sc nnlops}~\cite{NNLOPS,NNLOPS2} generators.
The {\sc nnlops}~simulation, which is the highest order parton shower matched \ggh simulation available, includes the
effects of finite quark masses.
Events are separated into 0, 1, 2, and $\ge$3 jet bins, where
the jets used for counting are clustered from all stable particles, excluding the decay products of the Higgs boson or associated vector bosons,
and have $\pt > 30\GeV$.
The sums of weights in each sample are first normalized to the inclusive N$^{3}$LO cross section. The ratio of the \pTH distribution from
the {\sc nnlops} generator to that from the \POWHEG or \MGvATNLO generators in each jet bin is applied to the \ggh signal samples.
The reweighting procedure has been checked against fully simulated {\sc nnlops} samples in the \hgg and \htt decay channels
and was found to give results compatible within the statistical uncertainty of the simulated samples.
The \hmm and boosted \hbb analyses, which are much less sensitive to \ggh production than other decay channels, use the NLO+PS simulation.

\subsection{Theoretical uncertainties in gluon fusion}
\label{ssec:gghunc}

The \ggh cross section uncertainty scheme for the \hzz and \htt decay channels has been updated to the one proposed in Ref.~\cite{YR4},
as already used in the \hgg and \hww analyses. This uncertainty scheme includes 9 nuisance parameters accounting
for uncertainties in the cross section prediction for exclusive jet bins (including the migration between the 0- and 1-jet, as well as between the 1- and $\ge$2-jet bins),
the 2-jet and $\ge$3-jet \vbf phase spaces, different \pTH regions, and the uncertainty in the \pTH distribution due to missing higher-order finite top quark mass corrections.
The boosted \hbb search, which is only sensitive to \ggh in the high \pTH tail, uses a dedicated prediction in this region, and hence the theoretical uncertainties assigned are
assumed to be uncorrelated with the other analyses.

\subsection{Statistical uncertainties in simulation}
\label{ssec:bbb}

In the combination, many of the nuisance parameters originate from the use of a limited number of Monte Carlo events to determine SM signal and background expectations.
Some of the input analyses have been modified to use the ``Barlow-Beeston lite'' approach, which assigns a single nuisance parameter per
bin that scales the total bin yield~\cite{BARLOW1993219,Conway2011}. This differs from the previous implementation, which utilized separate nuisance
parameters for each process per bin. With the Barlow-Beeston
approach, the maximum likelihood estimator for each of these nuisance parameters is independent from the others, and can be solved for analytically. This has been found to provide a significant reduction in the minimization time, while reproducing the results obtained with the full treatment to within 1\%.
\section{Combination procedure}
\label{sec:comb_procedure}

The overall statistical methodology used in this combination is the same as
the one developed by the ATLAS and CMS Collaborations, and described in Ref.~\cite{ATLASCMSRun1}.
The procedures used in this paper are described in more detail in Refs.~\cite{LHC-HCG, Chatrchyan:2012tx, CMSLong2013}
and are based on the standard LHC data modeling and handling toolkits {\sc RooFit}~\cite{Verkerke:2003ir} and {\sc RooStats}~\cite{Moneta:2010pm}.

{\tolerance=10000 The parameters of interest (POI) $\vec\alpha$ for a particular model are estimated with their corresponding confidence intervals using a
profile likelihood ratio test statistic $q(\vec\alpha)$~\cite{Cowan:2010st}, in which experimental or theoretical uncertainties
are incorporated via nuisance parameters (NP) {$\vec\theta$}:
\begin{equation}
  q(\vec\alpha) = -2\ln \left(\frac{L\big(\vec\alpha\,,\,{\hat{\vec\theta}}_{\vec\alpha}\big)}
                              {L(\hat{\vec\alpha},\hat{\vec\theta})\label{eq:LH}}\right).
\end{equation} \par}

The likelihood functions in the numerator and denominator of Eq.~(\ref{eq:LH}) are constructed using products
of probability density functions of signal and background for the various discriminating variables used in the input analyses, as well as constraint
terms for certain NPs.
The probability density functions are derived from simulation for the signal and from both data and simulation for the background.
The quantities $\hat{\vec\alpha}$ and
$\hat{\vec\theta}$ denote the unconditional maximum likelihood estimates of the parameter values,
while $\hat{\vec\theta}_{\vec\alpha}$ denotes the conditional maximum likelihood estimate for fixed values of the parameters
of interest $\vec\alpha$. The choice of the POIs, e.g., signal strengths ($\mu$), couplings modifiers, production cross sections,
branching fractions or related ratios of the above quantities, depends on the specific model under consideration, while the remaining
parameters are treated as NPs. An individual NP represents a single source of systematic uncertainty, and its effect is therefore
considered fully correlated between all of the input analyses included in the fit.

For each model considered, the maximum likelihood estimates $\hat{\vec\alpha}$ are identified as the best fit parameter values. The
$1\sigma$ and $2\sigma$ confidence level (\CL) intervals for one-dimensional measurements of each POI are determined as the union of intervals for which $q(\vec\alpha)<1$ and $q(\vec\alpha)<4$,
respectively, unless otherwise stated. In models with more than one POI, these intervals are determined treating the other POIs as NPs.
The differences between the boundaries of the $1\sigma$ and $2\sigma$ \CL intervals and the best fit value yield the $\pm1\sigma$ and $\pm2\sigma$ uncertainties
on the measurement.
In cases where a physical boundary restricts the interval,
we report a truncated interval and determine the uncertainty from that interval.
(See Fig.~\ref{fig:obsplot_A15PD} and Table~\ref{tab:results_A1_5PD}, for example).
In these cases, the intervals are not expected to maintain coverage.
In the case where the intervals are not contiguous, the interval that contains the best fit point
is used to determine these uncertainties.
The 2D $1\sigma$ and $2\sigma$ \CL regions are determined from the set of parameter values for
which $q(\vec\alpha)<2.30$ and $q(\vec\alpha)<6.18$, respectively, unless otherwise stated.

The likelihood functions are constructed with respect to either the observed data or an Asimov
data set~\cite{Cowan:2010st} constructed using the expected values of the POIs for the SM, in order to obtain the
observed or expected results, respectively.  Because of fluctuations in the observed data the observed intervals may differ from the expected ones.

Finally, the SM predictions for the production and decay rates of the Higgs boson depend on the mass of the Higgs boson, \mH.
For all measurements in this paper, the mass is taken to be $\mH =125.09\pm0.21\mathrm{(stat)}\pm0.11\mathrm{(syst)}\GeV$,
determined from the ATLAS and CMS combined measurement, from the LHC Run 1 data, using the high-resolution \hgg and \hzzllll decay channels \cite{LHC:RunI_Higgs_mass_combination}.
\section{Systematic uncertainties}
\label{sec:systematic_uncertainties}

For many of the POIs, the systematic uncertainties in their determination are expected to be as large
as, or larger than, the statistical uncertainties.
The theoretical uncertainties affecting the signal are among the most important contributions to the systematic uncertainties.
The uncertainties in the total cross section prediction for the signal processes arising from the parton distribution functions, the renormalization
and factorization scales used in the calculations and the branching fraction predictions
are correlated between all input analyses. Instead, theoretical uncertainties that affect
kinematic distributions and cause migrations between event categories are largely uncorrelated between the input analyses.
An exception is the set of theoretical uncertainties for the \ggh production mode, where the correlation scheme described in Section~\ref{ssec:gghunc}
is used to correlate both the normalization and shape uncertainties between input analyses.
The theoretical uncertainties affecting the background predictions, including the parton distribution function uncertainties,
are assumed to be uncorrelated
with those affecting the signal predictions~\cite{LHC-HCG}, with the exception of the uncertainties from the underlying event and parton shower model.

The majority of the systematic uncertainties arising from experimental sources are uncorrelated between the input analyses, with
a few exceptions. The uncertainties in the integrated luminosity measurement~\cite{CMS-PAS-LUM-17-001}, and in the modeling of additional collisions in the event (pileup),
are correlated between all of the input input analyses.
Certain analyses, namely the \htt, \vhbb, and \tthbb analyses, are able to further constrain the jet energy scale uncertainties determined in
auxiliary measurements. The jet energy scale uncertainty in these analyses is
decomposed into several nuisance parameters corresponding to different sources of uncertainty
(for example, different flavor composition and kinematic regions) that are correlated among these analyses but uncorrelated with the other analyses.
An independent jet energy scale uncertainty is assumed to be correlated between the input analyses that are not sensitive to the different sources of uncertainty.
The uncertainties in the $\PQb$ tagging efficiency are correlated between the \tth analyses, but are uncorrelated from the \vhbb analysis, which is
sensitive to different kinematic regions. A separate set of NPs is used to describe the uncertainty in the $\PQb$ tagging efficiency
 in the \hww, \hgg, and \hzz analyses. The uncertainty in the efficiency of the double-$\PQb$-tagger algorithm described in Section~\ref{sec:boostedhbb} is
taken to be uncorrelated from the single $\PQb$ tagging uncertainties.
Finally, the uncertainties in the lepton efficiency and misidentification rate in the \tth-$\tauh$ and \tth-$\Pe/\Pgm$ event classes
are correlated, since the same reconstruction and identification algorithms were used. In other input analyses, different algorithms were used and therefore the uncertainties
are assumed to be uncorrelated.

The free parameters describing the shapes and normalizations of the background models, and parameters
 that allow for the choice of the background parametrization in each of the
\hgg analysis categories are fully determined by the data without any additional constraints,
and are therefore assigned to the statistical uncertainty of a measurement. The remaining uncertainties are assigned to groups of
systematic uncertainties.

\section{Signal strength and cross section fits}
\label{sec:signalstrength}

The signal strength modifier $\mu$, defined as the ratio between the measured Higgs boson yield and its SM expectation, has been used extensively to characterize the Higgs boson yields. However, the specific
meaning of $\mu$ varies depending on the analysis. For a specific production and decay channel $i\to H\to f$, the signal strengths for the production, $\mu_i$, and for the decay, $\mu^f$, are defined as:
\begin{equation}
\mu_i = \frac{\sigma_i}{(\sigma_i)_\SM}\hspace*{0.5cm} {\rm and}\hspace*{0.5cm} \mu^f = \frac{\mathcal{B}^f}{(\mathcal{B}^f)_\SM},\,\,
\label{eq:mui_f}
\end{equation}
respectively. Here $\sigma_i\; (i=\ggh,\,\vbf,\,\wh,\,\zh,\,\tth)$
and $\BR^f\; (f = \cPZ\cPZ,\,\PW\PW,\,\gamgam,\,\tautau,\,\bb,\,\mumu)$ are,
respectively, the production cross section for $i\to \PH$ and the branching fraction for $\PH\to f$. The subscript~''SM''
refers to their respective SM~predictions, so by definition, the SM corresponds to $\mu_i=\mu^f=1$.
Since $\sigma_i$ and $\BR^f$ cannot be separately measured without additional assumptions, only the product of $\mu_i$ and $\mu^f$ can be extracted experimentally,
leading to a signal strength~$\mu_i^f$ for the combined production and decay:
\begin{equation}
 \mu_i^f =  \frac{\sigma_i  \BR^f}{(\sigma_i)_\SM   (\BR^f)_\SM} = \mu_i \mu^f.
\label{eq:muif}
\end{equation}
This parametrization makes use of the narrow width approximation, and the reliability of this approximation was studied in Ref.~\cite{PhysRevD.89.035001} and found to be adequate for global fits.

In this section, results are presented for several signal strength parametrizations starting with a single global signal strength $\mu$,
which is the most restrictive in terms of the number of assumptions. Further parametrizations are defined by relaxing the constraint
that all production and decay rates scale with a common signal strength modifier.

{\tolerance=800 The combined measurement of the common signal strength modifier at $m_{\PH} = 125.09\,\GeV$ is,
\begin{equation}
 \begin{split}
 \mu &= 1.17\pm{0.10} \\
     &= 1.17\pm{0.06}\,\mathrm{(stat)}\,^{+0.06}_{-0.05}\,\mathrm{(sig\,theo)}\,\pm{0.06}\,\mathrm{(other\,syst)},
 \end{split}
\end{equation}
where the total uncertainty has been decomposed into statistical, signal theoretical systematic, and other systematic components.
The largest single source of uncertainty apart from the signal theoretical systematic uncertainties is the integrated luminosity ($\Delta\mu/\mu=2.5\%$),
which is correlated between all of the input analyses. In this measurement and others, however, the other systematic uncertainty component is mostly dominated by
uncertainties that only affect a single input analysis.\par}

\begin{figure*}[hbtp]
\centering
\includegraphics[width=0.49\textwidth]{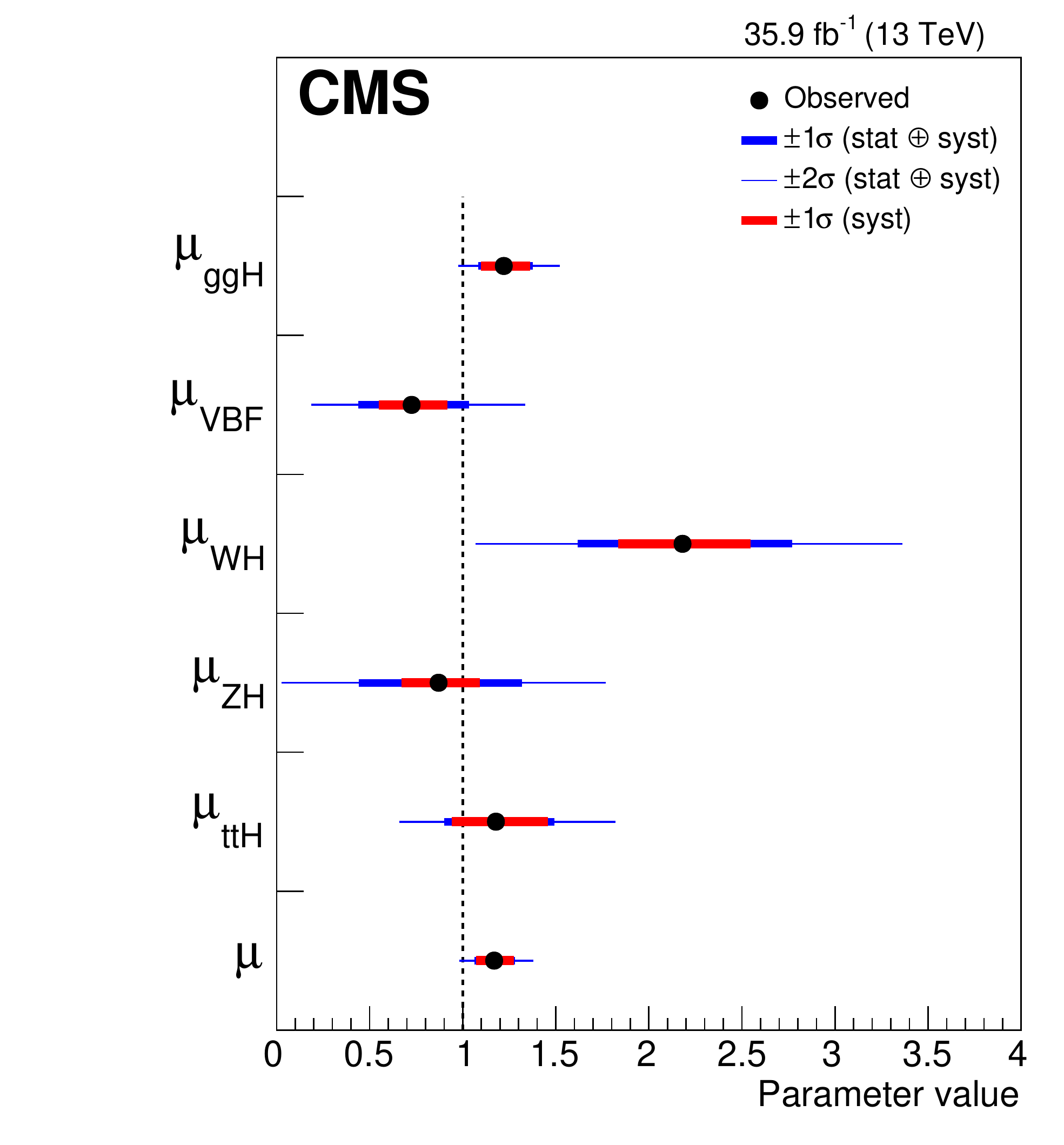}
\includegraphics[width=0.49\textwidth]{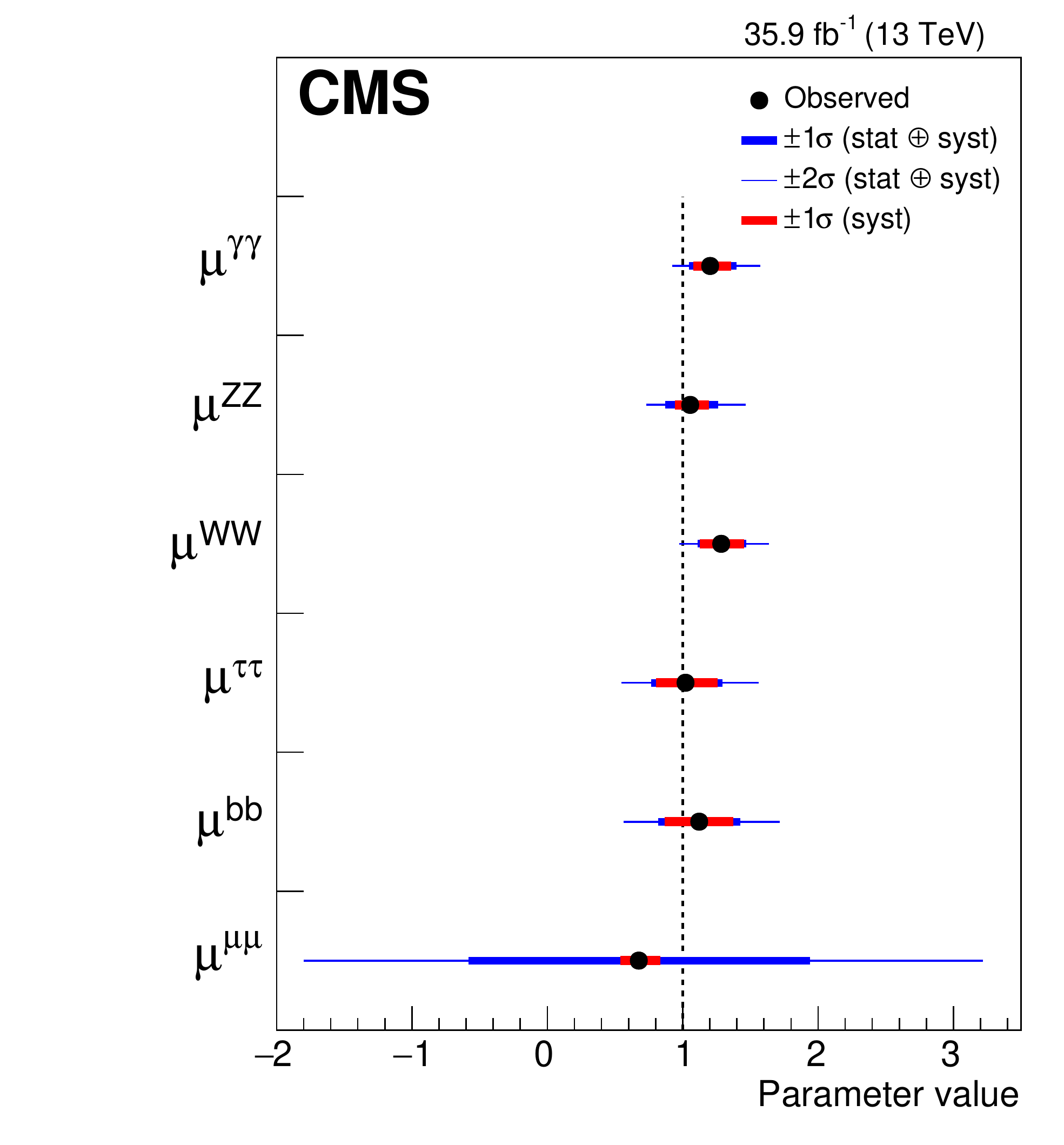}
\caption{Summary plot of the fit to the per-production mode (left) and per-decay mode (right) signal strength modifiers. The thick and thin horizontal bars indicate the $\pm1\sigma$ and $\pm2\sigma$ uncertainties, respectively.
Also shown are the $\pm1\sigma$ systematic components of the uncertainties.
The last point in the per-production mode summary plot is taken from a separate fit and indicates the result of the combined overall signal strength $\mu$.}
\label{fig:obsplot_A15P}
\end{figure*}

{\tolerance=800 Relaxing the assumption of a common production mode scaling, but still assuming the relative SM branching fractions,
leads to a parametrization with five production signal strength modifiers:
$\mu_{\ggh}$, $\mu_{\vbf}$, $\mu_{\wh}$, $\mu_{\zh}$, and $\mu_{\tth}$. In this parametrization, as well as all subsequent parametrizations involving
signal strengths or cross sections, the \tH production is assumed to scale like \tth.
Conversely, relaxing the common decay mode scaling, but assuming the relative SM production cross sections, leads to one
with the modifiers: $\mu^{\gamgam}$, $\mu^{\cPZ\cPZ}$, $\mu^{\PW\PW}$, $\mu^{\tautau}$, $\mu^{\mu\mu}$, and $\mu^{\bb}$.
Results of the fits in these two parametrizations
are summarized in Fig.~\ref{fig:obsplot_A15P}. The numerical values, including the decomposition of the uncertainties into statistical and
systematic components, and the corresponding expected uncertainties, are given in Table~\ref{tab:results_A1_5P}.\par}

\begin{table}[hbtp]
\centering
\topcaption{Best fit values and $\pm 1\sigma$ uncertainties for the parametrizations with per-production mode and per-decay mode signal
strength modifiers. The expected uncertainties are given in brackets. }
\setlength\extrarowheight{3pt}
\begin{tabular}{ccccc}
\hline
Production process & \multicolumn{2}{c}{Best fit value} & \multicolumn{2}{c}{Uncertainty} \\
    &  &  & stat. & syst. \\
\hline
  \ggh  & $1.22$ & $^{+0.14}_{-0.12}$ &  $^{+0.08}_{-0.08}$ & $^{+0.12}_{-0.10}$ \\
        & &  $({}^{+0.11}_{-0.11})$ &     $({}^{+0.07}_{-0.07})$ & $({}^{+0.09}_{-0.08})$ \\[\cmsTabSkip]
  \vbf  & $0.73$ & $^{+0.30}_{-0.27}$ &  $^{+0.24}_{-0.23}$ & $^{+0.17}_{-0.15}$ \\
        & &  $({}^{+0.29}_{-0.27})$ &     $({}^{+0.24}_{-0.23})$ & $({}^{+0.16}_{-0.15})$ \\[\cmsTabSkip]
  \wh   & $2.18$ & $^{+0.58}_{-0.55}$ &  $^{+0.46}_{-0.45}$ & $^{+0.34}_{-0.32}$ \\
        & &  $({}^{+0.53}_{-0.51})$ &     $({}^{+0.43}_{-0.42})$ & $({}^{+0.30}_{-0.29})$ \\[\cmsTabSkip]
  \zh   & $0.87$ & $^{+0.44}_{-0.42}$ &  $^{+0.39}_{-0.38}$ & $^{+0.20}_{-0.18}$ \\
        & &  $({}^{+0.43}_{-0.41})$ &     $({}^{+0.38}_{-0.37})$ & $({}^{+0.19}_{-0.17})$ \\[\cmsTabSkip]
  \tth  & $1.18$ & $^{+0.30}_{-0.27}$ &  $^{+0.16}_{-0.16}$ & $^{+0.26}_{-0.21}$ \\
        & &  $({}^{+0.28}_{-0.25})$ &     $({}^{+0.16}_{-0.15})$ & $({}^{+0.23}_{-0.20})$ \\[\cmsTabSkip]
\hline
Decay mode & \multicolumn{2}{c}{Best fit value} & \multicolumn{2}{c}{Uncertainty} \\
    &  &  & stat. & syst. \\
\hline
  \hbb  & $1.12$ & $^{+0.29}_{-0.29}$ &  $^{+0.19}_{-0.18}$ & $^{+0.22}_{-0.22}$ \\
        & &  $({}^{+0.28}_{-0.27})$ &     $({}^{+0.18}_{-0.18})$ & $({}^{+0.21}_{-0.20})$ \\[\cmsTabSkip]
  \htt  & $1.02$ & $^{+0.26}_{-0.24}$ &  $^{+0.15}_{-0.15}$ & $^{+0.21}_{-0.19}$ \\
        & &  $({}^{+0.24}_{-0.22})$ &     $({}^{+0.15}_{-0.14})$ & $({}^{+0.19}_{-0.17})$  \\[\cmsTabSkip]
  \hww  & $1.28$ & $^{+0.17}_{-0.16}$ &  $^{+0.09}_{-0.09}$ & $^{+0.14}_{-0.13}$ \\
        & &  $({}^{+0.14}_{-0.13})$ &     $({}^{+0.09}_{-0.09})$ & $({}^{+0.11}_{-0.10})$  \\[\cmsTabSkip]
  \hzz  & $1.06$ & $^{+0.19}_{-0.17}$ &  $^{+0.16}_{-0.15}$ & $^{+0.11}_{-0.08}$ \\
        & &  $({}^{+0.18}_{-0.16})$ &     $({}^{+0.15}_{-0.14})$ & $({}^{+0.10}_{-0.08})$  \\[\cmsTabSkip]
  \hgg  & $1.20$ & $^{+0.18}_{-0.14}$ &  $^{+0.13}_{-0.11}$ & $^{+0.12}_{-0.09}$ \\
        & &  $({}^{+0.14}_{-0.12})$ &     $({}^{+0.10}_{-0.10})$ & $({}^{+0.09}_{-0.07})$  \\[\cmsTabSkip]
  \hmm  & $0.68$ & $^{+1.25}_{-1.24}$ &  $^{+1.24}_{-1.24}$ & $^{+0.13}_{-0.11}$ \\
        & &  $({}^{+1.20}_{-1.17})$ &     $({}^{+1.18}_{-1.17})$ & $({}^{+0.19}_{-0.03})$ \\[\cmsTabSkip]
\hline
\end{tabular}
\label{tab:results_A1_5P}
\end{table}

The improvement in the precision of the measurement
of the \ggh production rate of $\sim$50\% (from $\sim$20\% to $\sim$10\%) compared to Ref.~\cite{CMSRun1} and
$\sim$33\% (from $\sim$15\% to $\sim$10\%) compared to Ref.~\cite{ATLASCMSRun1}, can be attributed to the combined
effects of an increased \ggh production cross section, and a reduction in the associated theoretical uncertainties.
The improvements in the precision are up to $\sim$20\% for the \vbf and \vh production rates compared to Ref.~\cite{CMSRun1} .
The uncertainty in the measurement of the \tth production rate is reduced by around 50\% compared to Ref.~\cite{ATLASCMSRun1}.  This is in part due to the increase in the
\tth cross section between 8 and 13\TeV, but also due to the inclusion of additional exclusive event categories for this production process.

\begin{figure*}[hbtp]
\centering
\includegraphics[width=0.6\textwidth]{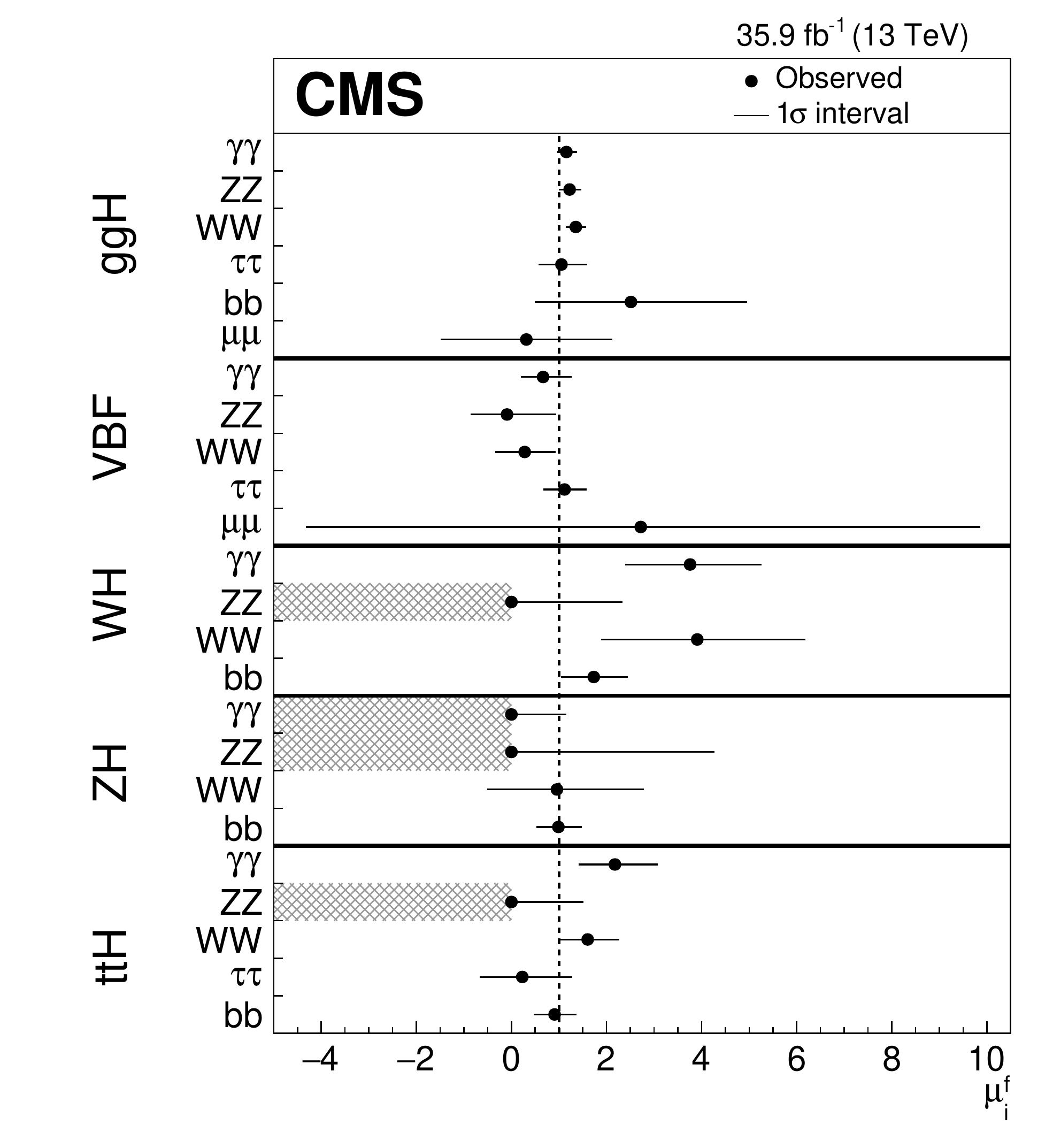}%
\caption{Summary plot of the fit to the production--decay signal strength products $\mu_{i}^{f}=\mu_{i} \mu^{f}$. The points indicate the
best fit values while the horizontal bars indicate the $1\sigma$ \CL intervals. The hatched areas indicate signal strengths that are restricted to nonnegative
values as described in the text.}
\label{fig:obsplot_A15PD}
\end{figure*}

The most generic signal strength parametrization has one signal strength parameter for each production and decay mode combination, $\mu_{i}^{f}$.
Given the five production and six decay modes listed above, this implies a model with 30 parameters of interest.
However not all can be experimentally constrained in this combination.
There is no dedicated analysis from CMS at $\sqrt{s}=13\TeV$ targeting \wh and \zh production with \htt decay, or \vbf production with \hbb decay,
therefore these signal strength modifiers are fixed to the SM expectation and are not included in the maximum likelihood fit.
Likewise, the \wh, \zh, and \tth production rates with \hmm decay are fixed to the SM expectation.
In the case of \wh, \zh, and \tth production with \hzz decay, as well as \zh production with \hgg decay, the background contamination is
sufficiently low so that a negative signal strength can result in an overall negative event yield. Therefore, these signal strengths
are restricted to nonnegative values.
Figure~\ref{fig:obsplot_A15PD} summarizes the results in this model along with the $1\sigma$ \CL intervals.
The numerical values, including the uncertainty decomposition into statistical and systematic parts, and the corresponding
expected uncertainties, are given in Table~\ref{tab:results_A1_5PD}.

\begin{table*}[hbtp!]
\centering
\topcaption{Best fit values and $\pm 1\sigma$ uncertainties for the parameters of the model with one signal strength parameter for each production and decay mode combination. The entries in the table represent the parameter
$\mu_{i}^{f}=\mu_{i} \mu^{f}$, where $i$ is indicated by the column and $f$ by the row. The expected uncertainties are given in brackets. Some of the
signal strengths are restricted to nonnegative values, as described in the text.}
\cmsTable{
\setlength\extrarowheight{3pt}
\setlength\tabcolsep{1pt}
\renewcommand{\arraystretch}{1.1}
\begin{tabular}{l@{\hskip 2pt}|rccc|rccc|rccc|rccc|rccc}
\multicolumn{21}{c}{~~~}\\ \hline
\multicolumn{1}{l|}{\multirow{2}{*}{\begin{tabular}{@{}l}Decay\\[-3pt] mode\\\mbox{}\end{tabular}}} &  \multicolumn{20}{c}{Production process}\\
\multicolumn{1}{l|}{} & \multicolumn{4}{c|}{\ggh}  &  \multicolumn{4}{c|}{\vbf}  &  \multicolumn{4}{c|}{\wh}  &  \multicolumn{4}{c|}{\zh} &  \multicolumn{4}{c}{\tth}  \\
\multicolumn{1}{l|}{} & \multicolumn{2}{c}{Best fit}  & \multicolumn{2}{c|}{Uncertainty}  &  \multicolumn{2}{c}{Best fit} & \multicolumn{2}{c|}{Uncertainty}  &  \multicolumn{2}{c}{Best fit}  & \multicolumn{2}{c|}{Uncertainty}  &  \multicolumn{2}{c}{Best fit}  & \multicolumn{2}{c|}{Uncertainty}  &  \multicolumn{2}{c}{Best fit}  & \multicolumn{2}{c}{Uncertainty}  \\[-3pt]
\multicolumn{1}{l|}{} & \multicolumn{2}{c}{value} &  stat  &  \multicolumn{1}{c|}{syst}  &  \multicolumn{2}{c}{value} &  stat  &  \multicolumn{1}{c|}{syst}  &  \multicolumn{2}{c}{value} &  stat  &  \multicolumn{1}{c|}{syst}  &  \multicolumn{2}{c}{value} &  stat  &  \multicolumn{1}{c|}{syst}  &  \multicolumn{2}{c}{value} &  stat  &  syst  \\
\hline
\renewcommand{\arraystretch}{1.8}
\hbb
 &  $2.51$ & $^{+2.43}_{-2.01}$ &  $^{+1.96}_{-1.92}$ & $^{+1.44}_{-0.59}$
 & \multicolumn{4}{c|}{\NA}
 &  $1.73$ & $^{+0.70}_{-0.68}$ &  $^{+0.53}_{-0.51}$ & $^{+0.46}_{-0.45}$
 &  $0.99$ & $^{+0.47}_{-0.45}$ &  $^{+0.41}_{-0.40}$ & $^{+0.23}_{-0.20}$
 &  $0.91$ & $^{+0.45}_{-0.43}$ &  $^{+0.24}_{-0.23}$ & $^{+0.38}_{-0.36}$
\\
 &  &  $({}^{+2.06}_{-1.86})$ &     $({}^{+1.85}_{-1.83})$ & $({}^{+0.89}_{-0.33})$
 & \multicolumn{4}{c|}{\NA}
 &  &  $({}^{+0.69}_{-0.67})$ &     $({}^{+0.52}_{-0.51})$ & $({}^{+0.45}_{-0.44})$
 &  &  $({}^{+0.46}_{-0.44})$ &     $({}^{+0.40}_{-0.39})$ & $({}^{+0.23}_{-0.20})$
 &  &  $({}^{+0.44}_{-0.42})$ &     $({}^{+0.23}_{-0.23})$ & $({}^{+0.37}_{-0.35})$
 \\[\cmsTabSkip]
\htt
 &  $1.05$ & $^{+0.53}_{-0.47}$ &  $^{+0.25}_{-0.25}$ & $^{+0.46}_{-0.40}$
 &  $1.12$ & $^{+0.45}_{-0.43}$ &  $^{+0.37}_{-0.35}$ & $^{+0.25}_{-0.25}$
 & \multicolumn{4}{c|}{\NA}
 & \multicolumn{4}{c|}{\NA}
 &  $0.23$ & $^{+1.03}_{-0.88}$ &  $^{+0.80}_{-0.71}$ & $^{+0.65}_{-0.52}$
\\
 &  &  $({}^{+0.45}_{-0.41})$ &     $({}^{+0.23}_{-0.23})$ & $({}^{+0.38}_{-0.34})$
 &  &  $({}^{+0.45}_{-0.43})$ &     $({}^{+0.37}_{-0.35})$ & $({}^{+0.25}_{-0.24})$
 & \multicolumn{4}{c|}{\NA}
 & \multicolumn{4}{c|}{\NA}
 &  &  $({}^{+0.98}_{-0.87})$ &     $({}^{+0.80}_{-0.73})$ & $({}^{+0.56}_{-0.47})$
 \\[\cmsTabSkip]
\hww
 &  $1.35$ & $^{+0.21}_{-0.19}$ &  $^{+0.12}_{-0.12}$ & $^{+0.17}_{-0.15}$
 &  $0.28$ & $^{+0.64}_{-0.60}$ &  $^{+0.58}_{-0.53}$ & $^{+0.28}_{-0.28}$
 &  $3.91$ & $^{+2.26}_{-2.01}$ &  $^{+1.89}_{-1.72}$ & $^{+1.24}_{-1.05}$
 &  $0.96$ & $^{+1.81}_{-1.46}$ &  $^{+1.74}_{-1.44}$ & $^{+0.50}_{-0.22}$
 &  $1.60$ & $^{+0.65}_{-0.59}$ &  $^{+0.40}_{-0.39}$ & $^{+0.52}_{-0.45}$
\\
 &  &  $({}^{+0.17}_{-0.16})$ &     $({}^{+0.10}_{-0.10})$ & $({}^{+0.13}_{-0.12})$
 &  &  $({}^{+0.62}_{-0.58})$ &     $({}^{+0.57}_{-0.53})$ & $({}^{+0.26}_{-0.25})$
 &  &  $({}^{+1.47}_{-1.19})$ &     $({}^{+1.32}_{-1.06})$ & $({}^{+0.64}_{-0.54})$
 &  &  $({}^{+1.67}_{-1.37})$ &     $({}^{+1.61}_{-1.35})$ & $({}^{+0.45}_{-0.20})$
 &  &  $({}^{+0.56}_{-0.53})$ &     $({}^{+0.38}_{-0.38})$ & $({}^{+0.41}_{-0.37})$
 \\[\cmsTabSkip]
\hzz
 &  $1.22$ & $^{+0.23}_{-0.21}$ &  $^{+0.20}_{-0.19}$ & $^{+0.12}_{-0.10}$
 &  $-0.09$ & $^{+1.02}_{-0.76}$ &  $^{+1.00}_{-0.72}$ & $^{+0.21}_{-0.22}$
 &  $0.00$ & $^{+2.33}_{+0.00}$ &  $^{+2.31}_{-0.00}$ & $^{+0.30}_{-0.00}$
 &  $0.00$ & $^{+4.26}_{+0.00}$ &  $^{+4.19}_{-0.00}$ & $^{+0.80}_{-0.00}$
 &  $0.00$ & $^{+1.50}_{+0.00}$ &  $^{+1.47}_{-0.00}$ & $^{+0.30}_{-0.00}$
\\
 &  &  $({}^{+0.22}_{-0.20})$ &     $({}^{+0.20}_{-0.19})$ & $({}^{+0.10}_{-0.07})$
 &  &  $({}^{+1.27}_{-0.99})$ &     $({}^{+1.25}_{-0.97})$ & $({}^{+0.23}_{-0.21})$
 &  &  $({}^{+4.46}_{-0.99})$ &     $({}^{+4.42}_{-0.99})$ & $({}^{+0.57}_{-0.00})$
 &  &  $({}^{+7.57}_{-1.00})$ &     $({}^{+7.45}_{-1.00})$ & $({}^{+1.33}_{-0.00})$
 &  &  $({}^{+2.95}_{-0.99})$ &     $({}^{+2.89}_{-0.99})$ & $({}^{+0.59}_{-0.00})$
 \\[\cmsTabSkip]
\hgg
 &  $1.16$ & $^{+0.21}_{-0.18}$ &  $^{+0.17}_{-0.15}$ & $^{+0.13}_{-0.10}$
 &  $0.67$ & $^{+0.59}_{-0.46}$ &  $^{+0.49}_{-0.42}$ & $^{+0.32}_{-0.18}$
 &  $3.76$ & $^{+1.48}_{-1.35}$ &  $^{+1.45}_{-1.33}$ & $^{+0.33}_{-0.24}$
 &  $0.00$ & $^{+1.14}_{+0.00}$ &  $^{+1.14}_{-0.00}$ & $^{+0.09}_{-0.00}$
 &  $2.18$ & $^{+0.88}_{-0.75}$ &  $^{+0.82}_{-0.74}$ & $^{+0.32}_{-0.14}$
\\
 &  &  $({}^{+0.17}_{-0.16})$ &     $({}^{+0.14}_{-0.14})$ & $({}^{+0.11}_{-0.08})$
 &  &  $({}^{+0.59}_{-0.48})$ &     $({}^{+0.48}_{-0.43})$ & $({}^{+0.34}_{-0.21})$
 &  &  $({}^{+1.28}_{-1.16})$ &     $({}^{+1.27}_{-1.16})$ & $({}^{+0.13}_{-0.06})$
 &  &  $({}^{+2.51}_{-1.04})$ &     $({}^{+2.50}_{-1.04})$ & $({}^{+0.25}_{-0.00})$
 &  &  $({}^{+0.74}_{-0.63})$ &     $({}^{+0.72}_{-0.63})$ & $({}^{+0.16}_{-0.06})$
 \\[\cmsTabSkip]
\hmm
 &  $0.31$ & $^{+1.80}_{-1.79}$ &  $^{+1.79}_{-1.78}$ & $^{+0.19}_{-0.19}$
 &  $2.72$ & $^{+7.12}_{-7.03}$ &  $^{+7.12}_{-7.04}$ & $^{+0.26}_{-0.00}$
 & \multicolumn{4}{c|}{\NA}
 & \multicolumn{4}{c|}{\NA}
 & \multicolumn{4}{c}{\NA}
\\
 &  &  $({}^{+1.69}_{-1.65})$ &     $({}^{+1.67}_{-1.67})$ & $({}^{+0.28}_{-0.00})$
 &  &  $({}^{+7.02}_{-6.94})$ &     $({}^{+7.01}_{-6.93})$ & $({}^{+0.38}_{-0.50})$
 & \multicolumn{4}{c|}{\NA}
 & \multicolumn{4}{c|}{\NA}
 & \multicolumn{4}{c}{\NA}
\\[2pt] \hline
\end{tabular}
}
\label{tab:results_A1_5PD}
\end{table*}

\subsection{Ratios of cross sections and branching fractions, relative to \texorpdfstring{$\ggh\to \cPZ\cPZ$}{gluon fusion production and ZZ decay}}
\label{ssec:B1ZZ}

\begin{figure}[hbtp]
\centering
\includegraphics[width=0.45\textwidth]{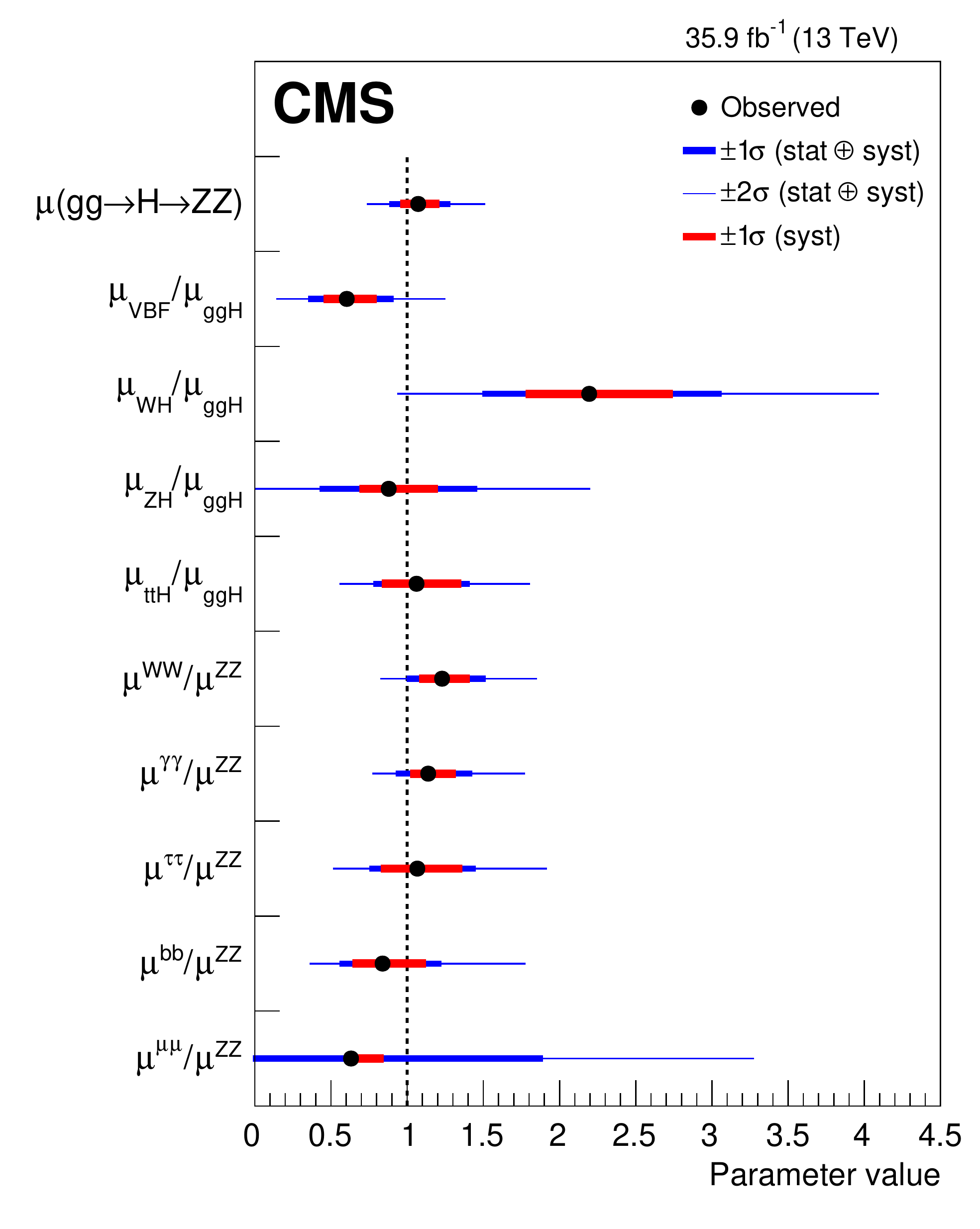}
\caption{Summary of the cross section and branching fraction ratio model. The thick and thin horizontal bars indicate the $\pm1\sigma$ and $\pm2\sigma$ uncertainties, respectively.
Also shown are the $\pm1\sigma$ systematic components of the uncertainties.
}
\label{fig:obsplot_B1ZZ}
\end{figure}

Results are presented for a model based on the ratios of cross sections and branching fractions. These are given relative to a well-measured reference process, chosen
to be $\ggh\to \cPZ\cPZ$ ($\mu_{\ggh}^{\cPZ\cPZ}$). Using ratios has the advantage that some systematic or theoretical uncertainties common to both the numerator and denominator cancel.
The following ratios are used: $\mu^{\gamgam}/\mu^{\cPZ\cPZ}$,
$\mu^{\PW\PW}/\mu^{\cPZ\cPZ}$, $\mu^{\tautau}/\mu^{\cPZ\cPZ}$,$\mu^{\mu\mu}/\mu^{\cPZ\cPZ}$, $\mu^{\bb}/\mu^{\cPZ\cPZ}$, $\mu_{\vbf}/\mu_{\ggh}$, $\mu_{\wh}/\mu_{\ggh}$, $\mu_{\zh}/\mu_{\ggh}$, and $\mu_{\tth}/\mu_{\ggh}$. These results are
summarized in Fig.~\ref{fig:obsplot_B1ZZ}, and the numerical values are given in~Table~\ref{tab:results_B1ZZ}.
The uncertainties in the SM predictions are included in the measurements.

\begin{table*}[hbtp!]
\centering
\topcaption{Best fit values and $\pm 1\sigma$ uncertainties for the parameters of the cross section and branching fraction ratio model. The expected uncertainties are given in brackets.}
\cmsTableX{0.8\textwidth}{
\begin{tabular}{@{} c r@{}c c c  c r@{}c c c @{}}
\hline
   &   &    & \multicolumn{2}{c}{Uncertainty} &   &    &   & \multicolumn{2}{c}{Uncertainty}  \\
  Parameter & \multicolumn{2}{c}{Best fit} & stat & syst  & Parameter & \multicolumn{2}{c}{Best fit} & stat & syst  \\
 \hline\\[-1.5ex]
\multirow{2}{*}{$\mu_{\ggh}^{\cPZ\cPZ}$} &  $1.07$ & $^{+0.20}_{-0.18}$ &  $^{+0.16}_{-0.15}$ & $^{+0.11}_{-0.09}$ &  \multirow{2}{*}{${\mathcal{B}}^{\bb}/{\mathcal{B}}^{\cPZ\cPZ}$} &   $0.84$ & $^{+0.37}_{-0.27}$ &  $^{+0.27}_{-0.21}$ & $^{+0.26}_{-0.17}$  \\[1pt]
  & &  $(^{+0.19}_{-0.16})$ &  $(^{+0.15}_{-0.14})$ & $(^{+0.11}_{-0.08})$ &  & &  $(^{+0.56}_{-0.37})$ &  $(^{+0.38}_{-0.28})$ & $(^{+0.40}_{-0.25})$ \\[10pt]
\multirow{2}{*}{$\mu_{\vbf}/\mu_{\ggh}$} &  $0.60$ & $^{+0.30}_{-0.24}$ &  $^{+0.24}_{-0.21}$ & $^{+0.17}_{-0.13}$ &  \multirow{2}{*}{${\mathcal{B}}^{\tautau}/{\mathcal{B}}^{\cPZ\cPZ}$} &   $1.07$ & $^{+0.37}_{-0.30}$ &  $^{+0.25}_{-0.21}$ & $^{+0.27}_{-0.22}$  \\[1pt]
  & &  $(^{+0.40}_{-0.32})$ &  $(^{+0.31}_{-0.27})$ & $(^{+0.24}_{-0.17})$ &  & &  $(^{+0.35}_{-0.28})$ &  $(^{+0.25}_{-0.20})$ & $(^{+0.25}_{-0.19})$ \\[10pt]
\multirow{2}{*}{$\mu_{\wh}/\mu_{\ggh}$} &  $2.19$ & $^{+0.86}_{-0.69}$ &  $^{+0.68}_{-0.56}$ & $^{+0.52}_{-0.39}$ &  \multirow{2}{*}{${\mathcal{B}}^{\PW\PW}/{\mathcal{B}}^{\cPZ\cPZ}$} &   $1.23$ & $^{+0.27}_{-0.22}$ &  $^{+0.22}_{-0.18}$ & $^{+0.16}_{-0.13}$  \\[1pt]
  & &  $(^{+0.65}_{-0.52})$ &  $(^{+0.53}_{-0.44})$ & $(^{+0.39}_{-0.29})$ &  & &  $(^{+0.24}_{-0.19})$ &  $(^{+0.19}_{-0.16})$ & $(^{+0.14}_{-0.11})$ \\[10pt]
\multirow{2}{*}{$\mu_{\zh}/\mu_{\ggh}$} &  $0.88$ & $^{+0.57}_{-0.44}$ &  $^{+0.49}_{-0.41}$ & $^{+0.30}_{-0.17}$ &  \multirow{2}{*}{${\mathcal{B}}^{\gamma\gamma}/{\mathcal{B}}^{\cPZ\cPZ}$} &   $1.14$ & $^{+0.28}_{-0.20}$ &  $^{+0.23}_{-0.18}$ & $^{+0.16}_{-0.09}$  \\[1pt]
  & &  $(^{+0.68}_{-0.47})$ &  $(^{+0.53}_{-0.41})$ & $(^{+0.43}_{-0.23})$ &  & &  $(^{+0.23}_{-0.18})$ &  $(^{+0.20}_{-0.16})$ & $(^{+0.11}_{-0.08})$ \\[10pt]
\multirow{2}{*}{$\mu_{\tth}/\mu_{\ggh}$} &  $1.06$ & $^{+0.34}_{-0.27}$ &  $^{+0.20}_{-0.18}$ & $^{+0.27}_{-0.20}$ &  \multirow{2}{*}{${\mathcal{B}}^{\mu\mu}/{\mathcal{B}}^{\cPZ\cPZ}$} &   $0.63$ & $^{+1.24}_{-1.21}$ &  $^{+1.24}_{-1.20}$ & $^{+0.15}_{-0.11}$  \\[1pt]
  & &  $(^{+0.36}_{-0.30})$ &  $(^{+0.23}_{-0.21})$ & $(^{+0.27}_{-0.21})$ &  & &  $(^{+1.26}_{-1.19})$ &  $(^{+1.25}_{-1.19})$ & $(^{+0.20}_{-0.03})$ \\[10pt]
\hline
\end{tabular}}
\label{tab:results_B1ZZ}
\end{table*}

\subsection{Stage 0 simplified template cross sections}

Measurements of production cross sections, which are complementary to the signal strength parametrization, are made for seven processes defined according to the
simplified template cross sections proposed in Ref.~\cite{YR4}. The results given here are for the stage 0 fiducial regions defined by the rapidity of the Higgs boson $\abs{y_{\PH}} < 2.5$. All input analyses have a negligible acceptance for $\abs{y_{\PH}} > 2.5$. Defining the fiducial region in this way reduces the theoretical uncertainty that would otherwise apply while extrapolating to the fully inclusive phase space. Subsequent stages involve splitting the fiducial regions into a number of smaller ones, for example based on ranges of the Higgs boson \pt. The measured cross sections are defined as:

{\tolerance=800 \begin{itemize}
\item $\STggH$: gluon fusion and $\PQb$-associated production. While Ref.~\cite{YR4} proposes separate bins for these modes, they are merged here because of the current lack of sensitivity to the associated production with $\PQb$ quarks.
\item $\STqqH$: \vbf production.
\item $\VHHQQ$: Associated production with a $\cPZ$ or $\PW$ boson, either quark or gluon initiated, in which the vector boson decays hadronically.
\item $\QQGGHLL$: Associated production with a $\cPZ$ boson, in which the $\cPZ$ boson decays leptonically. While Ref.~\cite{YR4} proposes separate bins for the quark- and gluon-initiated modes, they are merged here because they cannot easily be distinguished experimentally, and therefore, their measurements would be highly anticorrelated.
\item $\QQHLNU$: Associated production with a $\PW$ boson, in which the $\PW$ decays leptonically.
\item $\STttH$: Associated production with a pair of top quarks or a single top quark. While Ref.~\cite{YR4} proposes separate bins for these modes, they are merged here because of the lack of a dedicated analysis targeting $\tH$ production in this combination.
\end{itemize}\par}

\begin{figure*}[hbtp]
\centering
\includegraphics[width=0.6\textwidth]{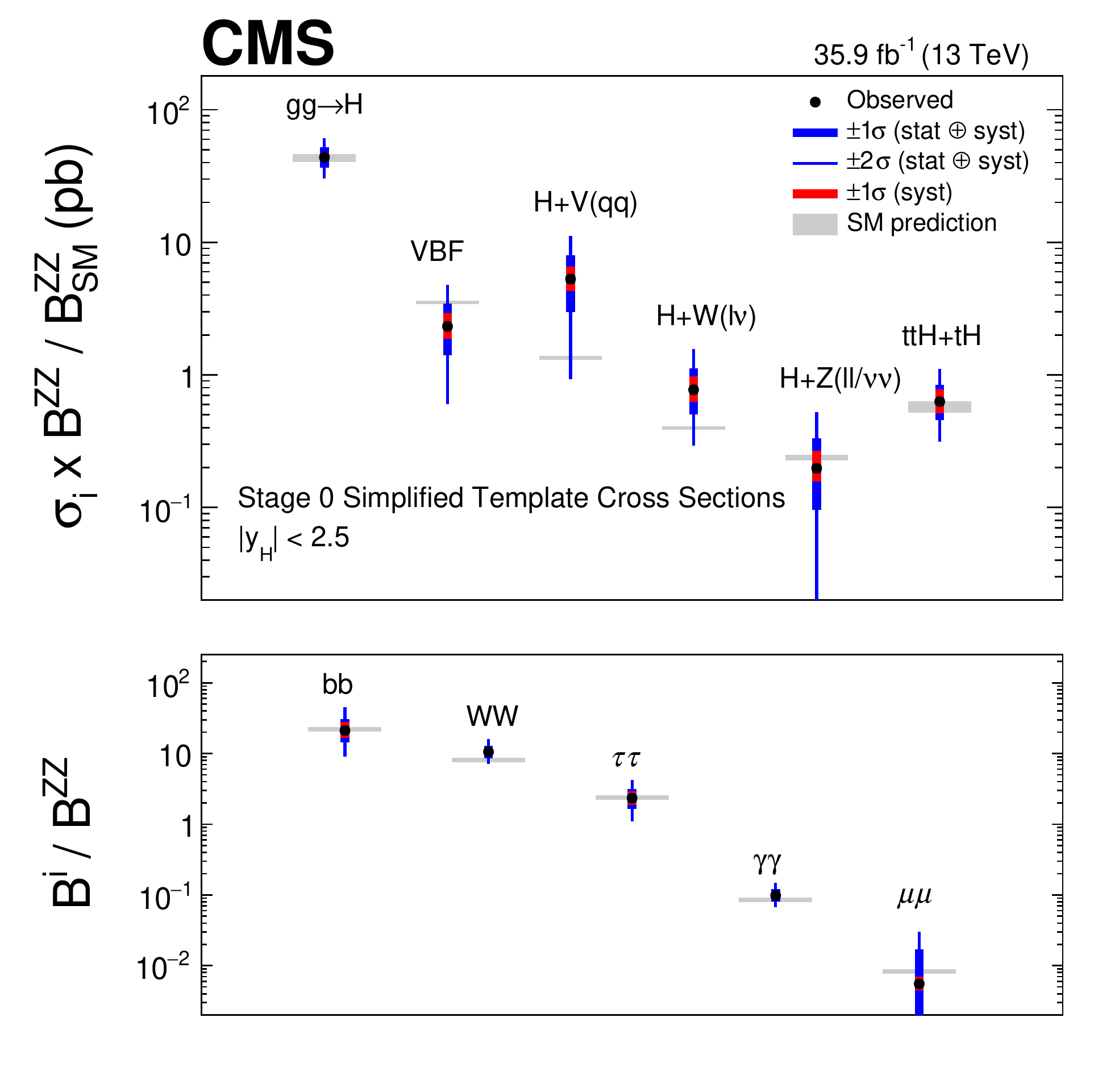}
\caption{Summary of the stage 0 model, ratios of cross sections and branching fractions. The points indicate the best fit values, while the error bars show the $\pm1\sigma$ and $\pm2\sigma$ uncertainties.
The $\pm1\sigma$ uncertainties on the measurements considering only the contributions from the systematic uncertainties are also shown. The uncertainties in the SM predictions are indicated.}
\label{fig:obsplot_stage0ZZ}
\end{figure*}

{\tolerance=800 In addition to the cross sections, the Higgs boson branching fractions are also included as POIs via ratios with respect to $\mathcal{B}^{\cPZ\cPZ}$. A summary of the
results in this model, normalized to the expected SM cross sections, is given in Fig.~\ref{fig:obsplot_stage0ZZ} and Table~\ref{tab:results_stage0ZZ}.
Since cross sections are measured and not signal strength modifiers, the theoretical uncertainties in these cross sections do not enter as sources of uncertainty.
In Fig.~\ref{fig:obsplot_stage0ZZ}, the uncertainties in the SM predictions are indicated by gray bands.\par}

\begin{table*}[hbtp!]
\centering
\topcaption{Best fit values and $\pm 1\sigma$ uncertainties for the parameters of the stage 0 simplified template cross section model. The values are all normalized to the SM predictions. The expected uncertainties are given in brackets.}
\cmsTableX{0.8\textwidth}{
\begin{tabular}{@{} c r@{}c c c  c r@{}c c c @{}}
\hline
   &   &    & \multicolumn{2}{c}{Uncertainty} &   &    &   & \multicolumn{2}{c}{Uncertainty}  \\
  Parameter & \multicolumn{2}{c}{Best fit} & stat & syst  & Parameter & \multicolumn{2}{c}{Best fit} & stat & syst  \\
 \hline\\[-1.5ex]
\multirow{2}{*}{$\sigma_{\ggh}  {\mathcal{B}}^{\cPZ\cPZ}$} &  $1.00$ & $^{+0.19}_{-0.16}$ &  $^{+0.16}_{-0.15}$ & $^{+0.09}_{-0.07}$ &  \multirow{2}{*}{${\mathcal{B}}^{\bb}/{\mathcal{B}}^{\cPZ\cPZ}$} &   $0.96$ & $^{+0.44}_{-0.31}$ &  $^{+0.32}_{-0.24}$ & $^{+0.30}_{-0.20}$  \\[1pt]
  & &  $(^{+0.18}_{-0.16})$ &  $(^{+0.16}_{-0.15})$ & $(^{+0.09}_{-0.07})$ &  & &  $(^{+0.57}_{-0.38})$ &  $(^{+0.40}_{-0.29})$ & $(^{+0.41}_{-0.25})$ \\[10pt]
\multirow{2}{*}{$\sigma_{\vbf}  {\mathcal{B}}^{\cPZ\cPZ}$} &  $0.66$ & $^{+0.32}_{-0.26}$ &  $^{+0.27}_{-0.22}$ & $^{+0.17}_{-0.13}$ &  \multirow{2}{*}{${\mathcal{B}}^{\tautau}/{\mathcal{B}}^{\cPZ\cPZ}$} &   $0.98$ & $^{+0.35}_{-0.28}$ &  $^{+0.24}_{-0.20}$ & $^{+0.25}_{-0.20}$  \\[1pt]
  & &  $(^{+0.40}_{-0.32})$ &  $(^{+0.33}_{-0.27})$ & $(^{+0.22}_{-0.16})$ &  & &  $(^{+0.36}_{-0.29})$ &  $(^{+0.26}_{-0.21})$ & $(^{+0.25}_{-0.19})$ \\[10pt]
\multirow{2}{*}{$\sigma_{\PH+\rm{V}(\Pq\Pq)}  {\mathcal{B}}^{\cPZ\cPZ}$} &  $3.93$ & $^{+2.00}_{-1.71}$ &  $^{+1.77}_{-1.53}$ & $^{+0.93}_{-0.75}$ &  \multirow{2}{*}{${\mathcal{B}}^{\PW\PW}/{\mathcal{B}}^{\cPZ\cPZ}$} &   $1.30$ & $^{+0.29}_{-0.24}$ &  $^{+0.24}_{-0.20}$ & $^{+0.17}_{-0.13}$  \\[1pt]
  & &  $(^{+1.66}_{-1.05})$ &  $(^{+1.49}_{-1.05})$ & $(^{+0.72}_{-0.00})$ &  & &  $(^{+0.24}_{-0.20})$ &  $(^{+0.20}_{-0.16})$ & $(^{+0.14}_{-0.11})$ \\[10pt]
\multirow{2}{*}{$\sigma_{\PH+\PW(\ell\nu)}  {\mathcal{B}}^{\cPZ\cPZ}$} &  $1.95$ & $^{+0.88}_{-0.68}$ &  $^{+0.72}_{-0.57}$ & $^{+0.51}_{-0.38}$ &  \multirow{2}{*}{${\mathcal{B}}^{\gamma\gamma}/{\mathcal{B}}^{\cPZ\cPZ}$} &   $1.14$ & $^{+0.26}_{-0.20}$ &  $^{+0.23}_{-0.18}$ & $^{+0.13}_{-0.09}$  \\[1pt]
  & &  $(^{+0.69}_{-0.52})$ &  $(^{+0.56}_{-0.44})$ & $(^{+0.40}_{-0.29})$ &  & &  $(^{+0.23}_{-0.19})$ &  $(^{+0.21}_{-0.17})$ & $(^{+0.11}_{-0.08})$ \\[10pt]
\multirow{2}{*}{$\sigma_{\PH+{\cPZ(\ell\ell/\nu\nu)}}  {\mathcal{B}}^{\cPZ\cPZ}$} &  $0.84$ & $^{+0.57}_{-0.43}$ &  $^{+0.49}_{-0.40}$ & $^{+0.29}_{-0.17}$ &  \multirow{2}{*}{${\mathcal{B}}^{\mu\mu}/{\mathcal{B}}^{\cPZ\cPZ}$} &   $0.67$ & $^{+1.40}_{-1.36}$ &  $^{+1.39}_{-1.35}$ & $^{+0.18}_{-0.13}$  \\[1pt]
  & &  $(^{+0.71}_{-0.46})$ &  $(^{+0.56}_{-0.41})$ & $(^{+0.44}_{-0.22})$ &  & &  $(^{+1.35}_{-1.28})$ &  $(^{+1.34}_{-1.28})$ & $(^{+0.17}_{-0.05})$ \\[10pt]
\multirow{2}{*}{$\sigma_{\tth}  {\mathcal{B}}^{\cPZ\cPZ}$} &  $1.08$ & $^{+0.37}_{-0.30}$ &  $^{+0.26}_{-0.22}$ & $^{+0.26}_{-0.19}$ &  \multirow{2}{*}{$$} & \multicolumn{4}{c}{~}  \\[1pt]
  & &  $(^{+0.38}_{-0.31})$ &  $(^{+0.28}_{-0.23})$ & $(^{+0.26}_{-0.20})$ &  \multirow{2}{*}{~} & \multicolumn{4}{c}{~} \\[10pt]
\hline
\end{tabular}}
\label{tab:results_stage0ZZ}
\end{table*}

\section{Measurements of Higgs boson couplings}
\label{sec:couplings}

In the $\kappa$-framework~\cite{LHCHXSWGYR3}, coupling modifiers are introduced in order to test for deviations in the couplings of the Higgs boson to other particles. In order to measure the individual Higgs couplings in this framework, some assumption must be made to constrain the total Higgs boson width since
it cannot be directly measured at the LHC.
Unless stated otherwise, it is assumed that there are no BSM contributions to the total Higgs boson width.
With this assumption, the cross section times branching fraction for a production process $i$ and decay $f$ can be expressed as,
\begin{equation}
\sigma_i  {\BR}^f = \frac{\sigma_{\mathit{i}}(\vec\kappa) \Gamma^{\mathit{f}}(\vec\kappa)}{\Gamma_{\PH}(\vec\kappa)},
\end{equation}
where $\Gamma_{\PH}(\vec\kappa)$ is the total width of the Higgs boson and $\Gamma^{\mathit{f}}(\vec\kappa)$ is the partial width of the Higgs boson decay to the final state~$f$.
A set of coupling modifiers,
$\vec\kappa$, is introduced to parameterize potential deviations in the bosonic and fermionic couplings of the Higgs boson from the SM predictions.
For a given production process or decay mode $j$, a coupling modifier $\kappa_j$ is defined such that,
\begin{equation}
\label{eq:kappa}
  \kappa_j^2=\sigma_j/\sigma_j^\SM\ \ \ \mathrm{or}\ \ \  \kappa_j^2=\Gamma^j/\Gamma^j_\SM.
\end{equation}
In the SM, all $\kappa_j$ values are positive and equal to unity.
In this parametrization it is assumed that the higher-order accuracy of the QCD and electroweak corrections to the SM cross sections and branching fractions is preserved when the values of $\kappa_j$ deviate from unity. While this does not hold in general, for the parameter ranges considered in this paper the dominant higher-order QCD corrections largely factorize from the rescaling of the couplings, therefore the approach is considered valid.
Individual coupling modifiers, corresponding to tree-level Higgs boson couplings to the different particles, are introduced, as well as effective coupling modifiers $\kappa_{\cPg}$ and $\kappa_\PGg$ that describe \ggh production and \hgg decay. This is possible because the presence of any BSM~particles in these loops is not expected to significantly change the corresponding kinematic properties of the processes. This approach is not possible for $\cPg\cPg\to\zh$ production, which occurs at leading order through box and triangular loop diagrams, because a tree-level contact interaction from BSM physics would likely exhibit a kinematic structure very different from the SM, and is expected to be highly suppressed~\cite{Englert:2013vua}.
Other possible BSM effects on the $\cPg\cPg\to\zh$ process are related to modifications of the $\PH\cPZ\cPZ$ and $\tth$ vertices, which are best taken into account, within the limitation of the framework,
by resolving the loop in terms of the corresponding coupling modifiers, $\kappa_{\cPZ}$ and $\kappa_{\PQt}$.
More details on the development of this framework as well as its theoretical and phenomenological foundations and extensions can be found, for example, in Refs.~\cite{PhysLettB.318.155,NuclPhysB.433.41,PhysRevD.62.013009,PhysRevD.67.115001,PhysRevD.70.113009,PhysLettB.636.107,JEHP04.2010.126,JEHP06.2012.140,PhysRevD.86.075013,NuclPhysB.868.416,PhysRevLett.109.101801,JHEP07.2013.035,PhysRevD.88.033004,JHEP05.2014.046,JHEP05.2015.175}.

The normalization scaling effects of each of the $\kappa$ parameters are given in Table~\ref{tab:kexpr}. Loop processes such as \ggh and \hgg can be
studied through either the effective coupling modifiers, thereby providing sensitivity to potential BSM physics in the loops, or the modifiers of the
SM particles themselves. Interference between different diagrams, such as those that contribute to $\cPg\cPg\to\zh$, provides some sensitivity
to relative signs between Higgs boson couplings to different particles.
Modifications to the kinematic distributions of the \tH production are also expected when the relative sign of $\kappa_{\PQt}$ and $\kappa_{\PW}$
is negative. These effects were studied and the distributions of the final observables were found to be insensitive with the present dataset
to the relative sign of $\kappa_{\PQt}$ and $\kappa_{\PW}$.

\begin{table*}[h!]
\centering
\topcaption{Normalization scaling factors for all relevant production cross sections and decay partial widths. For the $\kappa$ parameters representing loop processes, the resolved scaling in terms of the fundamental SM couplings is also given.}
\centering
\cmsTableX{0.95\textwidth}{
\begin{tabular}{lcccl}\hline
 & & & Effective &   \\
   & Loops & Interference & scaling factor & Resolved scaling factor \\
\hline
Production &&&&\\
\hspace*{5mm} $\sigma(\ggh)$ & $\checkmark$ & \Pg-\PQt &  $\kappa_{\Pg}^2 $&$ 1.04  \kappa_{\PQt}^2 + 0.002   \kappa_{\PQb}^2 - 0.038 \kappa_{\PQt}\kappa_{\PQb}$ \\
\hspace*{5mm} $\sigma(\vbf)$     & \NA            &  \NA    &   & $ 0.73   \kappa_{\PW}^2 + 0.27   \kappa_{\PZ}^2$ \\
\hspace*{5mm} $\sigma(\wh)$             & \NA            &  \NA    &  & $\kappa_{\PW}^2$\\
\hspace*{5mm} $\sigma(\Pq\Pq/\Pq\cPg \to \PZ\PH)$ & \NA            &  \NA    &  & $\kappa_{\PZ}^2$\\
\hspace*{5mm} $\sigma(\cPg\cPg \to \PZ\PH)$       & $\checkmark$ & \PZ-\PQt &  & $2.46 \kappa_{\PZ}^2 + 0.47  \kappa_{\PQt}^2 - 1.94   \kappa_{\PZ}\kappa_{\PQt} $\\
\hspace*{5mm} $\sigma(\tth)$            & \NA            &  \NA    &  & $\kappa_{\PQt}^2$ \\
\hspace*{5mm} $\sigma(\cPg\PQb \to \PW\PQt\PH)$     & \NA            & \PW-\PQt &    & $ 2.91   \kappa_{\PQt}^2 + 2.31   \kappa_{\PW}^2 - 4.22   \kappa_{\PQt}\kappa_{\PW}$ \\
\hspace*{5mm} $\sigma(\Pq\PQb \to \PQt\PH\Pq)$ & \NA          & \PW-\PQt &    & $ 2.63   \kappa_{\PQt}^2 + 3.58   \kappa_{\PW}^2 - 5.21   \kappa_{\PQt}\kappa_{\PW}$ \\
\hspace*{5mm} $\sigma(\bb\PH)$            & \NA            &  \NA    &  & $\kappa_{\PQb}^2$ \\[\cmsTabSkip]
Partial decay width \\
\hspace*{5mm} $\Gamma^{\PZ\PZ}$            & \NA             &  \NA    &  & $\kappa_{\PZ}^2$ \\
\hspace*{5mm} $\Gamma^{\PW\PW}$            & \NA             &  \NA    &  & $\kappa_{\PW}^2$ \\
\hspace*{5mm} $\Gamma^{\gamma\gamma}$  & $\checkmark$  & \PW-\PQt &  $\kappa_{\PGg}^2 $&$ 1.59   \kappa_{\PW}^2 + 0.07   \kappa_{\PQt}^2 -0.67   \kappa_{\PW} \kappa_{\PQt}$ \\
\hspace*{5mm} $\Gamma^{\tau\tau}$      & \NA             &  \NA    &  & $\kappa_{\tau}^2$ \\
\hspace*{5mm} $\Gamma^{\bb}$      & \NA             &  \NA    &  & $\kappa_{\PQb}^2$ \\
\hspace*{5mm} $\Gamma^{\mu\mu}$        & \NA             &  \NA    &  & $\kappa_{\mu}^2$ \\[\cmsTabSkip]
Total width for $\BRbsm=0$ \\
& & & & $0.58   \kappa_{\PQb}^2 + 0.22   \kappa_{\PW}^2 + 0.08   \kappa_{\Pg}^2 +$\\
\hspace*{5mm} $\Gamma_{\PH}$  & $\checkmark$  & \NA  &  $\kappa_{\PH}^2 $ & $+\,0.06   \kappa_{\PGt}^2 + 0.026   \kappa_{\PZ}^2 + 0.029   \kappa_{\PQc}^2 + $ \\
& & & & $+\,0.0023   \kappa_{\PGg}^2 +\,0.0015   \kappa_{\PZ\PGg}^2 +$\\
& & & & $+\,0.00025   \kappa_{\PQs}^2 + 0.00022   \kappa_{\PGm}^2$\\
\hline
\end{tabular}
}
\label{tab:kexpr}
\end{table*}

\subsection{Generic model within \texorpdfstring{$\kappa$}{kappa}-framework assuming resolved loops}
\label{ssec:resolvedkappas}

\begin{figure}[hbtp]
\centering
\includegraphics[width=0.45\textwidth]{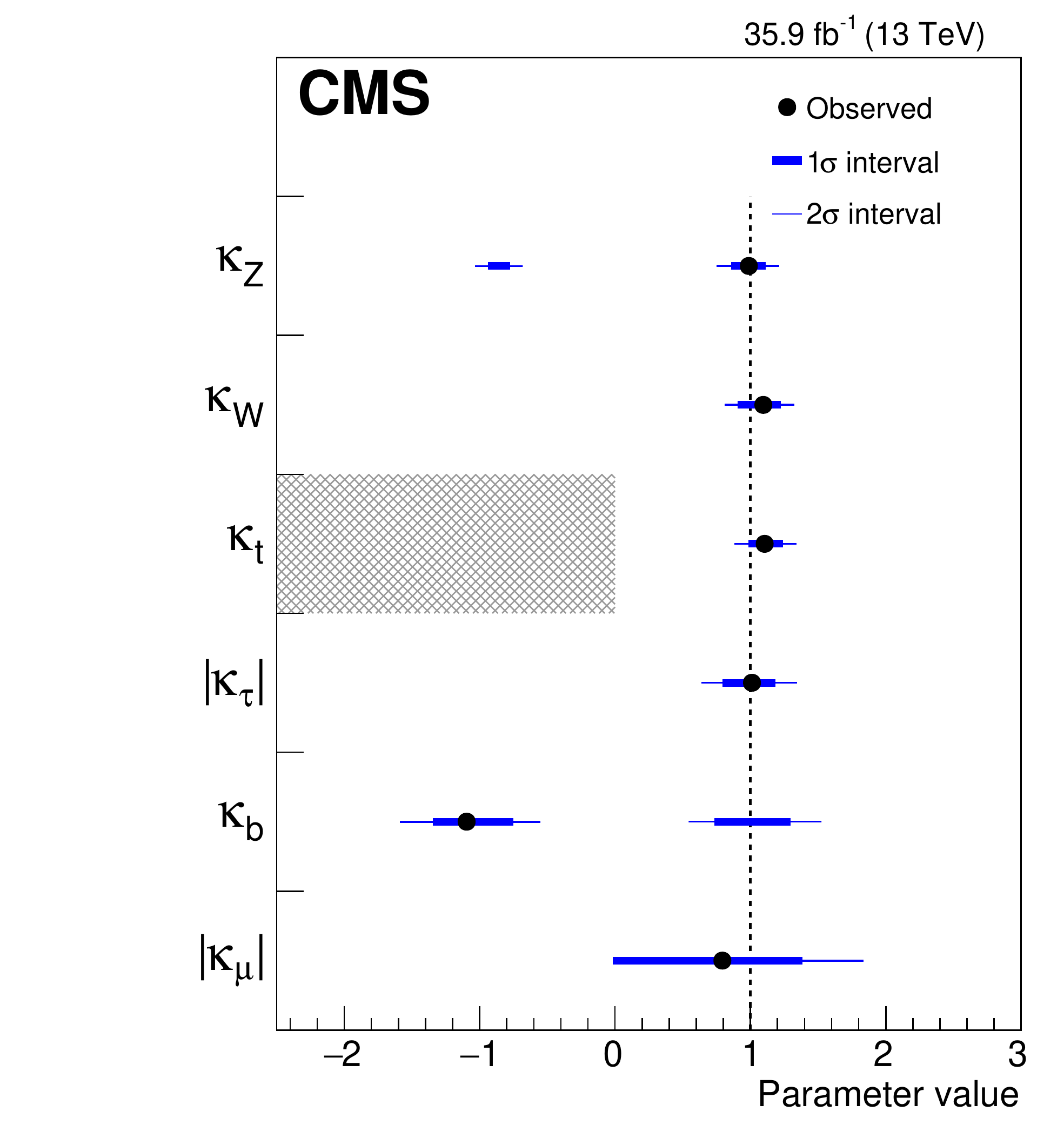}
\caption{Summary of the $\kappa$-framework model assuming resolved loops and $\BRbsm=0$. The points indicate the best fit values while the
thick and thin horizontal bars show the $1\sigma$ and $2\sigma$ \CL intervals, respectively.
In this model, the \ggh\ and \hgg\ loops are resolved in terms of the remaining coupling modifiers.
For this model, both positive and negative values of $\kappa_{\PW}$, $\kappa_{\cPZ}$, and $\kappa_{\PQb}$ are considered.
Negative values of $\kappa_{\PW}$ in this model are disfavored by more than $2\sigma$.
}
\label{fig:obsplot_K1}
\end{figure}

Under the assumption that there are no BSM particles contributing to the \ggh production or $\hgg$ decay loops, these processes can be expressed in terms of the coupling
modifiers to the SM particles as described previously.
There are six free coupling parameters: $\kappa_{\PW}$, $\kappa_{\cPZ}$, $\kappa_{\PQt}$, $\kappa_{\Pgt}$, $\kappa_{\PQb}$, and
$\kappa_{\Pgm}$.
Without loss of generality, the value of $\kappa_{\PQt}$ is restricted to be positive, while both negative and positive values of $\kappa_{\PW}$, $\kappa_{\cPZ}$ and $\kappa_{\PQb}$ are allowed.
In this model, the rates of the $\ggh$ and $\hgg$ processes, which occur through loop diagrams at leading order, are resolved, meaning that they are described by the functions of
$\kappa_{\PW}$, $\kappa_{\cPZ}$, $\kappa_{\Pgt}$, and $\kappa_{\PQb}$ given in Table~\ref{tab:kexpr}.
The results of the fits with this parametrization are given in Fig.~\ref{fig:obsplot_K1} and Table~\ref{tab:results_K1_mm}.

The rate of the \hzz decay and \zh production depend only on the absolute value of $\kappa_{\cPZ}$. The interference between the two
diagrams shown in Fig.~\ref{fig:feynam_diag_ZH}, however, allows contributions from the $\Pg\Pg\to\zh$ production mode
to break the degeneracy between the signs, leading to a positive value of $\kappa_{\cPZ}$ being preferred. As these contributions are typically
small compared to other production modes, the 1$\sigma$ and 2$\sigma$ intervals also include negative values of $\kappa_{\cPZ}$. Although a negative value of
$\kappa_{\PQb}$ is preferred in this model, the difference in $q$ between the best fit point and the minimum in the region $\kappa_{\PQb}>0$ is smaller than 0.1.

An additional fit is performed using a phenomenological
parametrization relating the masses of the fermions and vector bosons to the
corresponding $\kappa$ modifiers using two
parameters, denoted $M$ and $\epsilon$~\cite{EllisYou2012,EllisYou2013}.
In such a model one can relate the coupling modifiers to $M$ and $\epsilon$ as
$\kappa_{\mathrm{F}} = v \; m_\mathrm{f}^{\epsilon} / M^{1+\epsilon}$
for fermions and
$\kappa_{\mathrm{V}} = v \; m_\mathrm{V}^{2\epsilon} / M^{1+2\epsilon}$
for vector bosons.
Here, $v=246.22$\GeV, is the SM Higgs boson vacuum expectation value~\cite{Patrignani:2016xqp}. The SM expectation, $\kappa_{i}=1$, is recovered when
$(M,\epsilon)=(v,0)$.

The lepton and vector boson mass values are taken from Ref.~\cite{Patrignani:2016xqp}, while
the top quark mass is taken to be 172.5\GeV for consistency with theoretical calculations used in setting the SM predictions.
The bottom quark mass is evaluated at the scale of the Higgs boson mass,
$m_\PQb(\mH=125\GeV)=2.76\GeV$.

\begin{figure*}[hbtp]
\centering
\includegraphics[width=0.49\textwidth]{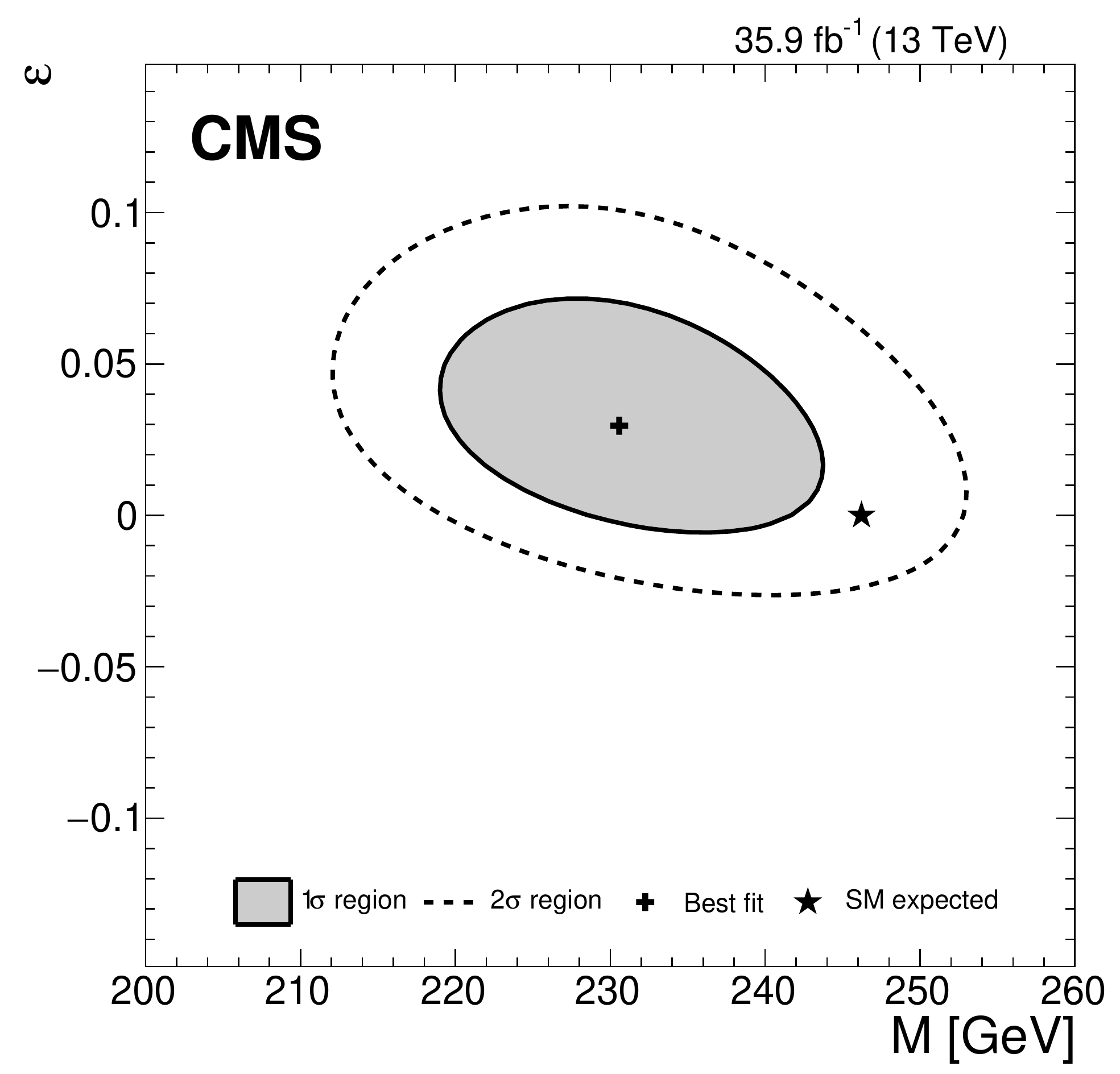}
\includegraphics[width=0.49\textwidth]{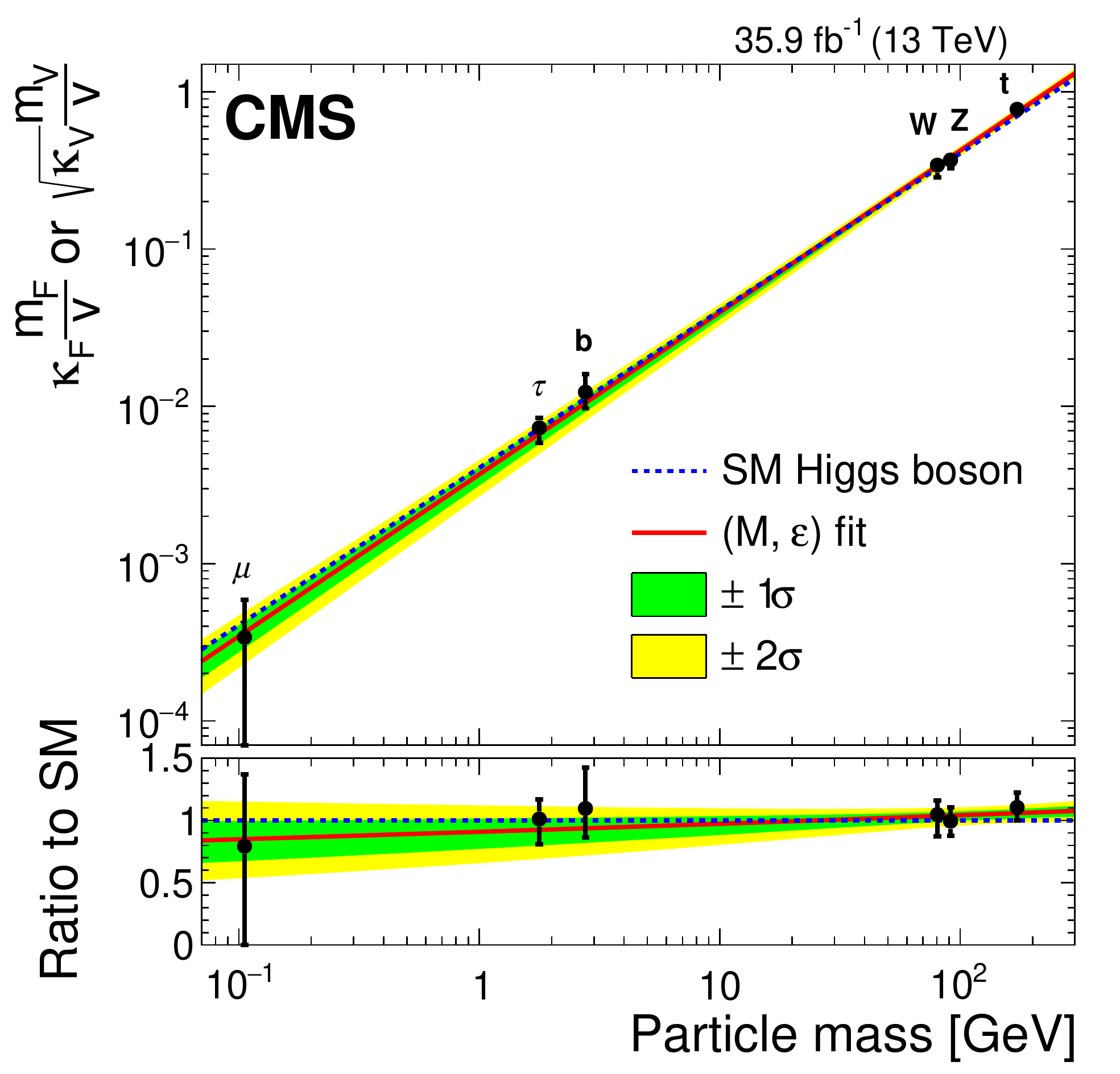}
\caption[]{Likelihood scan in the $M$-$\epsilon$ plane (left). The best fit point and the $1\sigma$ and $2\sigma$ \CL regions are shown, along with the
SM prediction. Result of the phenomenological $(M,\epsilon)$ fit overlayed with the resolved $\kappa$-framework model (right).}
\label{fig:obsplot_K1_meps}
\end{figure*}

The $1\sigma$ and $2\sigma$ \CL regions in the $(M,\epsilon)$ fit are shown in Fig.~\ref{fig:obsplot_K1_meps}~(left).
The results of the fit using the six parameter $\kappa$ model are plotted versus the particle masses in
Fig.~\ref{fig:obsplot_K1_meps}~(right), and the result of the $(M,\epsilon)$ fit is also shown for comparison. For the
$\PQb$ quark, since the best fit point for $\kappa_{\PQb}$ is negative, the absolute value of this coupling modifier is shown.
In order to show both the Yukawa and vector boson couplings in the same plot,
a ``reduced'' vector boson coupling $\sqrt{\kV}  m_\mathrm{V}/v$ is shown.

\begin{table}[hbtp]
\centering
\topcaption{Best fit values and $\pm 1\sigma$ uncertainties for the parameters of the $\kappa$ model in
which the loop processes are resolved. The expected uncertainties are given in brackets.}
\setlength\extrarowheight{3pt}
\begin{tabular}{ccccc}
\hline
Parameter & \multicolumn{2}{c}{Best fit value} & \multicolumn{2}{c}{Uncertainty} \\
    &  &  & stat. & syst. \\
\hline
  $\kappa_{\PW}$  & $1.10$ & $^{+0.12}_{-0.17}$ &  $^{+0.08}_{-0.16}$ & $^{+0.08}_{-0.06}$ \\
                  & &  $({}^{+0.11}_{-0.10})$ &     $({}^{+0.08}_{-0.08})$ & $({}^{+0.06}_{-0.06})$ \\[\cmsTabSkip]

  $\kappa_{\cPZ}$ & $0.99$ & $^{+0.11}_{-0.12}$ &  $^{+0.09}_{-0.10}$ & $^{+0.06}_{-0.07}$ \\
                  & &  $({}^{+0.11}_{-0.11})$ &     $({}^{+0.09}_{-0.09})$ & $({}^{+0.06}_{-0.06})$ \\[\cmsTabSkip]

  $\kappa_{\PQt}$ & $1.11$ & $^{+0.12}_{-0.10}$ &  $^{+0.07}_{-0.07}$ & $^{+0.09}_{-0.08}$ \\
                  & &  $({}^{+0.11}_{-0.12})$ &     $({}^{+0.07}_{-0.08})$ & $({}^{+0.09}_{-0.09})$ \\[\cmsTabSkip]

  $\kappa_{\PQb}$ & $-1.10$ & $^{+0.33}_{-0.23}$ &  $^{+0.29}_{-0.16}$ & $^{+0.15}_{-0.17}$ \\
                  & &  $({}^{+0.22}_{-0.22})$ &     $({}^{+0.15}_{-0.15})$ & $({}^{+0.17}_{-0.16})$ \\[\cmsTabSkip]

  $\kappa_{\Pgt}$ & $1.01$ & $^{+0.16}_{-0.20}$ &  $^{+0.11}_{-0.17}$ & $^{+0.12}_{-0.11}$ \\
                  & &  $({}^{+0.17}_{-0.15})$ &     $({}^{+0.12}_{-0.10})$ & $({}^{+0.12}_{-0.11})$ \\[\cmsTabSkip]

  $\kappa_{\mu}$ & $0.79$ & $^{+0.58}_{-0.79}$ &  $^{+0.56}_{-0.80}$ & $^{+0.14}_{-0.00}$ \\
                 & &  $({}^{+0.50}_{-1.01})$ &     $({}^{+0.50}_{-1.01})$ & $({}^{+0.08}_{-0.10})$ \\[\cmsTabSkip]
\hline
\end{tabular}
\label{tab:results_K1_mm}
\end{table}

\subsection{Generic model within \texorpdfstring{$\kappa$}{kappa}-framework with effective loops}

\begin{figure*}[hbtp]
\centering
\includegraphics[width=0.55\textwidth]{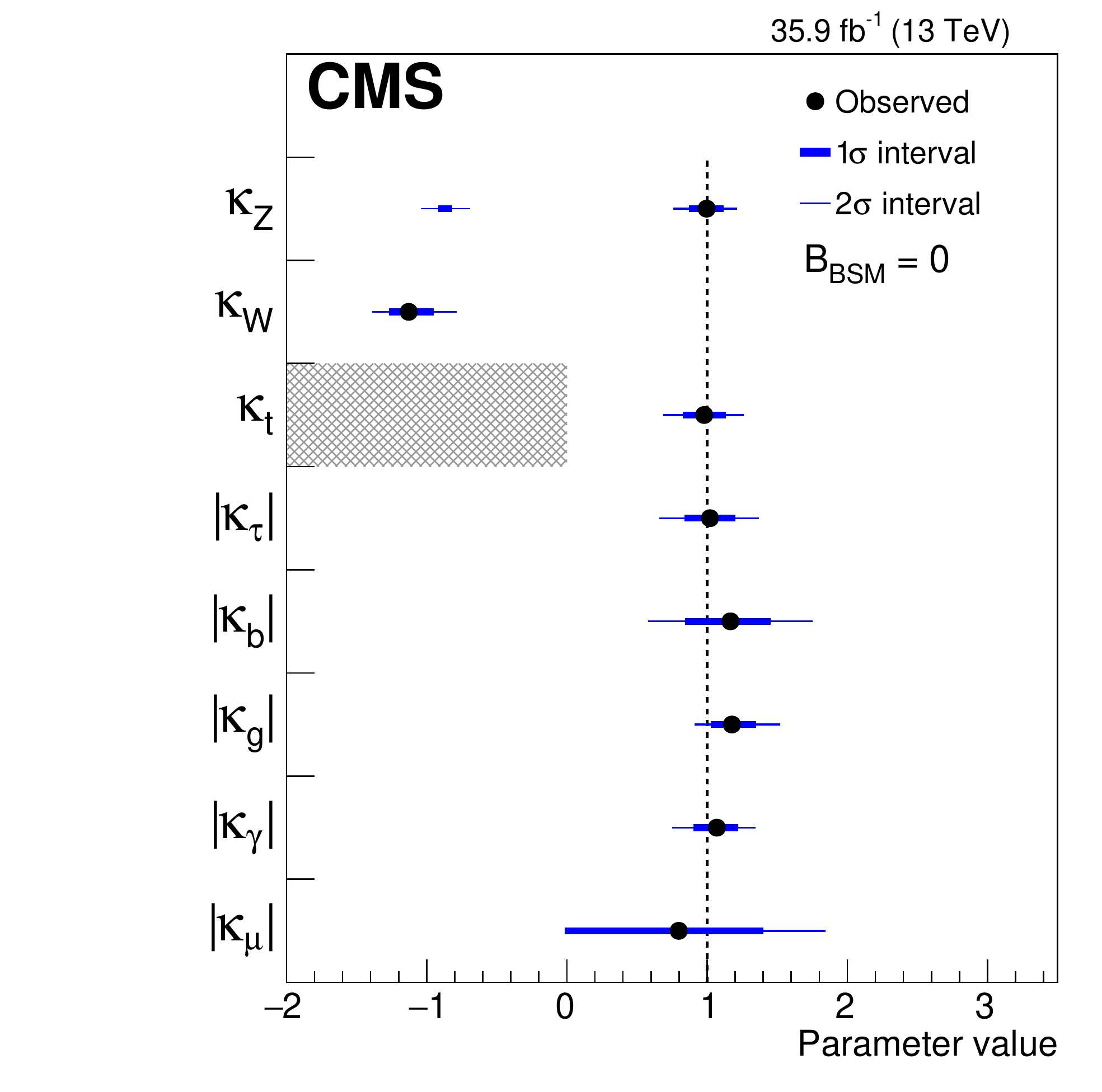}
\includegraphics[width=0.44\textwidth]{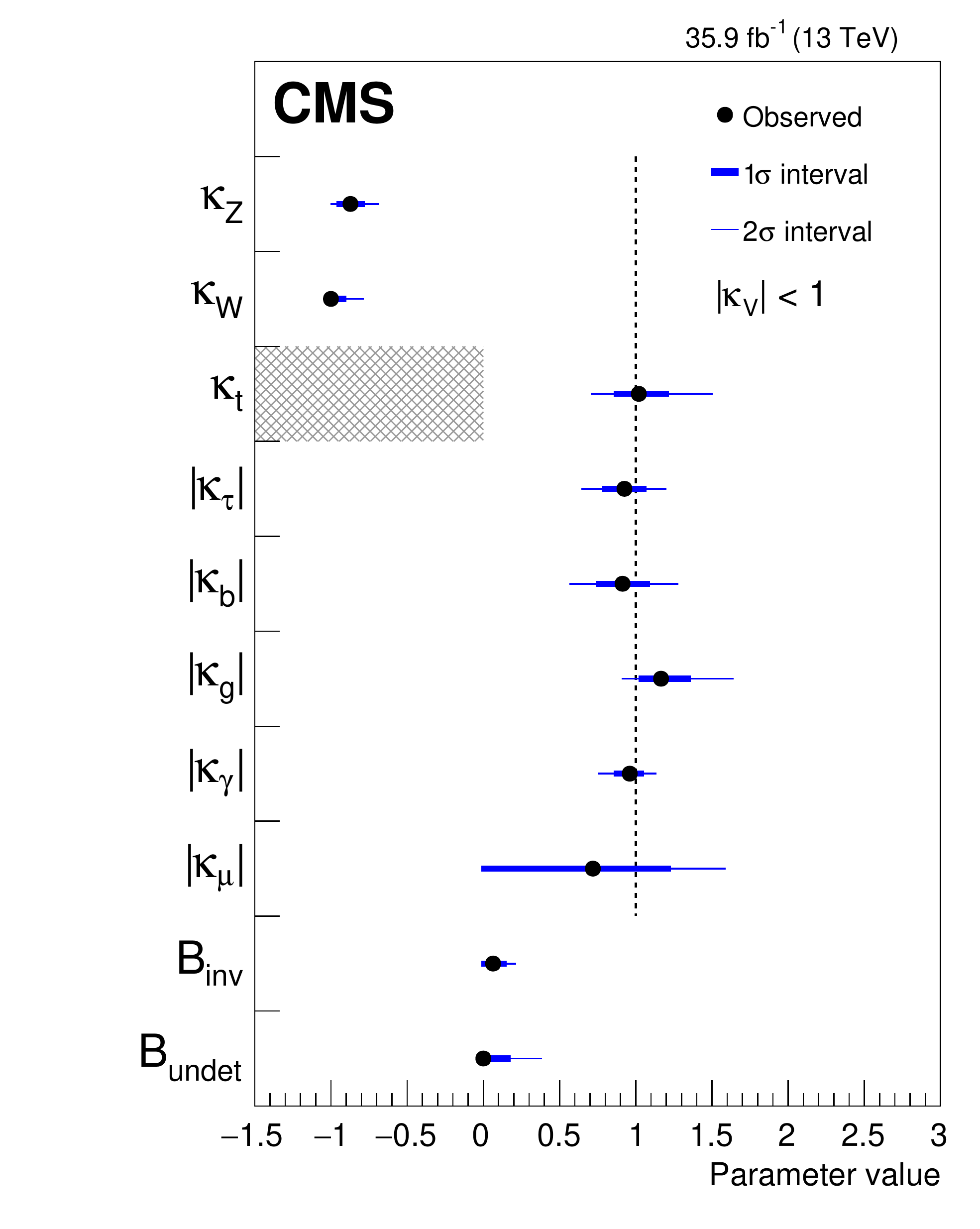}
\caption{Summary plots for the $\kappa$-framework model in which the \ggh\ and \hgg\ loops are scaled with effective couplings. The points indicate
the best fit values while the
thick and thin horizontal bars show the $1\sigma$ and $2\sigma$ \CL intervals, respectively.
In the left figure the constraint $\BRbsm=0$ is imposed, and both positive and negative values of $\kappa_{\PW}$ and $\kappa_{\cPZ}$ are
considered.
In the right figure a constraint $\abs{\kappa_{\PW}},\,\abs{\kappa_{\cPZ}}\le 1$ is imposed (same sign of $\kappa_{\PW}$ and $\kappa_{\cPZ}$), while $\BRinv>0$ and $\BRundet>0$ are free parameters. }
\label{fig:obsplot_K2}
\end{figure*}

The results of the fits to the generic $\kappa$ model where the \ggh and \hgg\ loops are scaled using the effective coupling modifiers
$\kappa_{\cPg}$ and $\kappa_{\PGg}$ are given in Fig.~\ref{fig:obsplot_K2} and Table~\ref{tab:results_K2}. In this parametrization, additional contributions
from BSM decays are allowed for by rewriting the total width of the Higgs boson, relative to its SM value, as,
\begin{equation}
\label{eq:width_bsm}
  \frac{\Gamma_{\PH}}{\Gamma_{\PH}^{\text{SM}}} = \frac{\kappa_{\PH}^{2}}{1-\left(\BRundet+\BRinv\right)},
\end{equation}
where $\kappa_{\PH}$ is defined in Table~\ref{tab:kexpr}.

\begin{table*}[hbtp]
\centering
\topcaption{Best fit values and $\pm 1\sigma$ uncertainties for the parameters of the $\kappa$-framework model with effective loops.
The expected uncertainties are given in brackets.}
\cmsTableX{0.8\textwidth}{
\begin{tabular}{@{} c r@{}c c c  c r@{}c c c @{}}
\hline
\multicolumn{5}{l}{$\BRbsm=0$} & \multicolumn{5}{l}{$\BRbsm>0$,~~$\abs{\kappa_{\mathrm{V}}}<1$} \\
    &            &               &  \multicolumn{2}{c}{Uncertainty} &               &             & & \multicolumn{2}{c}{Uncertainty}  \\
  Parameter & \multicolumn{2}{c}{Best fit} & stat & \multicolumn{1}{c}{syst}                   & Parameter & \multicolumn{2}{c}{Best fit} & stat & syst                    \\
 \hline\\[-1.5ex]
\multirow{2}{*}{$\kappa_{\cPZ}$} &  $1.00$ & $^{+0.11}_{-0.11}$ &  $^{+0.09}_{-0.09}$ & $^{+0.06}_{-0.07}$ &  \multirow{2}{*}{$\kappa_{\cPZ}$} &   $-0.87$ & $^{+0.08}_{-0.08}$ &  $^{+0.07}_{-0.06}$ & $^{+0.04}_{-0.04}$  \\[1pt]
  & &  $(^{+0.11}_{-0.11})$ &  $(^{+0.09}_{-0.09})$ & $(^{+0.06}_{-0.06})$ &  & &  $(^{+0.00}_{-0.12})$ &  $(^{+0.00}_{-0.10})$ & $(^{+0.00}_{-0.06})$ \\[10pt]
\multirow{2}{*}{$\kappa_{\PW}$} &  $-1.13$ & $^{+0.16}_{-0.13}$ &  $^{+0.15}_{-0.10}$ & $^{+0.06}_{-0.08}$ &  \multirow{2}{*}{$\kappa_{\PW}$} &   $-1.00$ & $^{+0.09}_{-0.00}$ &  $^{+0.07}_{-0.00}$ & $^{+0.05}_{-0.00}$  \\[1pt]
  & &  $(^{+0.12}_{-0.12})$ &  $(^{+0.09}_{-0.09})$ & $(^{+0.07}_{-0.07})$ &  & &  $(^{+0.00}_{-0.12})$ &  $(^{+0.00}_{-0.09})$ & $(^{+0.00}_{-0.07})$ \\[10pt]
\multirow{2}{*}{$\kappa_{\PQt}$} &  $0.98$ & $^{+0.14}_{-0.14}$ &  $^{+0.08}_{-0.08}$ & $^{+0.12}_{-0.11}$ &  \multirow{2}{*}{$\kappa_{\PQt}$} &   $1.02$ & $^{+0.19}_{-0.15}$ &  $^{+0.13}_{-0.09}$ & $^{+0.13}_{-0.13}$  \\[1pt]
  & &  $(^{+0.14}_{-0.15})$ &  $(^{+0.08}_{-0.09})$ & $(^{+0.12}_{-0.12})$ &  & &  $(^{+0.18}_{-0.15})$ &  $(^{+0.13}_{-0.09})$ & $(^{+0.13}_{-0.12})$ \\[10pt]
\multirow{2}{*}{$\kappa_{\Pgt }$} &  $1.02$ & $^{+0.17}_{-0.17}$ &  $^{+0.11}_{-0.13}$ & $^{+0.12}_{-0.10}$ &  \multirow{2}{*}{$\kappa_{\Pgt }$} &   $0.93$ & $^{+0.13}_{-0.13}$ &  $^{+0.08}_{-0.09}$ & $^{+0.11}_{-0.10}$  \\[1pt]
  & &  $(^{+0.16}_{-0.15})$ &  $(^{+0.11}_{-0.11})$ & $(^{+0.12}_{-0.11})$ &  & &  $(^{+0.14}_{-0.15})$ &  $(^{+0.09}_{-0.10})$ & $(^{+0.11}_{-0.11})$ \\[10pt]
\multirow{2}{*}{$\kappa_{\PQb}$} &  $1.17$ & $^{+0.27}_{-0.31}$ &  $^{+0.18}_{-0.29}$ & $^{+0.20}_{-0.10}$ &  \multirow{2}{*}{$\kappa_{\PQb}$} &   $0.91$ & $^{+0.17}_{-0.16}$ &  $^{+0.11}_{-0.12}$ & $^{+0.13}_{-0.11}$  \\[1pt]
  & &  $(^{+0.25}_{-0.23})$ &  $(^{+0.18}_{-0.17})$ & $(^{+0.17}_{-0.16})$ &  & &  $(^{+0.19}_{-0.22})$ &  $(^{+0.14}_{-0.16})$ & $(^{+0.13}_{-0.15})$ \\[10pt]
\multirow{2}{*}{$\kappa_{\cPg}$} &  $1.18$ & $^{+0.16}_{-0.14}$ &  $^{+0.10}_{-0.09}$ & $^{+0.12}_{-0.10}$ &  \multirow{2}{*}{$\kappa_{\cPg}$} &   $1.16$ & $^{+0.18}_{-0.13}$ &  $^{+0.14}_{-0.09}$ & $^{+0.12}_{-0.10}$  \\[1pt]
  & &  $(^{+0.14}_{-0.12})$ &  $(^{+0.10}_{-0.09})$ & $(^{+0.10}_{-0.09})$ &  & &  $(^{+0.17}_{-0.12})$ &  $(^{+0.13}_{-0.09})$ & $(^{+0.11}_{-0.09})$ \\[10pt]
\multirow{2}{*}{$\kappa_{\PGg }$} &  $1.07$ & $^{+0.14}_{-0.15}$ &  $^{+0.10}_{-0.14}$ & $^{+0.09}_{-0.05}$ &  \multirow{2}{*}{$\kappa_{\PGg }$} &   $0.96$ & $^{+0.09}_{-0.09}$ &  $^{+0.06}_{-0.08}$ & $^{+0.06}_{-0.06}$  \\[1pt]
  & &  $(^{+0.12}_{-0.12})$ &  $(^{+0.10}_{-0.09})$ & $(^{+0.07}_{-0.07})$ &  & &  $(^{+0.09}_{-0.11})$ &  $(^{+0.07}_{-0.09})$ & $(^{+0.05}_{-0.07})$ \\[10pt]
\multirow{2}{*}{$\kappa_{\Pgm}$} &  $0.80$ & $^{+0.59}_{-0.80}$ &  $^{+0.56}_{-0.81}$ & $^{+0.17}_{-0.00}$ &  \multirow{2}{*}{$\kappa_{\Pgm}$} &   $0.72$ & $^{+0.50}_{-0.72}$ &  $^{+0.50}_{-0.71}$ & $^{+0.00}_{-0.07}$  \\[1pt]
  & &  $(^{+0.51}_{-1.01})$ &  $(^{+0.50}_{-1.01})$ & $(^{+0.09}_{-0.09})$ &  & &  $(^{+0.49}_{-1.01})$ &  $(^{+0.48}_{-1.00})$ & $(^{+0.06}_{-0.08})$ \\[10pt]
 \multirow{2}{*}{~} & \multicolumn{4}{c}{~} &  \multirow{2}{*}{$\BRinv$} &   $0.07$ & $^{+0.08}_{-0.07}$ &  $^{+0.03}_{-0.03}$ & $^{+0.07}_{-0.06}$  \\[1pt]
  \multirow{2}{*}{~} & \multicolumn{4}{c}{~} &  & &  $(^{+0.09}_{+0.00})$ &  $(^{+0.04}_{-0.00})$ & $(^{+0.08}_{-0.00})$ \\[10pt]
 \multirow{2}{*}{~} & \multicolumn{4}{c}{~} &  \multirow{2}{*}{$\BRundet$} &   $0.00$ & $^{+0.17}_{+0.00}$ &  $^{+0.14}_{-0.00}$ & $^{+0.09}_{-0.00}$  \\[1pt]
  \multirow{2}{*}{~} & \multicolumn{4}{c}{~} &  & &  $(^{+0.20}_{+0.00})$ &  $(^{+0.17}_{-0.00})$ & $(^{+0.11}_{-0.00})$ \\[10pt]
\hline
\end{tabular}}
\label{tab:results_K2}
\end{table*}

\begin{figure*}[hbtp]
\centering
\includegraphics[width=0.48\textwidth]{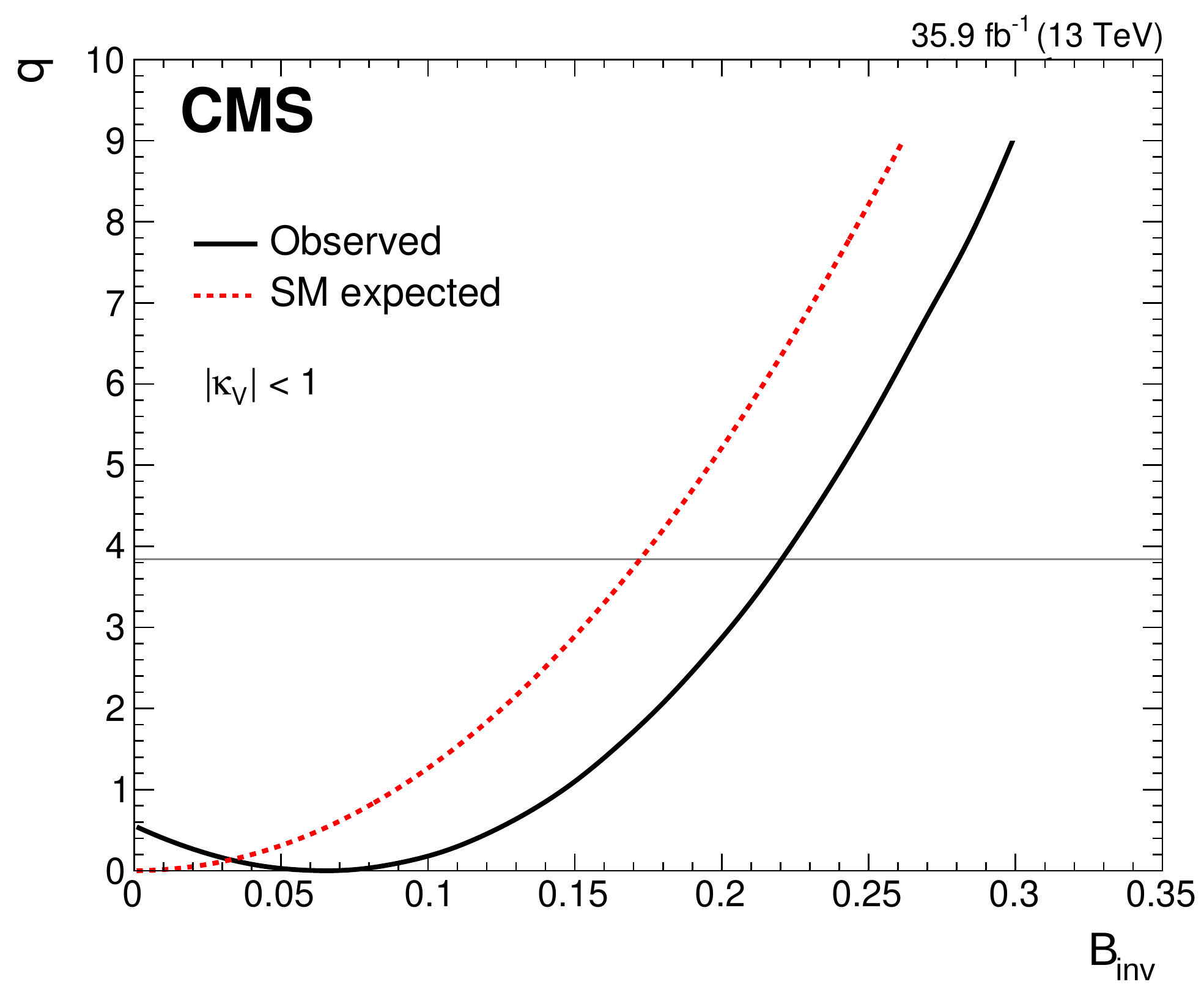}
\includegraphics[width=0.42\textwidth]{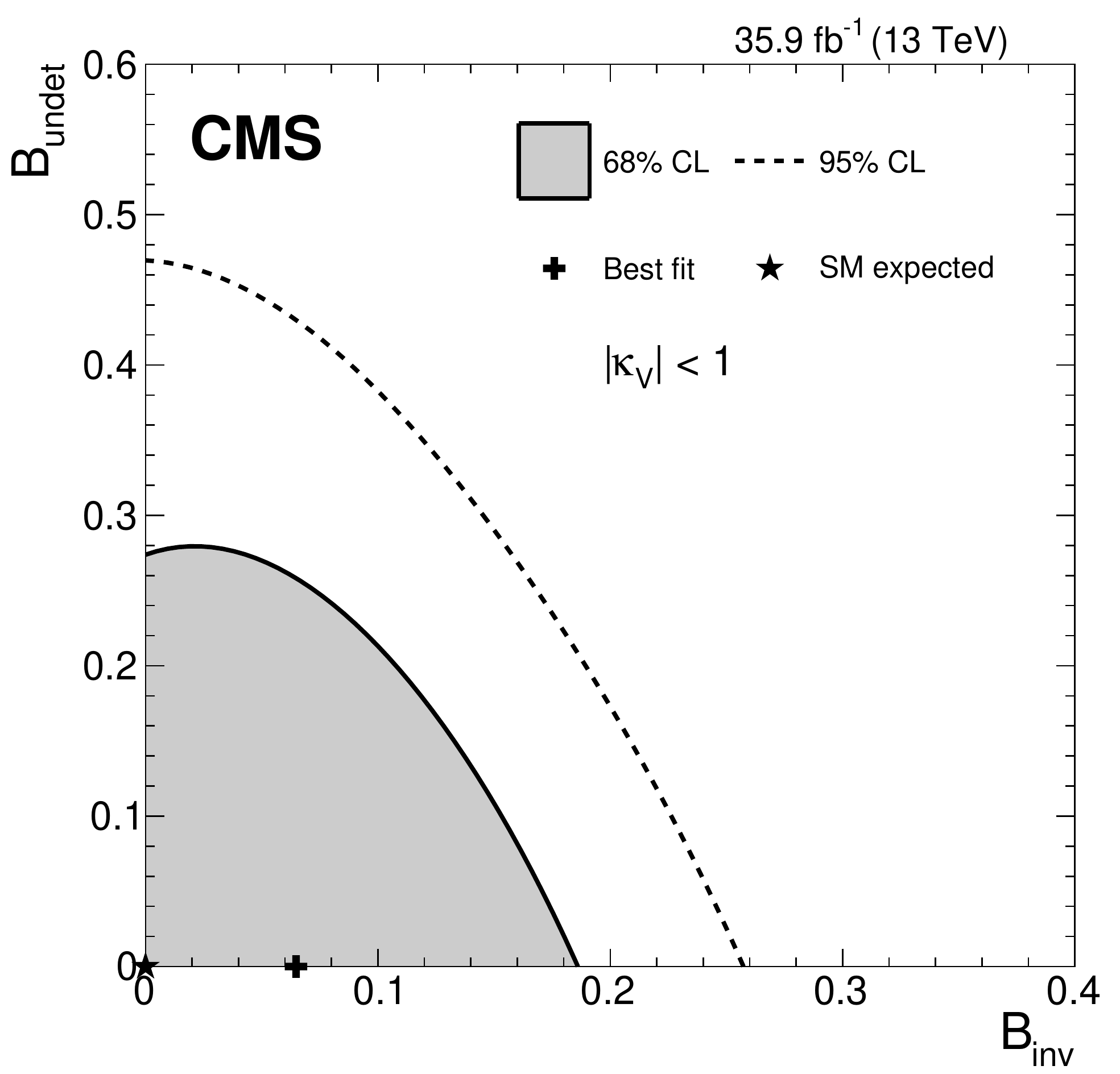}
\caption{Results within the generic $\kappa$-framework model with effective loops and with the constraint $\abs{\kappa_{\PW}},\,\abs{\kappa_{\cPZ}}\le 1$ (same sign of $\kappa_{\PW}$ and $\kappa_{\cPZ}$), and with $\BRinv>0$ and $\BRundet>0$ as free parameters.
Scan of the test statistic $q$ as a function of $\BRinv$ (left), and 68 and 95\% \CL regions for $\BRinv$ vs. $\BRundet$ (right).
The scan of the test statistic $q$ as a function of $\BRinv$ expected assuming the SM is also shown in the left figure. }
\label{fig:scan_K2Inv}
\end{figure*}

Two different model assumptions
are made concerning the BSM branching fraction. In the first parametrization, it is assumed that $\BRbsm=\BRinv+\BRundet=0$, whereas in the second, $\BRinv$ and
$\BRundet$ are allowed to vary as POIs, and instead the constraint $\abs{\kappa_{\PW}},\,\abs{\kappa_{\cPZ}} \le1$ is imposed. This avoids a complete degeneracy
in the total width where all of the coupling modifiers can be scaled equally to account for a non-zero $\BRundet$.
The parameter $\BRundet$ represents the total branching fraction to any final state that is not detected by the analyses included in this combined analysis.
The likelihood scan for the $\BRinv$ parameter in this model, and the 2D likelihood scan of $\BRinv$ vs. $\BRundet$ are given in Fig.~\ref{fig:scan_K2Inv}.
The 68 and 95\% \CL regions for Fig.~\ref{fig:scan_K2Inv}~(right) are determined as the regions for which $q(\BRundet,\BRinv)< 2.28$ and
$5.99$, respectively. The 95\% \CL upper limits of $\BRinv < 0.22$ and $\BRundet < 0.38$
are determined, corresponding to the value for which $q<3.84$~\cite{Cowan:2010st}. The uncertainty in the measurement of $\kappa_{\PQt}$ is reduced
by nearly 40\% compared to Ref.~\cite{ATLASCMSRun1}. This improvement is because of the improved sensitivity to the \tth production mode as described in Section~\ref{sec:signalstrength}.

\begin{figure*}[hbtp]
\centering
\includegraphics[width=0.49\textwidth]{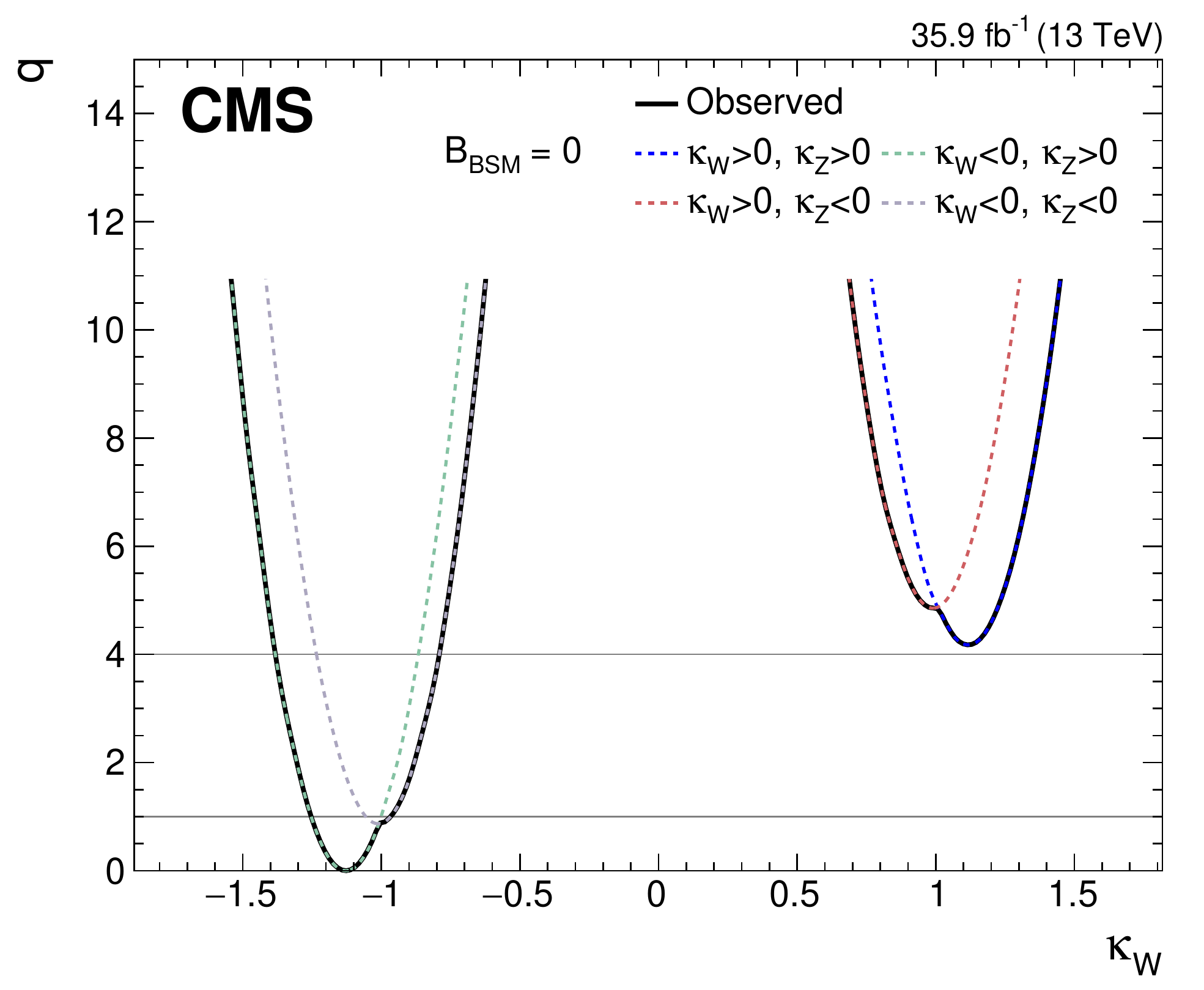}
\includegraphics[width=0.49\textwidth]{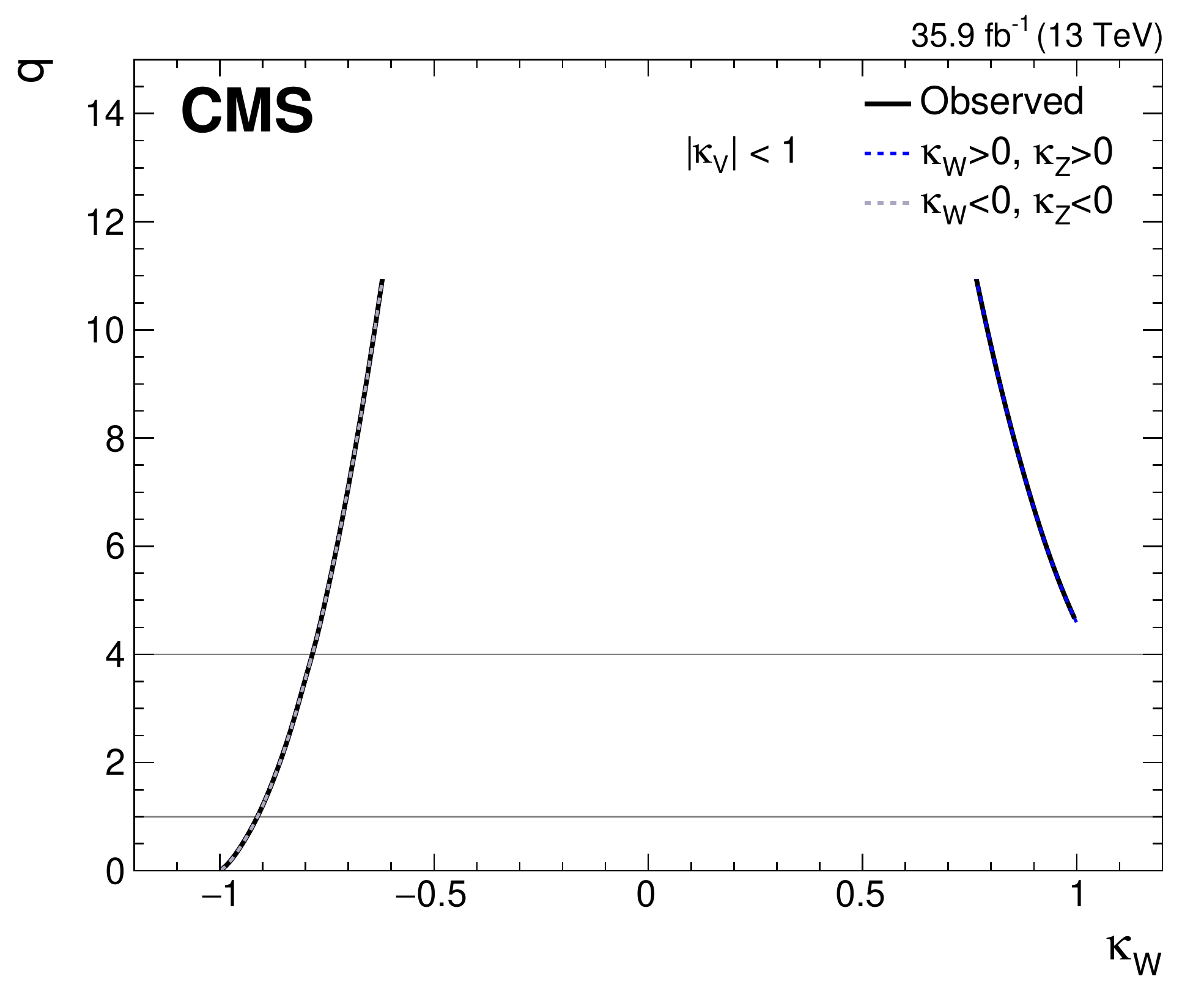}
\caption{Scan of the test statistic $q$ as a function of  $\kappa_{\PW}$ in the generic $\kappa$ model assuming $\BRbsm=0$ (left) and allowing $\BRinv$ and
$\BRundet$ to float (right). The different colored lines indicate the value of $q$ for different combinations of signs for $\kappa_{\PW}$ and $\kappa_{\cPZ}$.
The solid black line shows the minimum value of $q(\kappa_{\PW})$ in each case and is used to determine the best fit point and the $1\sigma$ and $2\sigma$ \CL regions. The scan in the right figure is truncated because of the constraints of $\abs{\kappa_{\PW}}\le1$ and $\abs{\kappa_{\cPZ}}\le1$, which are imposed in this model.}
\label{fig:K2_scan_kW}
\end{figure*}

In both of the generic $\kappa$ models, the best fit point for $\kappa_{\PW}$ is negative. The value of $q(\kappa_{\PW})$ as a function of $\kappa_{\PW}$ in the two cases
is shown in Fig.~\ref{fig:K2_scan_kW}. While different combinations of signs for $\kappa_{\PW}$ and $\kappa_{\cPZ}$ are shown, the minimum value of $q$ across
all combinations is used to determine the best fit point and the $1\sigma$ and $2\sigma$ \CL regions.

The preferred negative value of $\kappa_{\PW}$ is due
to the interference between some of the diagrams describing \tH production, which contributes
in several analyses entering the combination. In particular,
the excess in the \tth tagged categories of the $\hgg$ analysis can be accounted for by a negative value of $\kappa_{\PW}$
as this increases the contribution of \tH production. In these models, the $\hgg$ decay is treated as an effective coupling so that it has no dependence on
$\kappa_{\PW}$. This means that a negative value of $\kappa_{\PW}$ will not result in excesses in the other categories of the $\hgg$ analysis.

\begin{figure}[hbtp]
\centering
\includegraphics[width=0.45\textwidth]{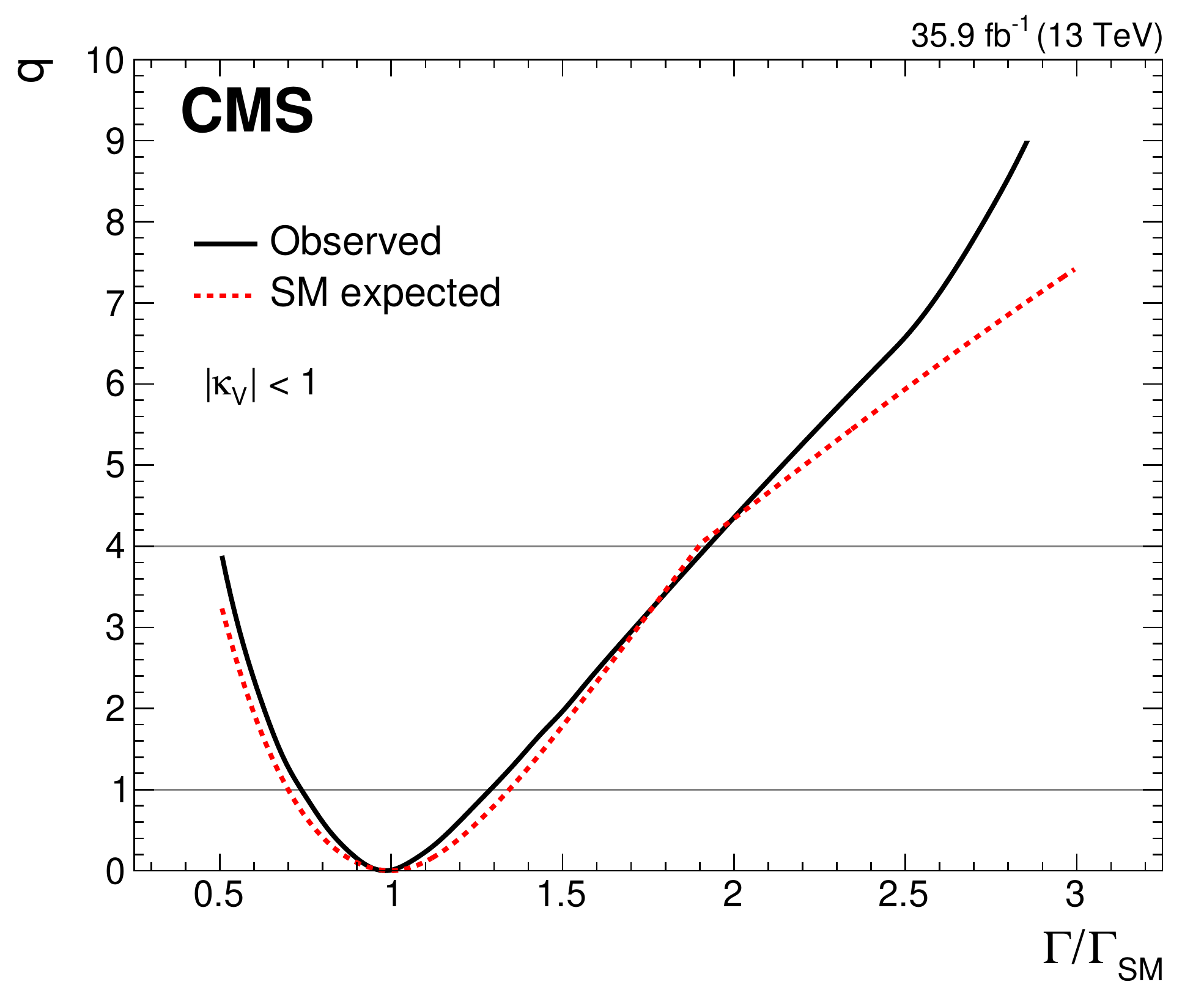}
\caption{The scan of the test statistic $q$ as a function of
$\Gamma_{\PH}/\Gamma_{\PH}^{\mathrm{SM}}$ obtained by reinterpreting the model allowing for BSM decays of the Higgs boson.
The expected scan of $q$ as a function of $\Gamma_{\PH}/\Gamma_{\PH}^{\mathrm{SM}}$ assuming the SM is also shown.}
\label{fig:scan_K2Undet_width}
\end{figure}

{\tolerance=800 Using Eq.~(\ref{eq:width_bsm}), this model is also reinterpreted as a constraint on the total Higgs boson width, and the corresponding likelihood scan is
shown in Fig.~\ref{fig:scan_K2Undet_width}.  Using this parametrization, the total Higgs boson width relative to the SM expectation is determined to be
$\Gamma_{\PH}/\Gamma_{\PH}^{\mathrm{SM}}=0.98^{+0.31}_{-0.25}$.
The different behavior between the observed and expected likelihood scans for large $\Gamma_{\PH}/\Gamma_{\PH}^{\mathrm{SM}}$ is due to the preference in data for the $\kappa_{\PQt} \kappa_{\PW}<0$ relative sign combination.\par}

\begin{figure}[hbtp]
\centering
\includegraphics[width=0.45\textwidth]{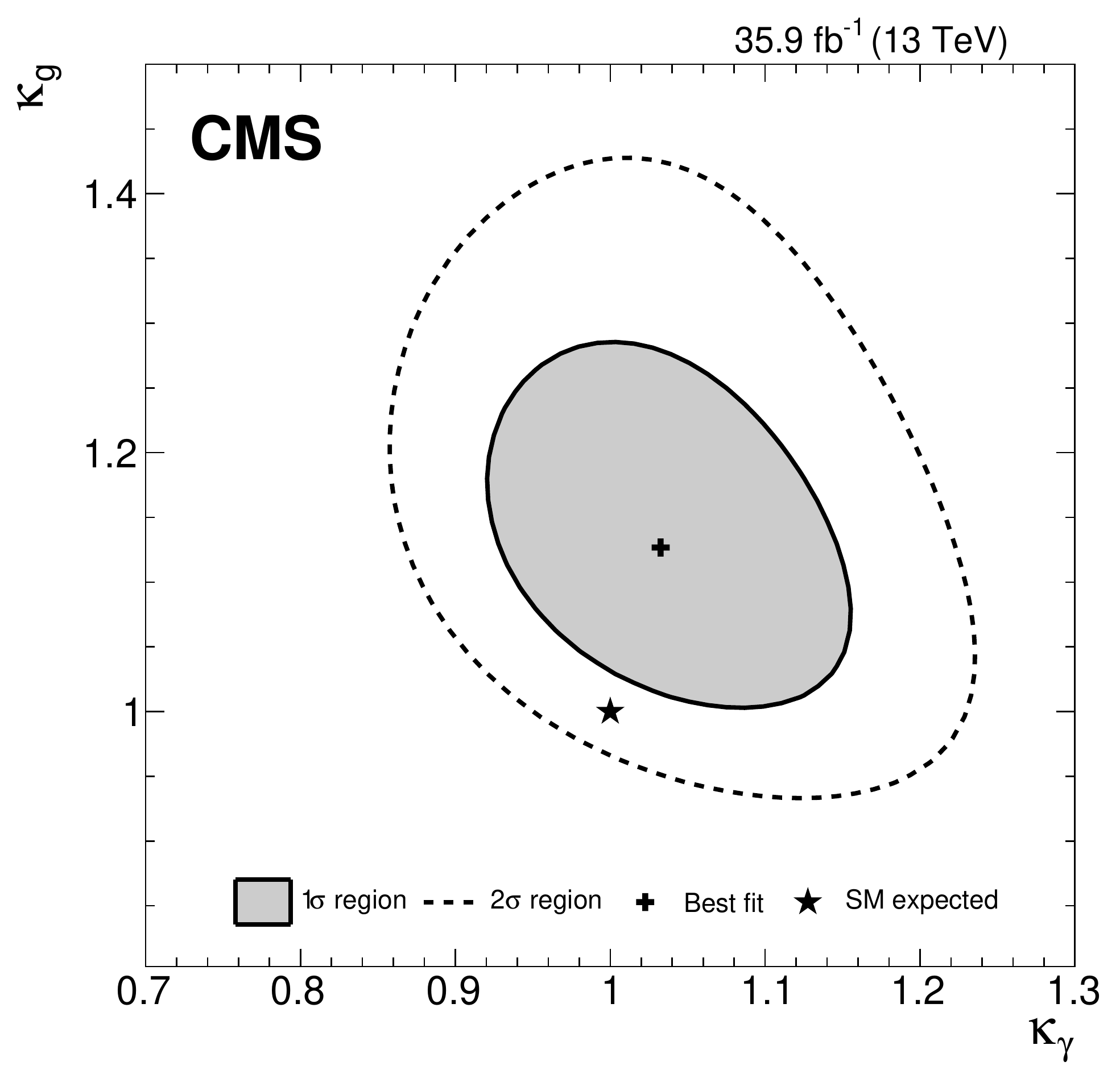}
\caption{The $1\sigma$ and $2\sigma$ \CL regions in the
$\kappa_{\Pg}$ vs.
$\kappa_{\PGg}$ parameter space for the model assuming the only BSM contributions to the Higgs boson couplings appear in the loop-induced processes or
in BSM Higgs decays.}
\label{fig:scan_K2Inv_2D}
\end{figure}

An additional fit is performed assuming that the only BSM contributions to the Higgs couplings appear in the loop-induced
\ggh and \hgg processes. In this fit, $\kappa_{\cPg}$ and $\kappa_{\PGg}$ are the POIs, $\BRinv$  and $\BRundet$ are floated,
and the other couplings are fixed to their SM predictions. The best fit point and the $1\sigma$ and $2\sigma$ \CL regions in the $\kappa_{\cPg}$-$\kappa_{\PGg}$
plane for this model are shown in Fig.~\ref{fig:scan_K2Inv_2D}.

\subsection{Generic model with effective loops and coupling modifier ratios}

\begin{figure}[hbtp]
\centering
\includegraphics[width=0.45\textwidth]{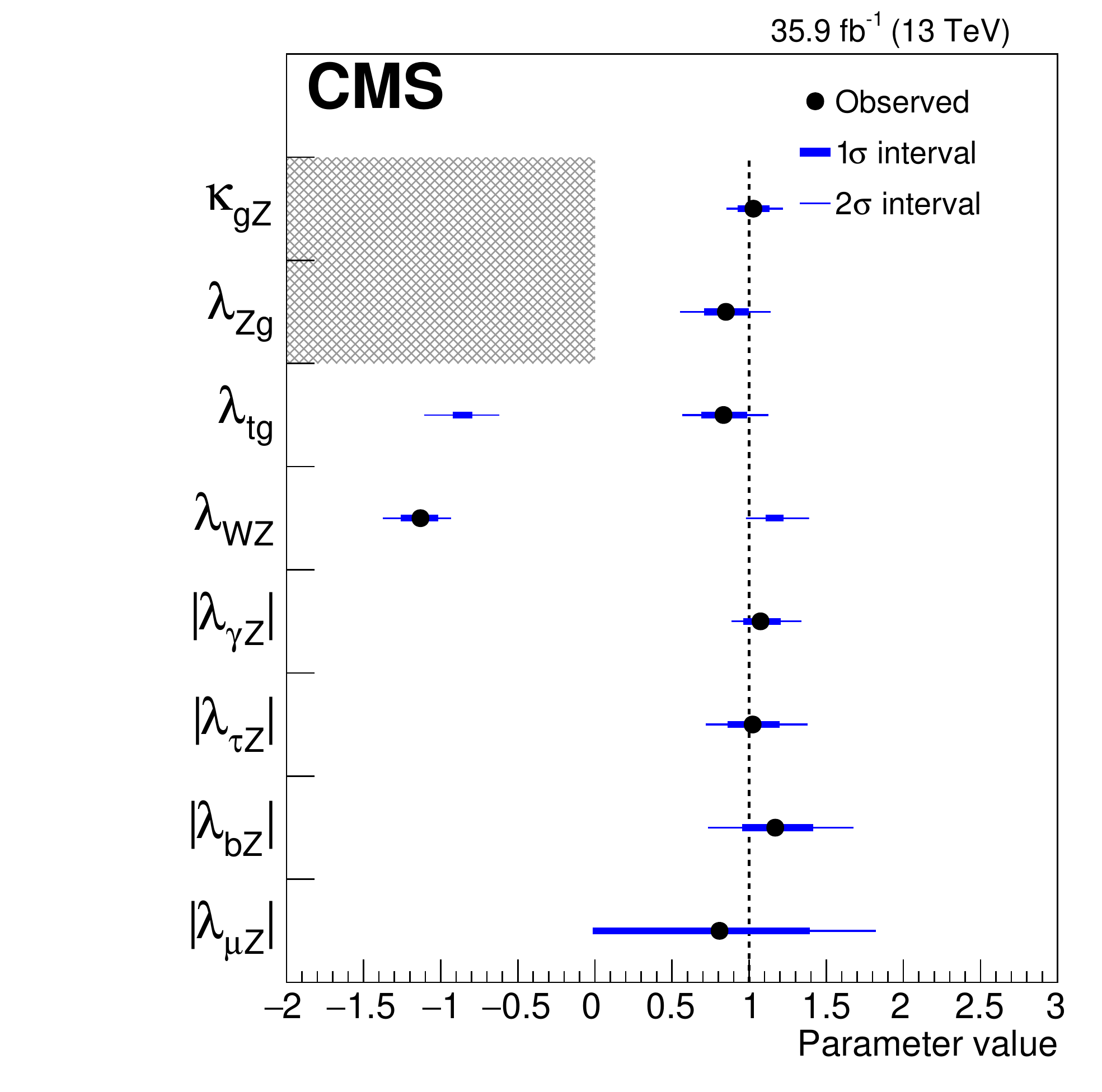}
\caption{Summary of the model with coupling ratios and effective couplings for the \ggh and \hgg loops. The points indicate the best fit values while the
thick and thin horizontal bars show the $1\sigma$ and $2\sigma$ \CL intervals, respectively. For this model, both positive and negative values of $\lambda_{\PW\cPZ}$ and $\lambda_{\PQt\cPg}$ are considered.}
\label{fig:obsplot_L1}
\end{figure}

An analogous parametrization to the ratios of cross sections and branching fractions described in the previous section can be derived in terms of ratios of the coupling modifiers ($\lambda_{ij} = \kappa_{i}/\kappa_{j}$). In this parametrization a reference combined coupling modifier is defined that accounts for modifications to the total event yield of a specific
production times decay process, thereby avoiding the need for assumptions on the total Higgs boson width.
The reference coupling modifier is taken to be $\kappa_{\Pg\cPZ} = \kappa_{\Pg} \kappa_{\cPZ}/\kappa_{\PH}$. The remaining parameters of
interest are ratios of the form: $\lambda_{\cPZ\Pg}$, $\lambda_{\PQt\Pg}$, $\lambda_{\PW\cPZ}$, $\lambda_{\PGg\cPZ}$, $\lambda_{\Pgt\cPZ}$, $\lambda_{\PQb\cPZ}$.
A summary of the results in this model is given in Fig.~\ref{fig:obsplot_L1}, and the numerical values along with the $\pm1\sigma$~uncertainties are shown in Table~\ref{tab:results_L1}.

\begin{table*}[hbtp]
\centering
\topcaption{Best fit values and $\pm 1\sigma$ uncertainties for the parameters of the coupling modifier ratio model. The expected uncertainties are given in brackets.}
\cmsTableX{0.8\textwidth}{
\begin{tabular}{@{} c r@{}c c c  c r@{}c c c @{}}
\hline
   &   &    & \multicolumn{2}{c}{Uncertainty} &   &    &   & \multicolumn{2}{c}{Uncertainty}  \\
  Parameter & \multicolumn{2}{c}{Best fit} & stat & syst  & Parameter & \multicolumn{2}{c}{Best fit} & stat & syst  \\
 \hline\\[-1.5ex]
\multirow{2}{*}{$\kappa_{\cPg\cPZ}$} &  $1.03$ & $^{+0.09}_{-0.09}$ &  $^{+0.07}_{-0.07}$ & $^{+0.05}_{-0.05}$ &  \multirow{2}{*}{$\lambda_{\PGg \cPZ}$} &   $1.07$ & $^{+0.12}_{-0.10}$ &  $^{+0.10}_{-0.08}$ & $^{+0.06}_{-0.05}$  \\[1pt]
  & &  $(^{+0.09}_{-0.09})$ &  $(^{+0.07}_{-0.07})$ & $(^{+0.05}_{-0.05})$ &  & &  $(^{+0.11}_{-0.09})$ &  $(^{+0.09}_{-0.08})$ & $(^{+0.05}_{-0.04})$ \\[10pt]
\multirow{2}{*}{$\lambda_{\PW\cPZ}$} &  $-1.13$ & $^{+0.10}_{-0.11}$ &  $^{+0.08}_{-0.09}$ & $^{+0.06}_{-0.06}$ &  \multirow{2}{*}{$\lambda_{\PQb\cPZ}$} &   $1.17$ & $^{+0.23}_{-0.20}$ &  $^{+0.16}_{-0.14}$ & $^{+0.16}_{-0.14}$  \\[1pt]
  & &  $(^{+0.11}_{-0.09})$ &  $(^{+0.09}_{-0.08})$ & $(^{+0.06}_{-0.05})$ &  & &  $(^{+0.22}_{-0.19})$ &  $(^{+0.16}_{-0.14})$ & $(^{+0.15}_{-0.13})$ \\[10pt]
\multirow{2}{*}{$\lambda_{\PQt\cPg}$} &  $0.83$ & $^{+0.14}_{-0.13}$ &  $^{+0.08}_{-0.08}$ & $^{+0.11}_{-0.10}$ &  \multirow{2}{*}{$\lambda_{\Pgt \cPZ}$} &   $1.02$ & $^{+0.16}_{-0.15}$ &  $^{+0.11}_{-0.10}$ & $^{+0.12}_{-0.11}$  \\[1pt]
  & &  $(^{+0.17}_{-0.16})$ &  $(^{+0.11}_{-0.11})$ & $(^{+0.12}_{-0.12})$ &  & &  $(^{+0.16}_{-0.14})$ &  $(^{+0.11}_{-0.10})$ & $(^{+0.11}_{-0.10})$ \\[10pt]
\multirow{2}{*}{$\lambda_{\cPZ\cPg}$} &  $0.85$ & $^{+0.14}_{-0.13}$ &  $^{+0.10}_{-0.12}$ & $^{+0.09}_{-0.05}$ &  \multirow{2}{*}{$\lambda_{\mu \cPZ}$} &   $0.81$ & $^{+0.57}_{-0.81}$ &  $^{+0.56}_{-0.82}$ & $^{+0.11}_{-0.00}$  \\[1pt]
  & &  $(^{+0.17}_{-0.16})$ &  $(^{+0.13}_{-0.13})$ & $(^{+0.11}_{-0.09})$ &  & &  $(^{+0.50}_{-1.01})$ &  $(^{+0.49}_{-1.01})$ & $(^{+0.07}_{-0.07})$ \\[10pt]
\hline
\end{tabular}}
\label{tab:results_L1}
\end{table*}

\subsection{Fits of vector boson and fermion coupling modifiers}
\label{ssec:kVkF}

\begin{figure}[hbtp]
\centering
\includegraphics[width=0.45\textwidth]{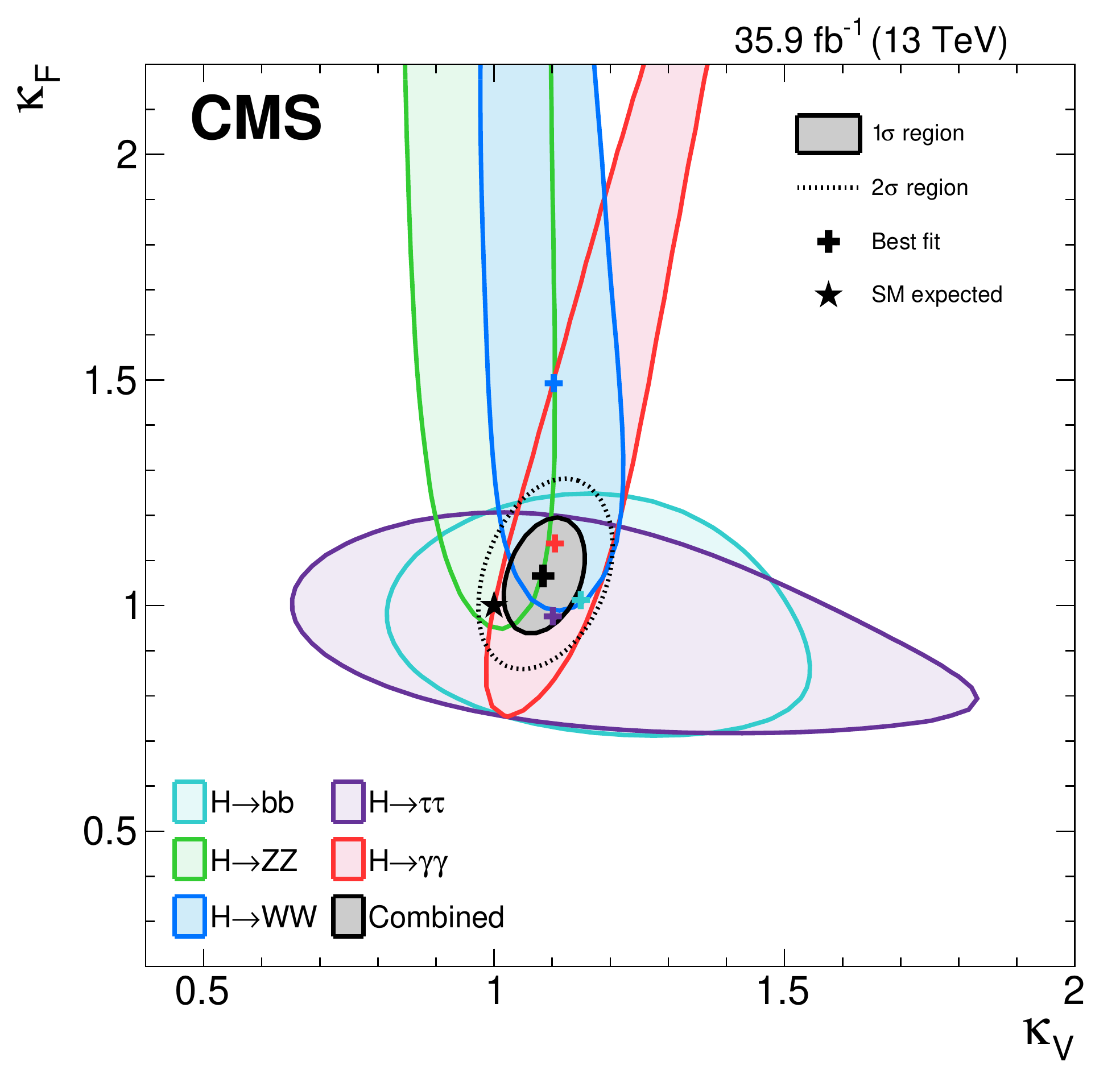}
\caption{The $1\sigma$ and $2\sigma$ \CL regions in the $\kappa_{\mathrm{F}}$ vs.
$\kappa_{\mathrm{V}}$ parameter space for the model assuming a common scaling of
all the vector boson and fermion couplings.}
\label{fig:scan_K3_5D_2D}
\end{figure}

A more constrained version of the loop-resolved $\kappa$ model is defined by assuming a common scaling of all
vector boson and fermion couplings, respectively. Two models are defined: one in which all signal processes are
scaled according to these two $\kappa_{\mathrm{V}}$ and $\kappa_{\mathrm{F}}$ parameters, and one in which separate
$\kappa_{\mathrm{V}}^{f}$ and $\kappa_{\mathrm{F}}^f$ parameters are defined for each of the five decay processes.
The best fit points and the $1\sigma$ and $2\sigma$ \CL regions in the $\kappa_{\mathrm{V}}$-$\kappa_{\mathrm{F}}$ plane
for both models are shown in Fig.~\ref{fig:scan_K3_5D_2D}, and the results are summarized in Table~\ref{tab:results_K3_5D}.
For large values of $\kappa_{\mathrm{F}}^{\cPZ\cPZ}$ the likelihood
becomes essentially flat, resulting in the best fit point for this parameter being beyond the scale of the axis shown.
The 1D 68\% \CL region for $\kappa_{\mathrm{F}}^{\cPZ\cPZ}$ can be expressed as $[1.22,\infty]$.

\begin{table*}[hbtp]
\centering
\topcaption{Best fit values and $\pm 1\sigma$ uncertainties for the parameters of the $\kappa_{\mathrm{V}},\kappa_{\mathrm{F}}$ model.
The expected uncertainties are given in brackets.}
\cmsTableX{0.8\textwidth}{
\begin{tabular}{@{} c r@{}c c c  c r@{}c c c @{}}
\hline
   &   &    & \multicolumn{2}{c}{Uncertainty} &   &    &   & \multicolumn{2}{c}{Uncertainty}  \\
  Parameter & \multicolumn{2}{c}{Best fit} & stat & syst  & Parameter & \multicolumn{2}{c}{Best fit} & stat & syst  \\
 \hline\\[-1.5ex]
\multirow{2}{*}{$\kappa_{\mathrm{V}}^{\PW\PW}$} &  $1.10$ & $^{+0.08}_{-0.08}$ &  $^{+0.06}_{-0.06}$ & $^{+0.06}_{-0.06}$ &  \multirow{2}{*}{$\kappa_{\mathrm{F}}^{\PW\PW}$} &   $1.49$ & $^{+1.55}_{-0.38}$ &  $^{+1.04}_{-0.33}$ & $^{+1.15}_{-0.18}$  \\[1pt]
  & &  $(^{+0.08}_{-0.08})$ &  $(^{+0.06}_{-0.06})$ & $(^{+0.05}_{-0.05})$ &  & &  $(^{+0.38}_{-0.20})$ &  $(^{+0.31}_{-0.17})$ & $(^{+0.21}_{-0.11})$ \\[10pt]
\multirow{2}{*}{$\kappa_{\mathrm{V}}^{\cPZ\cPZ}$} &  $0.96$ & $^{+0.09}_{-0.08}$ &  $^{+0.08}_{-0.07}$ & $^{+0.05}_{-0.04}$ &  \multirow{2}{*}{$\kappa_{\mathrm{F}}^{\cPZ\cPZ}$} &  \multicolumn{2}{c}{$\NA$} & $\NA$  &  $\NA$ \\[1pt]
  & &  $(^{+0.12}_{-0.11})$ &  $(^{+0.11}_{-0.10})$ & $(^{+0.05}_{-0.05})$ &  & &  $(^{+1.79}_{-0.31})$ &  $(^{+1.55}_{-0.30})$ & $(^{+0.91}_{-0.06})$ \\[10pt]
\multirow{2}{*}{$\kappa_{\mathrm{V}}^{\bb}$} &  $1.15$ & $^{+0.23}_{-0.22}$ &  $^{+0.18}_{-0.18}$ & $^{+0.15}_{-0.13}$ &  \multirow{2}{*}{$\kappa_{\mathrm{F}}^{\bb}$} &   $1.01$ & $^{+0.16}_{-0.18}$ &  $^{+0.09}_{-0.10}$ & $^{+0.13}_{-0.15}$  \\[1pt]
  & &  $(^{+0.22}_{-0.22})$ &  $(^{+0.17}_{-0.18})$ & $(^{+0.13}_{-0.12})$ &  & &  $(^{+0.16}_{-0.18})$ &  $(^{+0.09}_{-0.10})$ & $(^{+0.13}_{-0.15})$ \\[10pt]
\multirow{2}{*}{$\kappa_{\mathrm{V}}^{\tautau }$} &  $1.10$ & $^{+0.38}_{-0.29}$ &  $^{+0.26}_{-0.24}$ & $^{+0.27}_{-0.18}$ &  \multirow{2}{*}{$\kappa_{\mathrm{F}}^{\tautau }$} &   $0.98$ & $^{+0.15}_{-0.16}$ &  $^{+0.08}_{-0.09}$ & $^{+0.13}_{-0.14}$  \\[1pt]
  & &  $(^{+0.32}_{-0.29})$ &  $(^{+0.24}_{-0.23})$ & $(^{+0.20}_{-0.17})$ &  & &  $(^{+0.14}_{-0.14})$ &  $(^{+0.07}_{-0.08})$ & $(^{+0.11}_{-0.12})$ \\[10pt]
\multirow{2}{*}{$\kappa_{\mathrm{V}}^{\PGg \PGg }$} &  $1.10$ & $^{+0.14}_{-0.08}$ &  $^{+0.11}_{-0.07}$ & $^{+0.09}_{-0.05}$ &  \multirow{2}{*}{$\kappa_{\mathrm{F}}^{\PGg \PGg }$} &   $1.14$ & $^{+0.67}_{-0.29}$ &  $^{+0.53}_{-0.26}$ & $^{+0.41}_{-0.14}$  \\[1pt]
  & &  $(^{+0.10}_{-0.08})$ &  $(^{+0.08}_{-0.06})$ & $(^{+0.05}_{-0.04})$ &  & &  $(^{+0.47}_{-0.25})$ &  $(^{+0.41}_{-0.23})$ & $(^{+0.23}_{-0.10})$ \\[10pt]
\hline
\end{tabular}}
\label{tab:results_K3_5D}
\end{table*}

\subsection{Benchmark models with resolved loops to test the symmetry of fermion couplings}
\label{ssec:L2}

Several BSM models predict the existence of an extended Higgs sector. In such scenarios, the couplings to up-and down-type
fermions, or to leptons and quarks, can be separately modified.
In order to probe such models, parametrizations are introduced in which the couplings of the Higgs boson to fermions are scaled either by
separate common modifiers for up-type ($\kappa_{\PQu}$) and
down-type ($\kappa_{\PQd}$) fermions or by separate common modifiers for quarks $\kappa_{\Pq}$ and leptons $\kappa_{\text{l}}$ ($\text{l}=\Pe,\Pgm,\Pgt$).

\begin{figure*}[hbtp]
\centering
\includegraphics[width=0.49\textwidth]{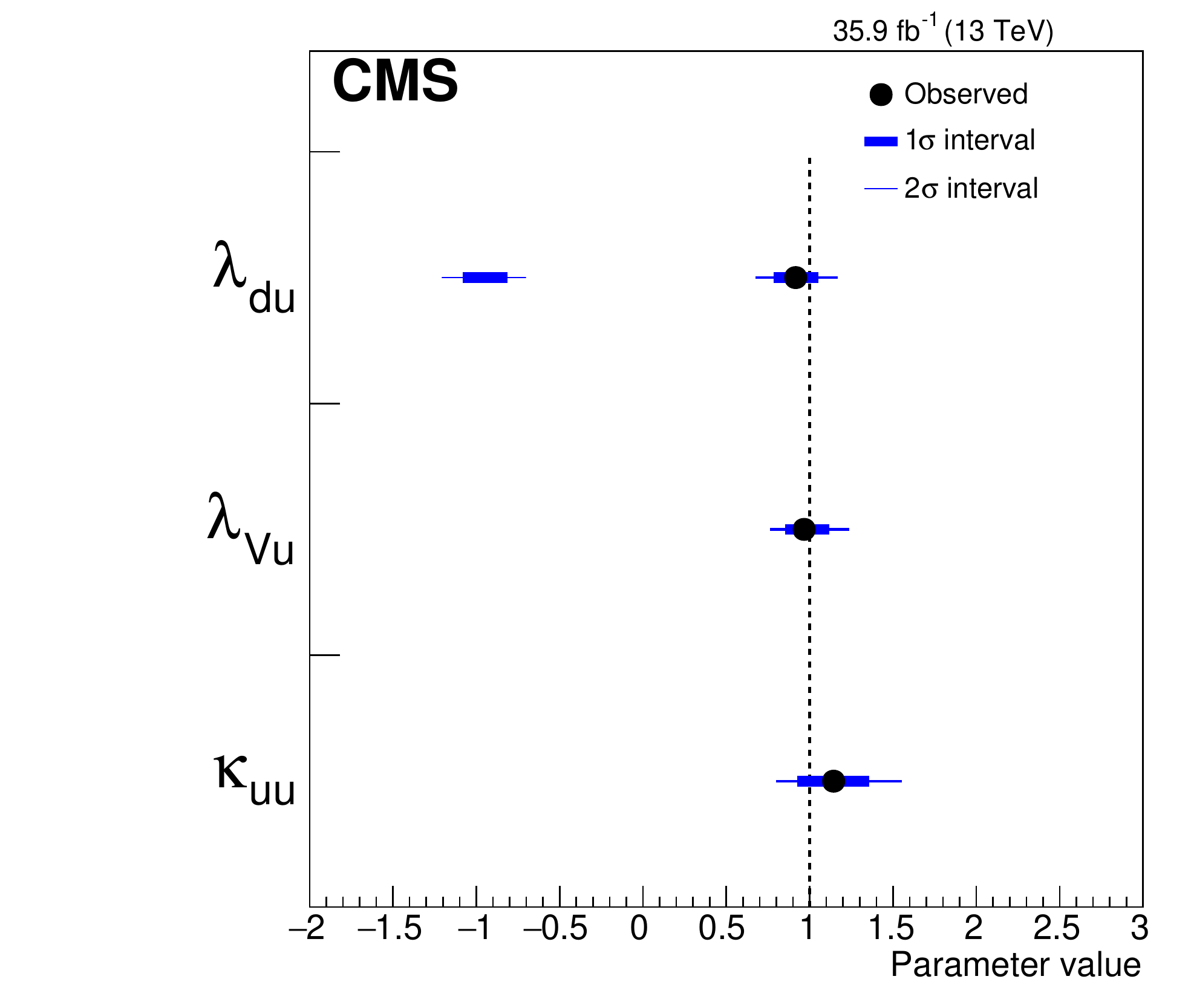}
\includegraphics[width=0.49\textwidth]{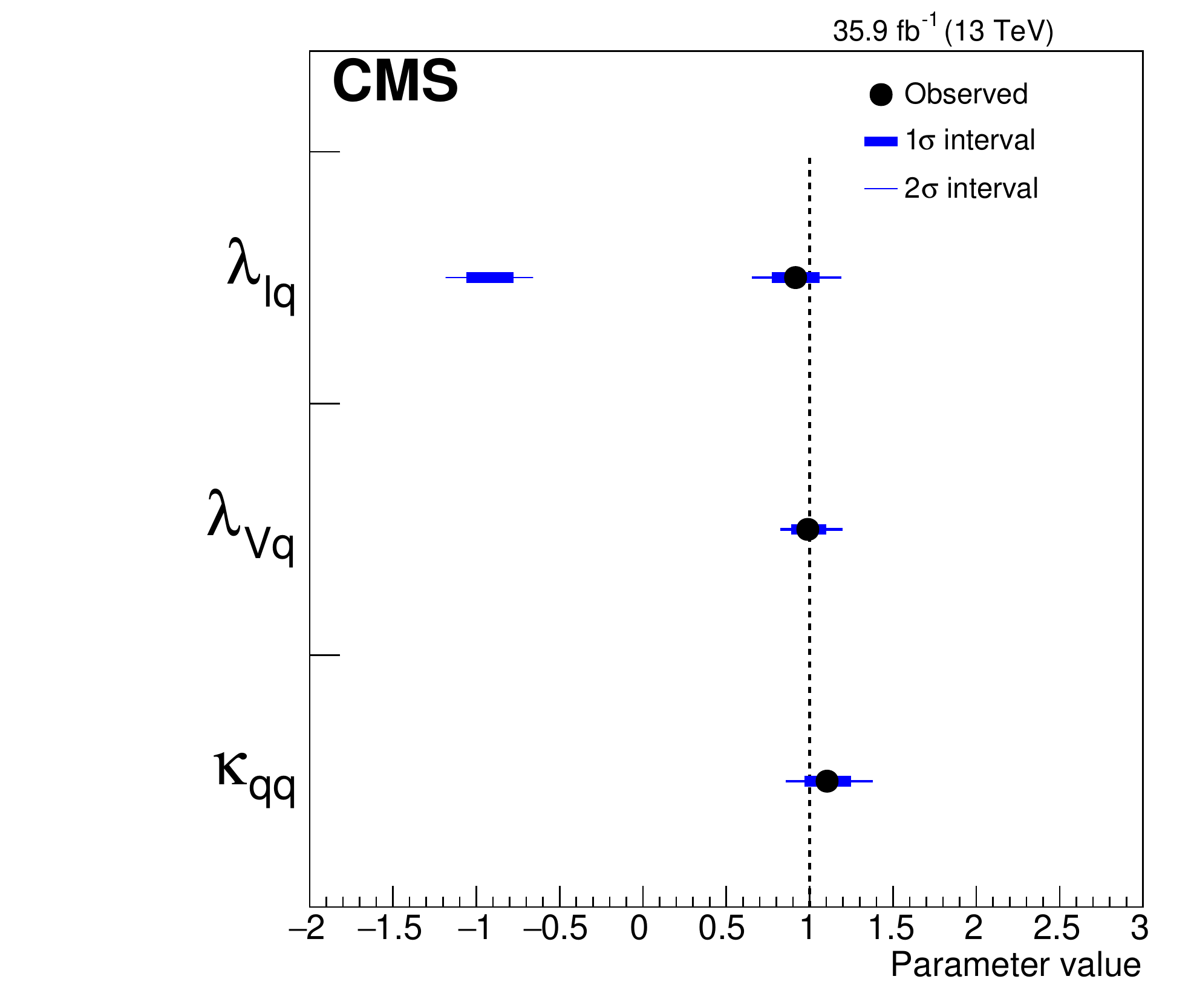}
\caption{Summary plots of the 3-parameter models comparing up- and down-type fermions, and floating the ratio of the vector coupling to the up-type coupling (left) and
comparing lepton and quark couplings (right). The points indicate the best fit values while the
thick and thin horizontal bars show the $1\sigma$ and $2\sigma$ \CL intervals, respectively.
Both positive and negative values of $\lambda_{\PQd\PQu}$, $\lambda_{\mathrm{V}\PQu}$, $\lambda_{\text{l}\Pq}$, and $\lambda_{\mathrm{V}\Pq}$ are considered.}
\label{fig:obsplot_L2}
\end{figure*}

Figure~\ref{fig:obsplot_L2} shows the results of the fits where the ratio of the couplings to
up- and down-type fermions $\lambda_{\PQd\PQu}=\kappa_{\PQd}/\kappa_{\PQu}$ is determined along with the ratio $\lambda_{\mathrm{V}\PQu}=\kappa_{\mathrm{V}}/\kappa_{\PQu}$ and $\kappa_{\PQu\PQu}=\kappa_{\PQu}^{2}/\Gamma_{\PH}$.
Also shown are the results of the fit where the ratio of the coupling to leptons and to quarks
$\lambda_{\text{l}\Pq}=\kappa_{\text{l}}/\kappa_{\Pq}$ is determined  along with the ratio $\lambda_{\mathrm{V}\Pq}=\kappa_{\mathrm{V}}/\kappa_{\Pq}$ and $\kappa_{\Pq\Pq}=\kappa_{\Pq}^{2}/\Gamma_{\PH}$. The results of these two parametrizations are summarized in Table~\ref{tab:results_L2_ldu}.

\begin{table*}[hbtp]
\centering
\topcaption{Best fit values and $\pm 1\sigma$ uncertainties for the parameters of the two benchmark models with
resolved loops to test the symmetry of fermion couplings. The expected uncertainties are given in brackets.}
\cmsTableX{0.7\textwidth}{
\setlength\extrarowheight{3pt}
\setlength\tabcolsep{1pt}
\renewcommand{\arraystretch}{1.1}
\begin{tabular}{rcccrcccrccc}
 \multicolumn{4}{c}{$\lambda_{\mathrm{V}\PQu}$} &  \multicolumn{4}{c}{$\lambda_{\PQd\PQu}$}  &  \multicolumn{4}{c}{$\kappa_{\PQu\PQu}$} \\
 \multicolumn{2}{c}{Best fit}  & \multicolumn{2}{c}{Uncertainty}  &  \multicolumn{2}{c}{Best fit} & \multicolumn{2}{c}{Uncertainty}  &  \multicolumn{2}{c}{Best fit}  & \multicolumn{2}{c}{Uncertainty}  \\[-3pt]
\multicolumn{2}{c}{value} &  stat  &  \multicolumn{1}{c}{syst}  &  \multicolumn{2}{c}{value} &  stat  &  \multicolumn{1}{c}{syst}  &  \multicolumn{2}{c}{value} &  stat  &  syst \\
\hline
\renewcommand{\arraystretch}{1.8}
  $0.97$ & $^{+0.14}_{-0.10}$ &  $^{+0.11}_{-0.08}$ & $^{+0.09}_{-0.06}$ &
  $0.92$ & $^{+0.12}_{-0.12}$ &  $^{+0.09}_{-0.09}$ & $^{+0.08}_{-0.08}$ &
  $1.14$ & $^{+0.20}_{-0.20}$ &  $^{+0.13}_{-0.16}$ & $^{+0.15}_{-0.12}$ \\
  &  $({}^{+0.15}_{-0.11})$ &     $({}^{+0.12}_{-0.09})$ & $({}^{+0.09}_{-0.07})$ &
  &  $({}^{+0.13}_{-0.13})$ &     $({}^{+0.10}_{-0.10})$ & $({}^{+0.08}_{-0.08})$ &
  &  $({}^{+0.16}_{-0.19})$ &     $({}^{+0.12}_{-0.16})$ & $({}^{+0.11}_{-0.11})$ \\ [\cmsTabSkip]
 \multicolumn{4}{c}{$\lambda_{\mathrm{V}\Pq}$}  &  \multicolumn{4}{c}{$\lambda_{\text{l}\Pq}$}  &  \multicolumn{4}{c}{$\kappa_{\Pq\Pq}$} \\
 \multicolumn{2}{c}{Best fit}  & \multicolumn{2}{c}{Uncertainty}  &  \multicolumn{2}{c}{Best fit} & \multicolumn{2}{c}{Uncertainty}  &  \multicolumn{2}{c}{Best fit}  & \multicolumn{2}{c}{Uncertainty}  \\[-3pt]
\multicolumn{2}{c}{value} &  stat  &  \multicolumn{1}{c}{syst}  &  \multicolumn{2}{c}{value} &  stat  &  \multicolumn{1}{c}{syst}  &  \multicolumn{2}{c}{value} &  stat  &  syst \\
\hline
\renewcommand{\arraystretch}{1.8}
  $0.99$ & $^{+0.10}_{-0.08}$ &  $^{+0.07}_{-0.06}$ & $^{+0.07}_{-0.06}$ &
  $0.92$ & $^{+0.13}_{-0.13}$ &  $^{+0.09}_{-0.08}$ & $^{+0.10}_{-0.10}$ &
  $1.10$ & $^{+0.13}_{-0.12}$ &  $^{+0.08}_{-0.08}$ & $^{+0.10}_{-0.09}$ \\
  &  $({}^{+0.12}_{-0.09})$ &     $({}^{+0.08}_{-0.07})$ & $({}^{+0.09}_{-0.07})$ &
  &  $({}^{+0.15}_{-0.14})$ &     $({}^{+0.10}_{-0.09})$ & $({}^{+0.11}_{-0.10})$ &
  &  $({}^{+0.13}_{-0.13})$ &     $({}^{+0.09}_{-0.09})$ & $({}^{+0.10}_{-0.10})$ \\[\cmsTabSkip]
\end{tabular}
}
\label{tab:results_L2_ldu}
\end{table*}

\subsection{Compatibility of measurements with the SM}
\label{ssec:smcompat}

Table~\ref{tab:compatibility-pvalues} shows a summary of the compatibility of the different models considered, as described in Sections~\ref{sec:signalstrength} and~\ref{sec:couplings}, with the SM
predictions. For each model, the value of $q$ at the values of the POIs for the SM expectation ($q_{\mathrm{SM}}$)
is converted to a $p$-value with respect to the SM. This is done assuming $q$ is distributed according to a $\chi^{2}$ function with
the number of degrees of freedom equal to the number of POIs. This $p$-value is found to be greater than 5\% for all parametrizations.

\begin{table*}[hbtp]
\centering
\topcaption{Compatibility of the fit results with the SM prediction under various signal parametrizations.
The value of $q$ at the values of the POIs for which
the SM expectation is obtained ($q_{\text{SM}}$) is shown along with the corresponding $p$-value,
with respect to the SM, assuming $q$ is distributed according to a $\chi^{2}$ function with
the specified number of degrees of freedom (DOF).
\label{tab:compatibility-pvalues}}
\centering
\setlength\extrarowheight{5pt}%
\setlength\tabcolsep{5pt}%
\cmsTable{
\begin{tabular}{p{0.28\textwidth}ccp{0.547\textwidth}} \\ \hline
      Parameterization                     & \makebox[7ex][c]{$p$-value} ($q_{\text{SM}}$) & \makebox[3ex][c]{DOF} & Parameters of interest\\
      \hline
      Global signal strength               & 6.28\% (3.46) &  1 & $\mu$ \\
      Production processes                 & 9.87\% (9.27) &  5 & $\mu_{\ggh}$, $\mu_{\vbf}$, $\mu_{\wh}$, $\mu_{\zh}$, $\mu_{\tth}$ \\
      Decay modes                          & 53.8\% (5.05) &  6 & $\mu^{\PGg\PGg}$, $\mu^{\cPZ\cPZ}$, $\mu^{\PW\PW}$, $\mu^{\Pgt\Pgt}$, $\mu^{\bb}$, $\mu^{\Pgm\Pgm}$ \\
      $\sigma_i   \BR^f$ products      & 61.2\% (21.5) & 24 & \mbox{$\sigma_{\ggh} \BR^{\bb}$}, \mbox{$\sigma_{\ggh} \BR^{\Pgt\Pgt}$}, \mbox{$\sigma_{\ggh} \BR^{\mu\mu}$}, \mbox{$\sigma_{\ggh} \BR^{\PW\PW}$}, \mbox{$\sigma_{\ggh} \BR^{\cPZ\cPZ}$}, \mbox{$\sigma_{\ggh} \BR^{\PGg\PGg}$}, \mbox{$\sigma_{\vbf} \BR^{\Pgt\Pgt}$}, \mbox{$\sigma_{\vbf} \BR^{\mu\mu}$}, \mbox{$\sigma_{\vbf} \BR^{\PW\PW}$}, \mbox{$\sigma_{\vbf} \BR^{\cPZ\cPZ}$}, \mbox{$\sigma_{\vbf} \BR^{\PGg\PGg}$}, \mbox{$\sigma_{\wh}  \BR^{\bb}$}, \mbox{$\sigma_{\wh}  \BR^{\PW\PW}$}, \mbox{$\sigma_{\wh}  \BR^{\cPZ\cPZ}$}, \mbox{$\sigma_{\wh}  \BR^{\PGg\PGg}$}, \mbox{$\sigma_{\zh}  \BR^{\bb}$}, \mbox{$\sigma_{\zh}  \BR^{\PW\PW}$}, \mbox{$\sigma_{\zh}  \BR^{\cPZ\cPZ}$}, \mbox{$\sigma_{\zh}  \BR^{\PGg\PGg}$}, \mbox{$\sigma_{\tth} \BR^{\Pgt\Pgt}$}, \mbox{$\sigma_{\tth} \BR^{\PW\PW}$}, \mbox{$\sigma_{\tth} \BR^{\cPZ\cPZ}$}, \mbox{$\sigma_{\tth} \BR^{\PGg\PGg}$}, \mbox{$\sigma_{\tth} \BR^{\bb}$} \\
      Ratios of $\sigma$ and $\BR$ relative to $\Pg\Pg\to\PH\to\cPZ\cPZ$ & 32.3\% (11.5) &  10 & \mbox{$\mu_{\ggh}^{\cPZ\cPZ}$}, \mbox{$\mu_{\vbf}/\mu_{\ggh}$}, \mbox{$\mu_{\wh}/\mu_{\ggh}$}, \mbox{$\mu_{\zh}/\mu_{\ggh}$}, \mbox{$\mu_{\tth}/\mu_{\ggh}$}, \mbox{$\mu^{\PW\PW}/\mu^{\cPZ\cPZ}$}, \mbox{$\mu^{\PGg\PGg}/\mu^{\cPZ\cPZ}$}, \mbox{$\mu^{\Pgt\Pgt}/\mu^{\cPZ\cPZ}$}, \mbox{$\mu^{\bb}/\mu^{\cPZ\cPZ}$}, \mbox{$\mu^{\bb}/\mu^{\mu\mu}$} \\
      Simplified template cross sections with branching fractions relative to $\BR^{\cPZ\cPZ}$ & 21.2\% (14.4) &  11 & \mbox{$\sigma_{\ggh} \BR^{\cPZ\cPZ}$}, \mbox{$\sigma_{\vbf} \BR^{\cPZ\cPZ}$}, \mbox{$\sigma_{\PH+\mathrm{V}(\Pq\Pq)} \BR^{\cPZ\cPZ}$}, \mbox{$\sigma_{\PH+\PW(\ell\nu)} \BR^{\cPZ\cPZ}$}, \mbox{$\sigma_{\PH+\cPZ(\ell\ell/\nu\nu)} \BR^{\cPZ\cPZ}$}, \mbox{$\sigma_{\tth} \BR^{\cPZ\cPZ}$}, \mbox{$\BR^{\bb}/\BR^{\cPZ\cPZ}$}, \mbox{$\BR^{\Pgt\Pgt}/\BR^{\cPZ\cPZ}$}, \mbox{$\BR^{\mu\mu}/\BR^{\cPZ\cPZ}$}, \mbox{$\BR^{\PW\PW}/\BR^{\cPZ\cPZ}$}, \mbox{$\BR^{\PGg\PGg}/\BR^{\cPZ\cPZ}$} \\
      Couplings, SM loops                  & 45.6\% (5.71) &  6 & $\kappa_{\cPZ}$, $\kappa_{\PW}$, $\kappa_{\PQt}$, $\kappa_{\Pgt}$, $\kappa_{\PQb}$, $\kappa_{\Pgm}$ \\
      Couplings vs. mass                    & 16.8\% (3.57) &  2 & $M$, $\epsilon$ \\
      Couplings, BSM loops                 & 18.5\% (11.3) &  8 & $\kappa_{\cPZ}$, $\kappa_{\PW}$, $\kappa_{\PQt}$, $\kappa_{\Pgt}$, $\kappa_{\PQb}$, $\kappa_{\Pgm}$, $\kappa_{\PGg}$, $\kappa_{\Pg}$\\
      Couplings, BSM loops and decays including $\hinv$ analyses     & 32.4\% (11.5) &  10 & $\kappa_{\cPZ}$, $\kappa_{\PW}$, $\kappa_{\PQt}$, $\kappa_{\Pgt}$, $\kappa_{\PQb}$, $\kappa_{\Pgm}$, $\kappa_{\PGg}$, $\kappa_{\Pg}$, $\BRinv$, $\BRundet$\\
      Ratios of coupling modifiers         & 18.1\% (11.4) & 8 & \mbox{$\kappa_{\Pg\cPZ}$}, \mbox{$\lambda_{\PW\cPZ}$}, \mbox{$\lambda_{\PGg\cPZ}$}, \mbox{$\lambda_{\PQt\Pg}$}, \mbox{$\lambda_{\PQb\cPZ}$}, \mbox{$\lambda_{\Pgt\cPZ}$}, \mbox{$\lambda_{\mu\cPZ}$}, \mbox{$\lambda_{\cPZ\Pg}$} \\
      Fermion and vector couplings         & 16.9\% (3.55) &  2 & $\kappa_{\mathrm{F}}$, $\kappa_{\mathrm{V}}$ \\
      Fermion and vector couplings, per decay mode         & 76.7\% (8.2) &  12 & \mbox{$\kappa^{\bb}_{\mathrm{F}}$}, \mbox{$\kappa^{\Pgt\Pgt}_{\mathrm{F}}$}, \mbox{$\kappa^{\mu\mu}_{\mathrm{F}}$}, \mbox{$\kappa^{\PW\PW}_{\mathrm{F}}$}, \mbox{$\kappa^{\cPZ\cPZ}_{\mathrm{F}}$}, \mbox{$\kappa^{\PGg\PGg}_{\mathrm{F}}$}, \mbox{$\kappa^{\bb}_{\mathrm{V}}$}, \mbox{$\kappa^{\Pgt\Pgt}_{\mathrm{V}}$}, \mbox{$\kappa^{\mu\mu}_{\mathrm{V}}$}, \mbox{$\kappa^{\PW\PW}_{\mathrm{V}}$}, \mbox{$\kappa^{\cPZ\cPZ}_{\mathrm{V}}$}, \mbox{$\kappa^{\PGg\PGg}_{\mathrm{V}}$}   \\
      Up vs. down-type couplings            & 25.5\% (4.06) & 3 & $\lambda_{\mathrm{V}\PQu}$, $\lambda_{\PQd\PQu}$, $\kappa_{\PQu\PQu}$\\
      Lepton vs. quark couplings            & 27.2\% (3.91) & 3 & $\lambda_{\text{l}\Pq}$, $\lambda_{\mathrm{V}\Pq}$, $\kappa_{\Pq\Pq}$\\
      \hline
\end{tabular}
}
\end{table*}

\section{Constraints on benchmark two Higgs doublet models}
\label{sec:2hdm}

The generic models described in Section~\ref{ssec:L2} can also be interpreted in the context of explicit
benchmark BSM models that contain a second Higgs doublet (2HDM)~\cite{BRANCO20121,PhysRevD.67.075019,MAIANI2013274}.
Only models with CP conservation and a discrete $\mathbb{Z}_{2}$ symmetry to prevent tree-level flavor changing neutral currents are considered.
Under these assumptions, four 2HDM types are possible, referred to as Types I, II, III, and IV.
Each of these 2HDMs contain seven free parameters.
Under the additional assumption that the Higgs boson with a mass of 125.09\GeV is
the lightest CP-even, neutral Higgs boson in the extended Higgs sector,
the predicted rates for its production and decay
are sensitive at leading order to only two 2HDM parameters: the angles $\alpha$ and $\beta$ that diagonalize the mass-squared matrices of the scalars and pseudoscalars.
These two parameters are conventionally substituted by $\cos(\beta-\alpha)$ and $\tan\beta$, without loss of generality.
In all of the 2HDMs, the coupling of the Higgs boson to vector bosons is modified by a factor $\sin(\beta-\alpha)$. The 2HDM types
differ in how the fermions couple to the Higgs doublets. In the Type I model, all fermions couple to just one of the Higgs doublets.
In Type II, the up-type fermions couple to one of the Higgs doublets, while the down-type fermions and the right-handed leptons
couple to the second. In Type III 2HDM, also referred to as ``lepton-specific'', the quarks couple to one of the Higgs
doublets and the right-handed leptons couple to the other. In the Type IV 2HDM, also referred to as ``flipped'',
the up-type fermions and right-handed leptons couple to one of the Higgs doublets, while the down-type quarks couple to the other.
Table~\ref{tab:LO-relations} shows the relation between the coupling modifiers to vector bosons, quarks and leptons and the 2HDM model parameters.

{\tolerance=10000 The minimal supersymmetric standard model (MSSM)~\cite{Djouadi:2005gj,GUNION19861,HABER198575,NILLES19841}
is a specific example of a 2HDM of
Type II that includes additional particle content compared to the SM.
The additional strong constraints given by the nontrivial fermion-boson symmetry fix all mass
relations between the Higgs bosons and the angle
$\alpha$, at tree-level, leaving only two free parameters to fully constrain the MSSM Higgs sector,
usually chosen to be $m_\PSA$ and $\tan\beta$. The hMSSM scenario~\cite{Djouadi:2013uqa,Djouadi:2015jea},
in particular, is an effective MSSM model, trading
the precise knowledge of $m_{\PH}$ against unknown higher-order corrections such that $m_{\PH}=125.09\GeV$ across the $m_\PSA$, $\tan\beta$ parameter space. Another requirement of the scenario is that
$\PH$ be identified as the lightest of the two neutral scalar Higgs bosons. Furthermore, one also obtains
relatively simple relations between $m_\PSA$, $\tan\beta$, and the Higgs boson coupling
modifiers~\cite{hMSSM:Aad2015}, which are shown in Table~\ref{tab:LO-relations} and completed by Eqs.~(\ref{eq:hMSSM-couplings1}) and (\ref{eq:hMSSM-couplings2}).
Although many other MSSM benchmark models have also been defined~\cite{mssmBenchmarks}, the lack of analytic expressions for the
Higgs boson couplings renders these models technically more challenging to consider and they are therefore beyond the scope of this paper.\par}

\begin{table*}[htbp]
    \centering
    \topcaption{
      Modifications to the couplings of the Higgs bosons to up-type ($\kappa_{\PQu}$) and down-type
      ($\kappa_{\PQd}$) fermions, and vector bosons ($\kappa_{\mathrm{V}}$), with respect to the SM expectation, in
      2HDM and for the hMSSM. The coupling modifications for the hMSSM are completed by the expressions
      for $s_{\PQu}$ and $s_{\PQd}$, as given by Eqs.~(\ref{eq:hMSSM-couplings1}) and (\ref{eq:hMSSM-couplings2}).}
      \label{tab:LO-relations}
    \begin{tabular}{lccccc}
      \hline
        & \multicolumn{4}{c}{{\bf 2HDM}}  & {\bf hMSSM} \\
        & Type I & Type II & Type III & Type IV & \\
        \hline
        $\kappa_{\mathrm{V}}$ & $\sin(\beta-\alpha)$ & $\sin(\beta-\alpha)$ & $\sin(\beta-\alpha)$ & $\sin(\beta-\alpha)$ & $\frac{s_{\PQd}+s_{\PQu}\tan\beta }{\sqrt{1+\tan^{2}\beta}}$ \\
        $\kappa_{\PQu}$ & $\cos(\alpha)/\sin(\beta)$ & $\cos(\alpha)/\sin(\beta)$ & $\cos(\alpha)/\sin(\beta)$ & $\cos(\alpha)/\sin(\beta)$ & $s_{\PQu} \frac{\sqrt{1+\tan^{2}\beta}}{\tan\beta}$ \\
        $\kappa_{\PQd}$ & $\cos(\alpha)/\sin(\beta)$ & $-\sin(\alpha)/\cos(\beta)$ & $\cos(\alpha)/\sin(\beta)$ & $-\sin(\alpha)/\cos(\beta)$ & $s_{\PQd} \sqrt{1+\tan^{2}\beta}$ \\
        $\kappa_{\text{l}}$ & $\cos(\alpha)/\sin(\beta)$ & $-\sin(\alpha)/\cos(\beta)$ & $-\sin(\alpha)/\cos(\beta)$ & $\cos(\alpha)/\sin(\beta)$ & $s_{\PQd} \sqrt{1+\tan^{2}\beta}$ \\
        \hline
    \end{tabular}
\end{table*}

\begin{eqnarray}
  s_{\PQu}  =  \frac{1}{\sqrt{1+ \frac{ (m_{\PSA}^{2}+m_{\cPZ}^{2})^{2}\tan^{2}\beta  }  { (m_{\cPZ}^{2}+m_{\PSA}^{2}\tan^{2}\beta - m_{\PH}^{2}(1+\tan^{2}\beta))^{2} } } }\,   \label{eq:hMSSM-couplings1}
\\
  s_{\PQd}  =  s_{\PQu} \frac{ (m_{\PSA}^{2}+m_{\cPZ}^{2}) \tan\beta }{m_{\cPZ}^{2}+m_{\PSA}^{2}\tan^{2}\beta - m_{\PH}^{2}(1+\tan^{2}\beta)}\,  \label{eq:hMSSM-couplings2}
\end{eqnarray}

To set constraints on the 2HDM model parameters, 3-dimensional likelihood scans of the parametrizations described in
Section~\ref{ssec:L2} (with necessary modifications to the lepton coupling modifiers to describe the Type IV 2HDM) are performed.
A test-statistic is then defined, for example in the Types I, II and hMSSM scenarios,
\begin{equation}
q(\lambda_{\PQd\PQu},\lambda_{\mathrm{V}\PQu},\kappa_{\PQu\PQu}) = -2\ln \left( \frac{{L}(\lambda_{\PQd\PQu},\lambda_{\mathrm{V}\PQu},\kappa_{\PQu\PQu}) }{{L}(\hat{\lambda}_{\PQd\PQu},\hat{\lambda}_{\mathrm{V}\PQu},\hat{\kappa}_{\PQu\PQu}) }\right),
\end{equation}
where $\hat{\lambda}_{\PQd\PQu},\hat{\lambda}_{\mathrm{V}\PQu},\hat{\kappa}_{\PQu\PQu}$ are the values of the POIs that maximize
the likelihood.
An interpolation scheme is used to determine the value of $q$ as a function of $\cos(\beta-\alpha)$ and $\tan\beta$,
or $m_\PSA$ and $\tan\beta$, for the Types I and II, or hMSSM scenarios, respectively, using the
relations in Table~\ref{tab:LO-relations}.

A second quantity $q\prime$ is defined as,
\begin{equation}
q\prime = -2\ln \left( \frac{L(\hat{\lambda}_{\PQd\PQu},\hat{\lambda}_{\mathrm{V}\PQu},\hat{\kappa}_{\PQu\PQu})}
{L_{\mathrm{max}}}\right),
\end{equation}
where $L_{\mathrm{max}}$ is the maximum likelihood value attained in the planes of $\cos(\beta-\alpha)$--$\tan\beta$,
or $m_\PSA$--$\tan\beta$.
The allowed regions are determined as the points in each plane for which the difference between $q$ and $q\prime$ ($\Delta q$) is less than 5.99.
This value corresponds to the 95\% confidence region assuming $\Delta q$ is distributed as a $\chi^{2}$ function with 2 degrees of freedom.
A similar procedure is performed using the
model with the parameters $\lambda_{\text{l}\Pq}$, $\lambda_{\mathrm{V}\Pq}$ and $\kappa_{\Pq\Pq}$, to determine the
allowed region for the Type III scenario.

\begin{figure*}[hbtp]
\centering
\includegraphics[width=0.42\textwidth]{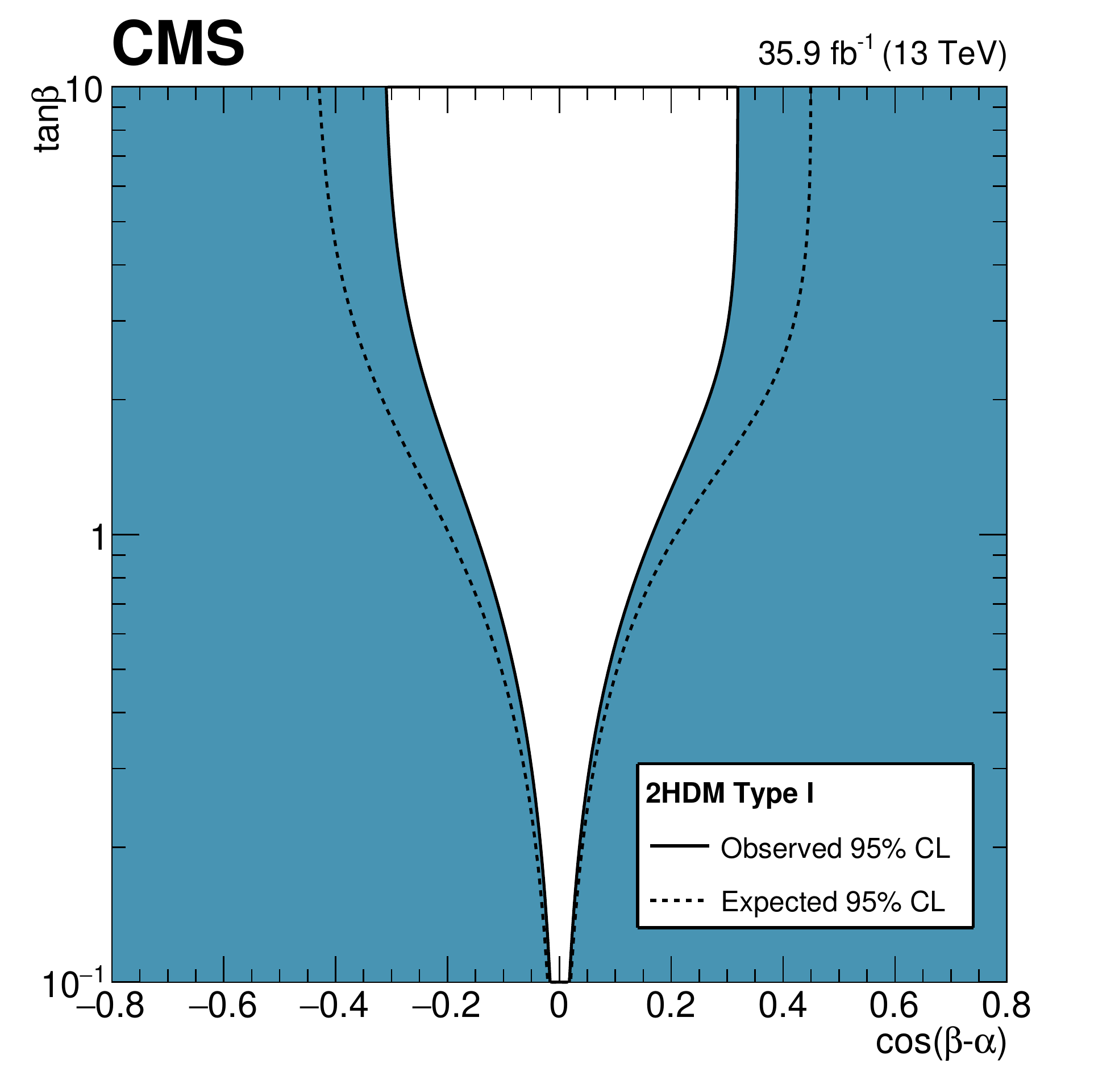}
\includegraphics[width=0.42\textwidth]{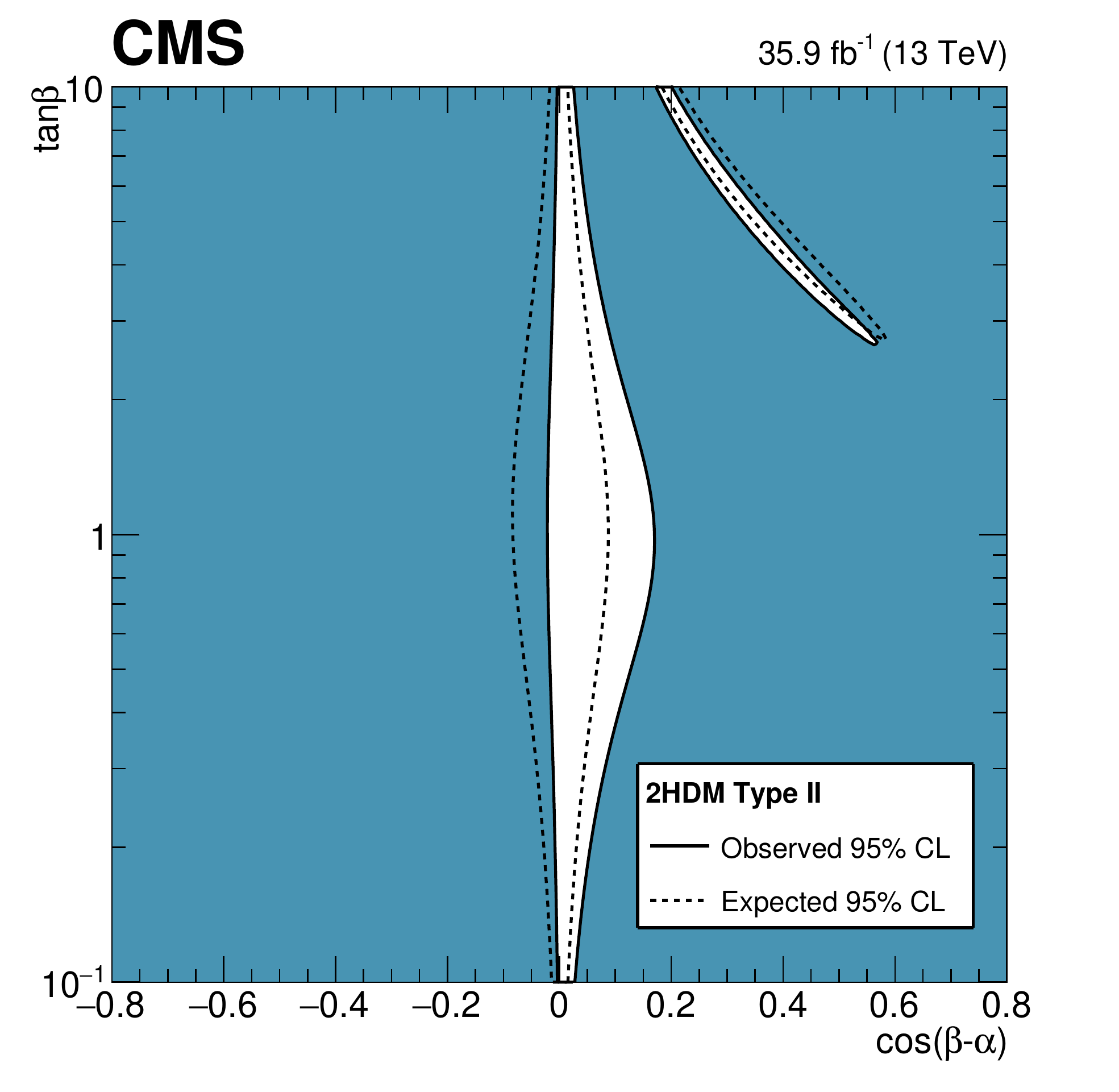} \\
\includegraphics[width=0.42\textwidth]{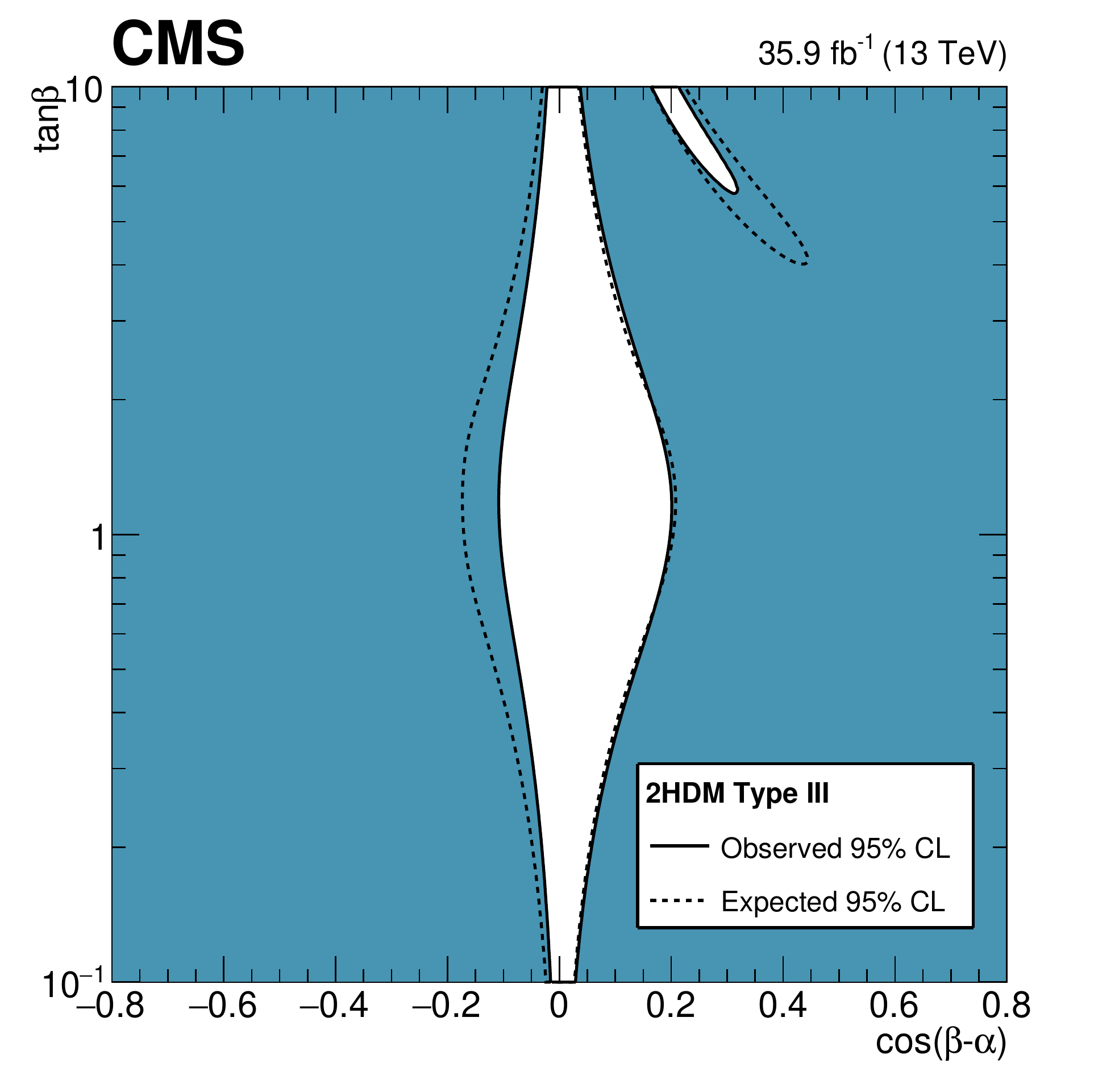}
\includegraphics[width=0.42\textwidth]{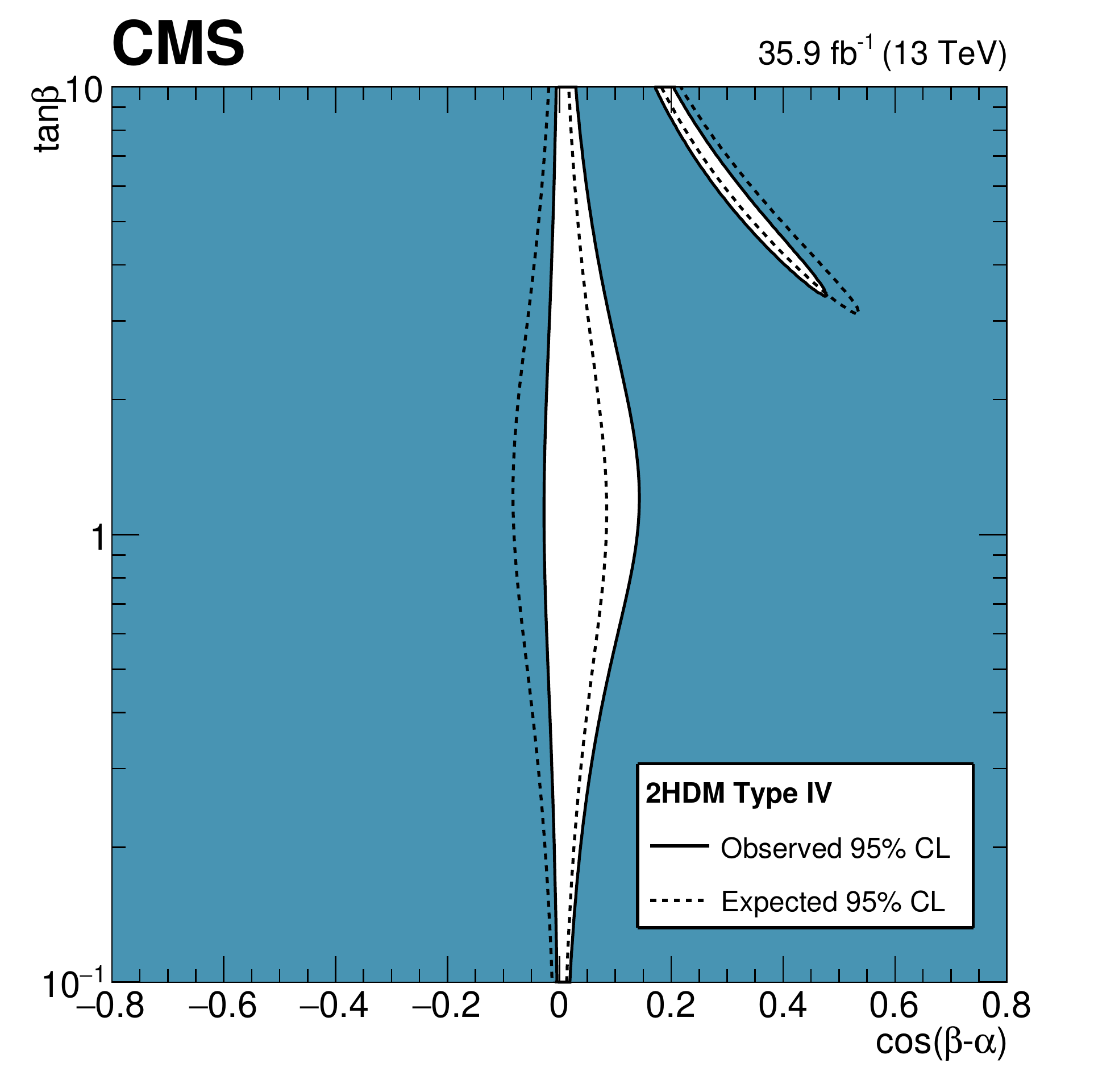} \\
\includegraphics[width=0.42\textwidth]{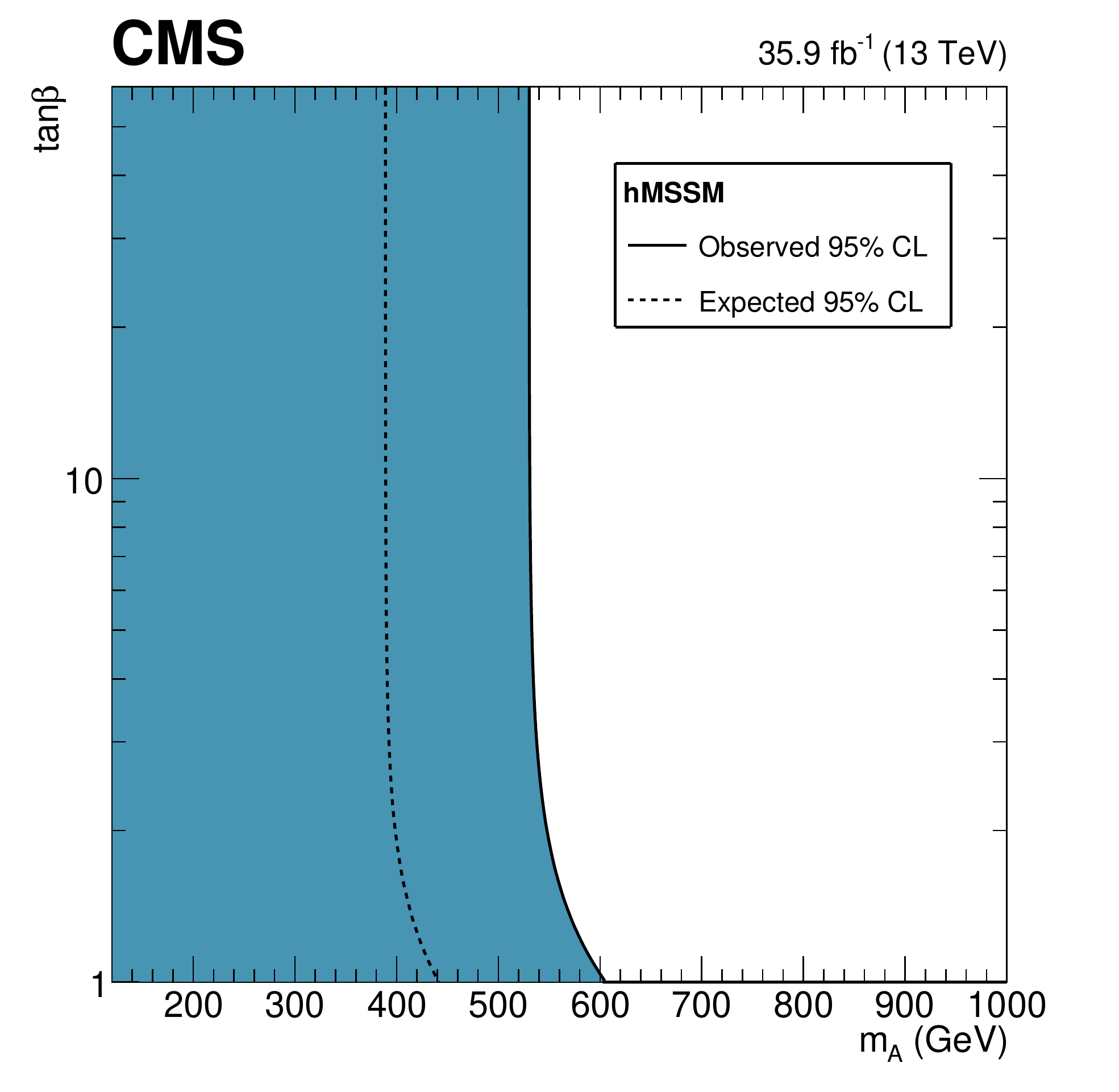}
\caption{Constraints in the $\cos({\beta-\alpha})$ vs. $\tan\beta$ plane for the Types I, II, III, and IV 2HDM, and constraints in the $m_\PSA$ vs. $\tan\beta$ plane for the hMSSM.
The white regions, bounded by the solid black lines, in each plane represent the regions
of the parameter space that are allowed at the 95\% \CL, given the data observed.
The dashed lines indicate the boundaries of the allowed regions expected for the SM Higgs boson.}
\label{fig:scan_2HDM}
\end{figure*}

Figure~\ref{fig:scan_2HDM} shows the results of the fits for the
different 2HDM benchmark scenarios. The lobe features that can be seen in
the Types~II, III, and IV constraints for $\cos(\beta-\alpha)>0$ are due
to negative values of $\kappa_{\PQd}$, $\kappa_{\Pgt}$, and $\kappa_{\PQb}$, which are not excluded with
the current sensitivity. In all of these 2HDM models, the Higgs boson couplings are
the same as those predicted in the SM for $\cos(\beta-\alpha)=0$.

The results for the hMSSM scenario are also shown in Fig.~\ref{fig:scan_2HDM}.
The constraints observed are more stringent than those expected under the SM.
This is due to the best fit value of $\lambda_{\PQd\PQu}$ being smaller than 1,
while in the hMSSM for $\tan\beta>1$, $\lambda_{\PQd\PQu}$ is strictly greater than 1
and asymptotically approaches unity only at large $m_{\PSA}$.
Therefore the observed data disfavors small values of $m_{\PSA}$, leading to the stronger constraint.

The constraints in the 2HDM and hMSSM scenarios are complementary to those obtained from direct
searches for additional Higgs bosons~\cite{Aaboud:2017gsl,Aaboud:2017sjh,Aaboud:2016dig,Khachatryan:2014wca,Chatrchyan:2013yoa}.

\section{Summary}
\label{sec:summary}

A set of combined measurements of Higgs boson production and decay rates has been presented,
along with the consequential constraints placed on its couplings to standard model (SM) particles, and on the parameter spaces
of several beyond the standard model (BSM) scenarios. The combination is based on analyses targeting the gluon fusion and
vector boson fusion production modes, and associated production with a vector
boson or a pair of top quarks. The analyses included in the combination target Higgs boson production
in the $\PH \to \cPZ\cPZ,\,\PW\PW,\,\gamgam,\,\tautau$, $\bb$, and $\mumu$ decay channels, using $13\TeV$ proton-proton collision data collected in 2016 and corresponding
to an integrated luminosity of 35.9\fbinv.
Additionally, searches for invisible Higgs boson decays are included to increase the sensitivity to potential interactions with BSM particles.

Measurements of the Higgs boson production cross section times branching fractions are presented, along with a generic
parametrization in terms of ratios of production cross sections and branching fractions, which makes no assumptions about the Higgs boson total width.
The combined signal yield relative to the SM prediction has been measured as $1.17\pm0.10$ at $m_{\PH} = 125.09\GeV$. An improvement in the measured
precision of the gluon fusion production
rate of around $\sim$50\% is achieved compared to previous ATLAS and CMS measurements.
Additionally, a set of fiducial Higgs boson cross sections, in the context of the simplified template cross section framework, is presented for the first time from a combination
of six decay channels. Furthermore, interpretations are provided in the context of a leading-order coupling modifier framework,
including variants for which effective couplings to the photon and gluon are introduced. All of the results presented are compatible with the SM prediction.
The invisible (undetected) branching fraction of the Higgs boson is constrained to be less than 22 (38\%) at 95\%~Confidence Level. The results are additionally interpreted in two
BSM models, the minimal supersymmetric model and the generic two Higgs doublet model. The constraints placed on the parameter spaces of these models are
complementary to those that can be obtained from direct searches for additional Higgs bosons.

\begin{acknowledgments}

\hyphenation{Bundes-ministerium Forschungs-gemeinschaft Forschungs-zentren Rachada-pisek} We congratulate our colleagues in the CERN accelerator departments for the excellent performance of the LHC and thank the technical and administrative staffs at CERN and at other CMS institutes for their contributions to the success of the CMS effort. In addition, we gratefully acknowledge the computing centres and personnel of the Worldwide LHC Computing Grid for delivering so effectively the computing infrastructure essential to our analyses. Finally, we acknowledge the enduring support for the construction and operation of the LHC and the CMS detector provided by the following funding agencies: the Austrian Federal Ministry of Science, Research and Economy and the Austrian Science Fund; the Belgian Fonds de la Recherche Scientifique, and Fonds voor Wetenschappelijk Onderzoek; the Brazilian Funding Agencies (CNPq, CAPES, FAPERJ, and FAPESP); the Bulgarian Ministry of Education and Science; CERN; the Chinese Academy of Sciences, Ministry of Science and Technology, and National Natural Science Foundation of China; the Colombian Funding Agency (COLCIENCIAS); the Croatian Ministry of Science, Education and Sport, and the Croatian Science Foundation; the Research Promotion Foundation, Cyprus; the Secretariat for Higher Education, Science, Technology and Innovation, Ecuador; the Ministry of Education and Research, Estonian Research Council via IUT23-4 and IUT23-6 and European Regional Development Fund, Estonia; the Academy of Finland, Finnish Ministry of Education and Culture, and Helsinki Institute of Physics; the Institut National de Physique Nucl\'eaire et de Physique des Particules~/~CNRS, and Commissariat \`a l'\'Energie Atomique et aux \'Energies Alternatives~/~CEA, France; the Bundesministerium f\"ur Bildung und Forschung, Deutsche Forschungsgemeinschaft, and Helmholtz-Gemeinschaft Deutscher Forschungszentren, Germany; the General Secretariat for Research and Technology, Greece; the National Research, Development and Innovation Fund, Hungary; the Department of Atomic Energy and the Department of Science and Technology, India; the Institute for Studies in Theoretical Physics and Mathematics, Iran; the Science Foundation, Ireland; the Istituto Nazionale di Fisica Nucleare, Italy; the Ministry of Science, ICT and Future Planning, and National Research Foundation (NRF), Republic of Korea; the Lithuanian Academy of Sciences; the Ministry of Education, and University of Malaya (Malaysia); the Mexican Funding Agencies (BUAP, CINVESTAV, CONACYT, LNS, SEP, and UASLP-FAI); the Ministry of Business, Innovation and Employment, New Zealand; the Pakistan Atomic Energy Commission; the Ministry of Science and Higher Education and the National Science Centre, Poland; the Funda\c{c}\~ao para a Ci\^encia e a Tecnologia, Portugal; JINR, Dubna; the Ministry of Education and Science of the Russian Federation, the Federal Agency of Atomic Energy of the Russian Federation, Russian Academy of Sciences, the Russian Foundation for Basic Research and the Russian Competitiveness Program of NRNU ``MEPhI''; the Ministry of Education, Science and Technological Development of Serbia; the Secretar\'{\i}a de Estado de Investigaci\'on, Desarrollo e Innovaci\'on, Programa Consolider-Ingenio 2010, Plan Estatal de Investigaci\'on Cient\'{\i}fica y T\'ecnica y de Innovaci\'on 2013-2016, Plan de Ciencia, Tecnolog\'{i}a e Innovaci\'on 2013-2017 del Principado de Asturias and Fondo Europeo de Desarrollo Regional, Spain; the Swiss Funding Agencies (ETH Board, ETH Zurich, PSI, SNF, UniZH, Canton Zurich, and SER); the Ministry of Science and Technology, Taipei; the Thailand Center of Excellence in Physics, the Institute for the Promotion of Teaching Science and Technology of Thailand, Special Task Force for Activating Research and the National Science and Technology Development Agency of Thailand; the Scientific and Technical Research Council of Turkey, and Turkish Atomic Energy Authority; the National Academy of Sciences of Ukraine, and State Fund for Fundamental Researches, Ukraine; the Science and Technology Facilities Council, UK; the US Department of Energy, and the US National Science Foundation.

Individuals have received support from the Marie-Curie programme and the European Research Council and Horizon 2020 Grant, contract No. 675440 (European Union); the Leventis Foundation; the A. P. Sloan Foundation; the Alexander von Humboldt Foundation; the Belgian Federal Science Policy Office; the Fonds pour la Formation \`a la Recherche dans l'Industrie et dans l'Agriculture (FRIA-Belgium); the Agentschap voor Innovatie door Wetenschap en Technologie (IWT-Belgium); the F.R.S.-FNRS and FWO (Belgium) under the ``Excellence of Science - EOS'' - be.h project n. 30820817; the Ministry of Education, Youth and Sports (MEYS) of the Czech Republic; the Lend\"ulet (``Momentum'') Programme and the J\'anos Bolyai Research Scholarship of the Hungarian Academy of Sciences, the New National Excellence Program \'UNKP, the NKFIA research grants 123842, 123959, 124845, 124850 and 125105 (Hungary); the Council of Scientific and Industrial Research, India; the HOMING PLUS programme of the Foundation for Polish Science, cofinanced from European Union, Regional Development Fund, the Mobility Plus programme of the Ministry of Science and Higher Education, the National Science Center (Poland), contracts Harmonia 2014/14/M/ST2/00428, Opus 2014/13/B/ST2/02543, 2014/15/B/ST2/03998, and 2015/19/B/ST2/02861, Sonata-bis 2012/07/E/ST2/01406; the National Priorities Research Program by Qatar National Research Fund; the Programa de Excelencia Mar\'{i}a de Maeztu and the Programa Severo Ochoa del Principado de Asturias; the Thalis and Aristeia programmes cofinanced by EU-ESF and the Greek NSRF; the Rachadapisek Sompot Fund for Postdoctoral Fellowship, Chulalongkorn University and the Chulalongkorn Academic into Its 2nd Century Project Advancement Project (Thailand); the Welch Foundation, contract C-1845; and the Weston Havens Foundation (USA).
\end{acknowledgments}

\bibliography{auto_generated}

\cleardoublepage \appendix\section{The CMS Collaboration \label{app:collab}}\begin{sloppypar}\hyphenpenalty=5000\widowpenalty=500\clubpenalty=5000\vskip\cmsinstskip
\textbf{Yerevan Physics Institute, Yerevan, Armenia}\\*[0pt]
A.M.~Sirunyan, A.~Tumasyan
\vskip\cmsinstskip
\textbf{Institut f\"{u}r Hochenergiephysik, Wien, Austria}\\*[0pt]
W.~Adam, F.~Ambrogi, E.~Asilar, T.~Bergauer, J.~Brandstetter, M.~Dragicevic, J.~Er\"{o}, A.~Escalante~Del~Valle, M.~Flechl, R.~Fr\"{u}hwirth\cmsAuthorMark{1}, V.M.~Ghete, J.~Hrubec, M.~Jeitler\cmsAuthorMark{1}, N.~Krammer, I.~Kr\"{a}tschmer, D.~Liko, T.~Madlener, I.~Mikulec, N.~Rad, H.~Rohringer, J.~Schieck\cmsAuthorMark{1}, R.~Sch\"{o}fbeck, M.~Spanring, D.~Spitzbart, A.~Taurok, W.~Waltenberger, J.~Wittmann, C.-E.~Wulz\cmsAuthorMark{1}, M.~Zarucki
% ...
\vskip\cmsinstskip
% ...
1:  Also at Vienna University of Technology, Vienna, Austria\\
% ...
\end{sloppypar}
\end{document}